\documentclass[twocolumn,showpacs]{revtex4}
\usepackage{bm}
\usepackage{latexsym}
\usepackage{dcolumn}
\usepackage{amsfonts,amssymb}
\usepackage{graphicx,epsfig}
\usepackage{psfrag}
\usepackage{amsmath}
\begin{document}

\def\apjl{Astrophys. J. Lett.}
\def\mnras{Mon. Not. Roy. Astron. Soc.}
\def\mnrasl{Mon. Not. Roy. Astron. Soc. Lett.}
\def\physrep{Phys. Rept.}
\def\apjs{Astrophys. J. Suppl.}
\def\aap{Astron. Astrophys.}
\def\araa{Annu. Rev. Astron. and Astrophys.}
\def\aj{Astron. J.}

\title{Observational constraints on low redshift evolution of dark energy:
  How consistent are different observations?} 

\author{H.~K.~Jassal $^a$, J.~S.~Bagla$^b$ and T.~Padmanabhan$^c$} 

\email[$^a$]{Email: hkj@hri.res.in}

\email[$^b$]{Email: jasjeet@hri.res.in}

\email[$^c$]{Email: nabhan@iucaa.ernet.in}

\affiliation{$^{a,b}$ Harish-Chandra Research Institute, Chhatnag Road,\\
Jhunsi, Allahabad 211 019, India.\\
$^{c}$Inter University Centre for Astronomy and Astrophysics,\\
Post Bag 4, Ganeshkhind, Pune 411~007, India. \\  
}

\begin{abstract}
The dark energy component of the universe is often interpreted either in terms
of a cosmological constant or as a scalar field. 
A generic feature of the scalar field models is that the equation of state
parameter $w\equiv P/\rho$ for the dark energy need not satisfy $w=-1$ and,
in general, it can be a function of time. 
Using the Markov chain Monte Carlo method we perform a critical analysis
of the cosmological parameter space, allowing for a varying $w$. 
We use constraints on $w(z)$ from the observations of high redshift supernovae 
(SN), the WMAP observations of CMB anisotropies and abundance of rich clusters
of galaxies.  
For models with a constant $w$, the $\Lambda$CDM model is allowed with a
probability of about $6\%$ by the SN observations while it is allowed
with a probability of $98.9\%$ by WMAP observations. 
The $\Lambda$CDM model is allowed even within the context of models with
variable $w$: 
WMAP observations allow it with a probability of $99.1\%$  whereas
SN data  allows it with $23\%$ probability.
The SN data, on its own, favors phantom like equation of state ($w<-1$) and
high values for $\Omega_{NR}$.  
It does not distinguish between constant $w$ (with $w<-1$) models and those
with varying $w(z)$ in a statistically significant manner. 
The SN data allows a very wide range for variation of dark energy density,
e.g., a variation by factor ten in the dark energy density between $z=0$ and
$z=1$ is allowed at $95\%$ confidence level. 
WMAP observations provide a better constraint and the corresponding allowed
variation is less than a factor of three. 
Allowing for variation in $w$ has an impact on the values
for other cosmological parameters in that the allowed range often becomes
larger.   
There is significant tension between SN and WMAP observations; the
best fit model for one is often ruled out by the other at a very high
confidence limit. 
Hence results based on only one of these can lead to unreliable conclusions.  
Given the divergence in models favored by individual observations, and the
fact that the best fit models are ruled out in the combined analysis, there is
a distinct possibility of the existence of systematic errors which are not
understood.   
\end{abstract}


\maketitle

\section{Introduction}

Observational evidence for accelerated expansion in the universe has been
growing in the last two decades
\cite{crisis2}.
Observations of high redshift supernovae
\cite{nova_data1,nova_data3} provided an independent confirmation.
Using these along with observations of cosmic microwave
background radiation (CMB) \cite{boomerang,wmap_params} and large scale
structure \cite{2df,sdss}, we can construct a ``concordance'' model for
cosmology and study variations around it (e.g., see
\cite{wmap_params,2003Sci...299.1532B,2004PhRvD..69j3501T}; for an overview of
our current understanding, see \cite{TPabhay}). 

Observations indicate that dark energy should have an equation of state
parameter $w\equiv P/\rho < -1/3$ for the universe to undergo accelerated
expansion. 
Indeed, observations show that dark energy is the dominant component of our
universe. 
The cosmological constant is the simplest explanation for accelerated
expansion \cite{ccprob,review3} and it is known to be consistent with
observations.   
In order to avoid theoretical problems related to cosmological constant
\cite{ccprob}, other scenarios have been investigated.  
In these models one can have $w \ne -1$ and in general $w$ varies with
redshift.  
These models include quintessence \cite{quint1}, k-essence
\cite{2001PhRvD..63j3510A}, tachyons \cite{tachyon1,2003PhRvD..67f3504B},
phantom fields \cite{2002PhLB..545...23C}, branes \cite{brane1}, etc. 
There are also some phenomenological models \cite{water}, field theoretical
and renormalisation group based models  (see e.g. \cite{tp173}), models that
unify dark matter and dark energy \cite{unified_dedm1} and many others like
those based on horizon thermodynamics (e.g. see \cite{2005astro.ph..5133S}).   
Even though these models have been proposed to overcome the fine tuning
problem for cosmological constant, most of these  require similar fine tuning
of parameter(s) to be consistent with observations. 
Nevertheless, they raise the possibility of $w(z)$ evolving with time (or being
different from $-1$), which --- in principle --- can be tested by 
observations. 

Given that $w < -1/3 $ for dark energy for the universe to undergo accelerated
expansion, the energy density of this component changes at a much slower rate
than that of matter and radiation.  
Indeed, $w=-1$ for cosmological constant and in this case the energy density
is a constant.
Unless $w$ is a rapidly varying function of redshift and becomes $w \sim 0$ at
($z \leq 1$), the energy density of the dark energy component should be
negligible at high redshifts ($z \gg 1$) as compared to that of
non-relativistic matter.   
If dark energy evolves in a manner such that its energy density is comparable 
to, or greater than the matter density in the universe at high redshifts then
the basic structure of the cosmological model needs to be modified. 
We do not consider such models here.
We confine our attention to models with dark energy density being an
insignificant component of the universe at $z \gg 1$ and choose observations
which are sensitive to evolution of $w(z)$ at redshifts $z \lesssim  1$.  

To put the present work in context, we recall that combining supernova
observations with the WMAP data provides strong constraints on the variation
of dark energy density \cite{2005MNRAS.356L..11J}. 
(A review of relevant observations for constraining dark energy models along
with a summary of the previous work in this area is given in Section
\ref{sec:prevwork}.)  
Reproducing the location of acoustic features requires the angular diameter
distance to the last scattering surface to be in the correct range. 
This analysis showed that while the data from SN observations allows for a
large range in parameters of dark energy, combining with WMAP data limits this
range significantly.   
However, in that work, we did not explore the cosmological parameter space
widely and had fixed nearly all parameters other than those used to describe
evolution of dark energy.  
In the present work, we allow many cosmological parameters to vary
and include constraints from cluster abundance in addition to the supernova
and WMAP constraints.  

In addition to obtaining quantitative bounds on parameters in different
contexts, we address the following key issues in this paper: 
\begin{itemize}
\item
Does allowing cosmological parameters to vary weaken the constraints on
variation of dark energy?  
\item
Conversely, how does the allowed range for different cosmological parameters
change when we allow for a epoch  dependent $w(z)$ ? 
\item
Do the observational constraints agree with each other?  
In particular, what kind of cosmological models are preferred by SN and WMAP
observations individually ? 
\end{itemize}
The last point is important and requires elaboration. 
Different observational sets are combined together precisely because these
observations are sensitive to different combinations of cosmological
parameters and facilitate in breaking degeneracies between parameters.   
If we consider $\Lambda$CDM models then the SN observations, for example,
broadly depend on the combination ($0.85\Omega_{NR} -0.53 \Omega_{V}$)
\cite{2005A&A...429..807C} while WMAP is sensitive to ($\Omega_{NR} +
\Omega_{V})$ \cite{wmap_params}, a feature which was originally highlighted in
the literature as `cosmic complementarity'.    
Therefore, we cannot expect constraints from different observations to agree
over the entire parameter space.  
At the same time, we do not expect models favored by one observation to be
ruled out by another when such a divergence is not expected.
This divergence may point to some shortcomings in the model, or to
systematic errors in observations, or even to an incorrect choice of priors. 
If all observational sets are consistent then we should be able to derive
similar constraints using subsets of observations, even though the final
constraints may not be as tight as with the full set of observations.

In order to address the questions listed above in a systematic manner, we
proceed in three steps. 
We choose a `base' reference model with cold dark matter and cosmological
constant, with neutrinos contributing a negligible amount to the energy
density of the Universe. 
We assume that the Universe is flat and restrict ourselves to an unbroken
power law for the primordial power spectrum of density fluctuations and we
assume that the perturbations are adiabatic.
Another assumption is that the perturbations in tensor mode are
negligible and we take $r=0$ \cite{2005PhRvD..71j3515S}.
We choose this to be our standard model as this can be described by a compact
set of parameters.  

Next, we generalize from $\Lambda$CDM models ($w=-1$) to
study a wider class of dark energy models with a \textit{constant}~$w$ and
address the issues listed above.      
In this case, we also study the effect of perturbations in dark energy.
Finally, we generalize to models in which $w$ is allowed to vary with
$z$ in a parameterized form.   
This approach allows us to delineate changes that come about from choosing a
constant $w \neq -1$, from those allowed by a varying dark energy.  
We do not impose theoretically motivated constraints on models, e.g. we do
\textit{not} require $w \geq -1$ as the present work is focused on
understanding the nature of models favored by observations.

The paper is organized as follows: 
In section \ref{sec:bkground} we discuss the background cosmological equations
followed by a brief review on the various observation used to constrain dark
energy equation of state and the observations we concentrate on. 
The Markov Chain Monte Carlo method is discussed in section \ref{sec:mcmc} and
detailed results are presented in section \ref{sec:details}.  
We conclude with a discussion of the results and future prospects for
constraining dark energy models in section \ref{sec:conclude}.


\section{Theoretical background}\label{sec:bkground}

\subsection{Cosmological equations}

If we assume that each of the constituents of the homogeneous and isotropic
universe can be considered to be an ideal fluid, and  that the
space is flat, the Friedman equations become:
\begin{eqnarray}
\left(\frac{\dot a}{a}\right)^2 &=& \frac{8 \pi G}{3} \rho \\
\frac{\ddot{a}}{a} &=& -\frac{4 \pi G}{3} (\rho + 3P)
\end{eqnarray}
where $P$ is the pressure and $\rho = \rho_{NR}+\rho_{\gamma} + \rho_{_{\rm
 DE}}$ with the respective terms denoting energy densities for nonrelativistic
 matter, for radiation/relativistic matter and for dark energy. 
Pressure is zero for the non-relativistic component, whereas radiation and
relativistic matter have $P_\gamma = \rho_\gamma / 3$.  
If the cosmological constant is the source of acceleration then $\rho_{_{\rm
 DE}} = $~constant and $P_{_{\rm DE}} = - \rho_{_{\rm DE}}$.  
 
An obvious generalization is to consider models with a constant equation of
state  parameter $w \equiv P/\rho = $~\textit{constant}.    
One can, in fact, further generalize to models with a varying equation of
state parameter $w(z)$.
Since a function is equivalent to an infinite set of numbers (defined e.g. by
a Taylor-Laurent series coefficients), it is clearly not possible to constrain 
the form of an arbitrary function $w(z)$ using finite number of observations. 
One possible way of circumventing this issue is to parameterize the function
$w(z)$ by a finite number of parameters and try to constrain these parameters
by observations. 
There have been many attempts to describe varying dark energy with different
parameterizations
\cite{constraints_2,2004ApJ...617L...1B,2005MNRAS.356L..11J,constraints_10,holographic,Hannestad:2004cb}  
where the functional form of $w(z)$ is fixed and the variation is described
with a small number of parameters.   
Observational constraints depend on the specific parameterization chosen, but
it should be possible to glean some parameterization independent results from
the analysis.     

To model varying dark energy we use two parameterizations
\begin{equation}
w(z) = w_0 +  w'(z=0) \frac{z}{(1+z)^p} ~ ; ~~~~~~~~ p=1,~2 \label{taylor}
\end{equation}
These are chosen so that, among other things, the high redshift behavior is
completely different in these two parameterizations
\cite{2005MNRAS.356L..11J}.  
If $p=1$ \cite{p1}, the asymptotic value $w(\infty) = w_0 + w'(z=0)$ and  for
$p=2$, $w(\infty) = w_0$. 
For both $p=1, 2$, the present value $w(0)=w_0$.
Clearly, we must have $w(z \gg 1) \leq -1/3$ for the standard cosmological
models with a hot big bang to be valid. 
This restriction is imposed over and above the priors used in our study. 

Integrating $d(\rho a^3)=-w(z)\rho da^3$, the energy density can be expressed
as  
\begin{equation}
\frac{\rho_{_{DE}}}{\rho_{_{DE_0}}} = \left( 1 + z
\right)^{3\left(1+w_0+w'_0\right)} \exp{\left[-\frac{3 w'_0)
      z}{1+z}\right]}  
\end{equation}
for  $p=1$ (in Eq.~\ref{taylor}) and
\begin{equation}
\frac{\rho_{_{DE}}}{\rho_{_{DE_0}}} = \left( 1 + z \right)^{
      3\left(1+w_0\right)} \exp{\left[\frac{3 w'_0}{2}
      \left(\frac{z}{1+z}\right)^2 \right]}  
\end{equation}
for $p=2$.
The allowed range of parameters $w_0$ and $w'_0\equiv w'(z=0)$ is likely to be
different for different $p$. 
However, the allowed variation at low redshifts in $\rho_{_{DE}}$ should
be similar in both models as observations actually probe the variation of dark
energy density. 
Indeed, in an earlier study \cite{2005MNRAS.356L..11J} where we had studied a
restricted class of models, we found this to be the case.   
For example, $\rho_{_{DE}}$ can vary by at most a factor two up to $z=2$ when
both the WMAP and SN data are taken into account \cite{2005MNRAS.356L..11J}.  
This reaffirms the expectation that the results are parameterization
independent at some level. 

\subsection{Observational constraints} \label{sec:prevwork}

In this subsection, we briefly review potential observational constraints on 
dark energy and we also summarize previous work in this area.

Constraints on dark energy models essentially arise as follows: To begin with,
dark energy affects the rate of expansion of the universe and thus the
luminosity distance and also the angular diameter distance.  
Constraints from observations of high redshift supernovae
\cite{1998ApJ...509...74G,nova_data1,dynamic_de1,2005A&A...429..807C,2004MNRAS.354..275A,dynamic_de5}
and the location of peaks in the angular power spectrum of CMB anisotropies
mainly use this feature \cite{2004PhRvD..70j3523E}.  
The signature of acoustic peaks in correlation function of galaxies also
provides a similar geometric constraint \cite{2005astro.ph..1171E}. 
There is also an effect on weak lensing statistics through changes in
distance-redshift relation \cite{2002PhRvD..65f3001H,2003ApJ...583..566M,2003MNRAS.341..251W,2003PhRvL..91d1301A,2003PhRvL..91n1302J}. 

Second, the rate of expansion influences the growth of perturbations in the
universe and this leads to another set of probes of dark energy
\cite{2001PhRvD..64h3501B}.   
Abundance of rich clusters of galaxies, their evolution and the integrated
Sachs-Wolfe (ISW) effect belong to this category of constraints, along with
constraints from redshift space distortions
\cite{1999ApJ...517...40B,cluster2,cluster4,2004astro.ph..8252V,2004astro.ph..9207Y,Calvao:2002zr}.
All these constraints are sensitive to different aspects of dark
energy and a combination of all of these can put tight limits on
models.  
The redshift space distortions are a local effect as these are sensitive
to the rate of growth of density perturbations at a given epoch. 
The abundance of rich clusters of galaxies, the ISW effect and distances are
integrated effects in that the effect of dark energy is averaged over a range
of redshifts in some sense. 

Observations of high redshift supernovae of type Ia provide the most
unambiguous evidence for accelerated expansion
\cite{1998ApJ...509...74G,nova_data1,dynamic_de1,2005A&A...429..807C,2004MNRAS.354..275A}.
Assuming these sources to be  standard candles, observations spanning a
range of redshifts can be used to study the change in rate of expansion and
this  imposes  direct constraints on the variation  of dark
energy density. 
Supernovae have been observed up to a redshift of $z_{max} \simeq 1.8$ and
hence can be used to constrain models of dark energy up to this redshift.
Constraints from SN observations alone however, permit a large variation in the
dark energy parameters \cite{2004MNRAS.354..275A} and in particular
favor models with $w<-1$ at the present epoch
\cite{nova_data3,dynamic_de1,2005A&A...429..807C}.   

Baryon oscillations in the matter-radiation fluid prior to decoupling provide
a standard scale and the angle at which the acoustic peaks occur in the angular
power spectrum of temperature anisotropies in the CMB fixes the distance to
the surface of last scattering.   
This provides a useful constraint on models of dark energy
\cite{2004PhRvD..70j3523E} as long as dark energy does not affect the dynamics
of universe at the time of decoupling of matter and radiation. 
Unlike supernovae that are observed over a range of redshifts where dark
energy is dominant, the surface of last scattering is at $z \simeq 1100$.  
However, the exquisite quality of CMB anisotropy measurements makes this a
very useful constraint and these observations offer a constraint that is
different from SN observations \cite{2005MNRAS.356L..11J}.  
Indeed, as we shall see, WMAP and SN observations often favor models that are
mutually unacceptable. 

Recent detection of the baryon acoustic peak in galaxy correlation function
using the luminous red galaxy sample of the SDSS survey has provided an
additional handle to constrain cosmological parameters
\cite{2005astro.ph..1171E}.  
The geometric constraint from these observations can, in principle, constrain
models of dark energy. 
A measurement of the angular scale corresponding to the peak at different
redshifts can indeed be a powerful constraint.

If we consider a given cosmological model that is consistent with observations
of CMB anisotropy then the amplitude of fluctuations at the time of decoupling
is fixed, and its linearly extrapolated value today can be computed using
linear perturbation theory. 
The abundance of rich clusters of galaxies is related to the amplitude of
perturbations in dark matter at a scale of about $8$h$^{-1}$Mpc. 
If we study different models for dark energy while other parameters are not
changed, the abundance of rich clusters constrains the net growth of
structures between the epoch of decoupling and the present epoch
\cite{lss_de}.   

Redshift space distortions due to kinematics and the Alcock-Paczynski effect
are also  potential probes of dark energy 
\cite{1979Natur.281..358A,2004astro.ph..9207Y}. 
Ongoing surveys like the SDSS and future surveys will be able to distinguish
between different dark energy models through these effects
\cite{2003NewAR..47..775W}.   
However, this method does not provide useful constraints at present. 

Dependence of the distance-redshift relation and a different rate of growth
for perturbations, as well as changes in the matter power spectrum are also
reflected in weak lensing statistics.   
Several studies have been carried out on the potential constraints that can be
put on dark energy models from weak lensing observations and their degeneracy
with other parameters. 
These studies indicate that future surveys will be able to put strong
constraints on dark energy models
\cite{2003ApJ...583..566M,2003PhRvL..91d1301A}.    

\begin{figure*}
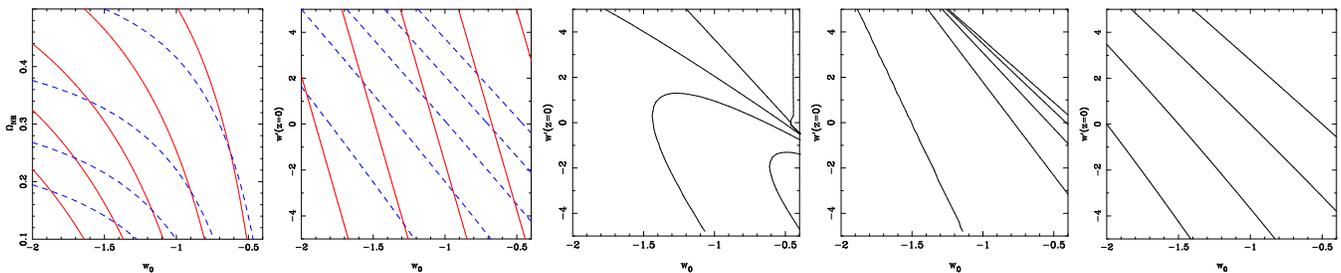

\begin{tabular}{ccccc}
\begin{minipage}{1.35in}
\centering
\includegraphics[width=1.35in]{fig1aa.ps} 
\end{minipage}
&
\begin{minipage}{1.35in}
\centering
\includegraphics[width=1.35in]{fig1ab.ps} 
\end{minipage}
&
\begin{minipage}{1.35in}
\centering
\includegraphics[width=1.35in]{fig1a.ps} 
\end{minipage}
&
\begin{minipage}{1.35in}
\centering
\includegraphics[width=1.35in]{fig1b.ps}
\end{minipage}
&
\begin{minipage}{1.35in}
\centering
\includegraphics[width=1.35in]{fig1c.ps}
\end{minipage}
\\
\end{tabular}
\caption{The left figure in this panel shows contours of constant luminosity
  distance $H_0 D_L(z)/c$ in $\Omega_{NR}-w$ plane. 
The red/solid contours are for luminosity distance at redshift $z=0.17$ and
 blue/dashed ones are for luminosity distance at redshift $z=1.17$.
The red/solid contours from top right to bottom left correspond to values
 between $H_0 D_L(z=0.17)/c =0.19-0.21$ and blue/dashed lines are for
 $H_0 D_L(z=1.17)/c=1.52-2.45 $. 
The other plots show contours in the $w_0-w'(z=0)$ plane.  
The second figure shows contours of constant $H_0 D_L(z=0.17)/c$ and
 $H_0 D_L(z=1.17)/c$. 
Starting from upper right the contours are for  $H_0 D_L(z=0.17)/c=0.18-0.2$
and  $H_0 D_L(z=1.17)/c=1.63 - 2.1$.
The third panel shows contours of redshift at which the expansion of the
  universe starts to accelerate.  
Starting from upper right corner in counter-clockwise direction,
the contours are for $z=0$, $0.2$, 0$.4$, $0.6$ and $0.8$.  
The constraint from SN observations essentially constrains this
 redshift and hence the allowed region can be expected to follow these
contours.  
The next panel shows red-shift at which the energy density in matter and dark
energy are equal.  
The curves, starting from left are for $z=0.2$, $0.4$, $0.6$, $0.8$ and $1.0$.
Structure formation constraints are likely to follow these contours as 
the rate of growth for density perturbations is significant only in the matter
dominated era.  
The right panel shows contours of $w_{eff}$ at the surface of
last scattering (see Eq. \ref{eqn:weff}).  From top right towards
bottom left the values are $w_{eff}=-0.5$, $-1.0$, $-1.5$ and $-2.0$ If
the location of acoustic peaks provides the main constraint then the
region allowed by CMB observations should follow these contours.  We
have used $\Omega_{NR}=0.3$ for these plots.}   
\label{fig:w_p2}
\end{figure*}

Growth of perturbations also leaves a signature in the CMB anisotropy spectrum
at large angular scales. 
The ISW effect leads to an enhancement in the angular power spectrum at these
scales.  
The detailed form of this enhancement depends on the equation of state
parameter $w$ and its variation. 
This effect can be detected by cross-correlation of temperature anisotropies
with the foreground distribution of matter \cite{isw0}. 
It is difficult to distinguish the ISW effect from the effect of a small but
non-zero optical depth $\tau$ due to re-ionisation by using only the 
temperature anisotropies in the CMB, cross-correlation with the matter
distribution or polarization anisotropies in the CMB must be used.

Redshift surveys of galaxy clustering do not constrain properties of
dark energy directly, however the shape of the power spectrum constrains the
combination $\Gamma = \Omega_{NR} h$ \cite{1992MNRAS.258P...1E}.  
This provides an indirect constraint on dark energy through the well known
degeneracy between $\Omega_{NR}$ and $w_0$  (e.g. see
\cite{2005MNRAS.356L..11J}). 

The large number of different observations that can be used to constrain dark
energy models is encouraging. 
Indeed, many attempts have been made to use some of these observations to put
constraints on models 
\cite{lss_de,2004PhRvD..69j3501T,2005MNRAS.356L..11J,constraints_2,constraints_3}.   


\subsection{A choice of three observations}

In this work, we concentrate on SN, WMAP and cluster abundance observations.  
We briefly explain the reason for this choice and the kind of constraints one
can expect.   

The left panel in Fig.~\ref{fig:w_p2} shows the degeneracy in $\Omega_{NR}$
and $w_0$ (for models with constant $w$; i.e., with $w'_0=0$ in
Eq.~\ref{taylor}).  
The figure shows contours of constant luminosity distance $H_0 d_l(z) $ at
$z=0.17$ (red/solid curves) and $z=1.17$ (blue/dashed curves). 
The second panel displays the constant luminosity distance contours in the
$w'(z=0)-w_0$ plane for $p=2$ if $\Omega_{NR}=0.3$.  
Given that SN observations constrain luminosity distance as a function of
redshifts, these figures illustrate the shape of the allowed region that we
are likely to get and also demonstrates the degeneracies between different
parameters. 
The third panel shows how the redshift at which the expansion of the
universe begins to accelerate depends on the parameters $w_0$ and $w'(z=0)$
for $p=2$. 
This epoch is constrained by SN observations and hence the allowed
region in parameter space should lie between contours of this nature.
Clearly, regions with a late onset of acceleration (upper right corner) as
well as a very early onset of acceleration (lower right corner) will be ruled
out by observations of supernovae. 

The fourth panel of this figure shows the redshift at which matter and dark
energy contribute equally in terms of the energy density of the universe.  
Structure formation constraints are likely to follow these contours as the
rate of growth for density perturbations is significant only in the matter
dominated era. 
Too little structure formation (upper right corner) as well as too much
structure formation (lower left corner) are likely to restrict the allowed
models along a diagonal (upper left to lower right) in this plane.

Lastly, the location of acoustic peaks in the angular power spectrum of
temperature anisotropies in the CMB is the most significant constraint
provided by CMB observations. 
This essentially constrains the distance to the surface of last scattering and
hence  a suitably defined  (see eqn.(\ref{eqn:weff})) mean value ($w_{eff}$) for $w$. 
The right panel shows contours of $w_{eff}$, which run almost diagonal in this
plane. 
Thus a band of allowed models is the likely outcome of comparison with
observations. 
The contours of $w_{eff}$ are the same as contours of equal distance to the
surface of last scattering, or the $l$ corresponding to the first peak in the
angular power spectrum of CMB temperature fluctuations. 

SN data provides geometric constraints for dark energy evolution.
These constraints are obtained by comparing the predicted luminosity
distance to the SN with the observed one.
The theoretical model and observations are compared for luminosity
measured in magnitudes:
\begin{equation}
m_{B}(z)={\mathcal M} + 5 log_{10}(D_{L})
\end{equation}
where ${\mathcal M}=M-5log_{10}(H_0)$ and $D_{L}=H_{0}d_{L}$, $M$
being the absolute magnitude of the object and $d_L$ is the luminosity
distance  
\begin{equation}
d_{L}=(1+z) a(t_0)r(z);~~~~r(z)=c \int \frac{dz}{(1+z) ~ H(z)}  
\end{equation}
where $z$ is the redshift.
This depends on evolution of dark energy through $H(z)$. 
For our analysis we use the combined {\it gold} and {\it silver}
SN data set in \cite{nova_data3}.
This data is a  collection of supernova observations from
\cite{nova_data1,1998ApJ...509...74G} and many other sources with $16$
supernovae discovered with Hubble space telescope \cite{nova_data3}. 
The parameter space for comparison of models with SN observations is
small and we do a dense sampling of the parameter space.

CMB anisotropies constrain dark energy in two ways, through the distance
to the last scattering surface and through the ISW effect.  
Given that the physics of recombination and evolution of perturbations
does not change if $w(z)$ remains within some {\it safe limits}, any
change in the location of peaks will be due to dark energy
\cite{2004PhRvD..70j3523E}.   
The angular size $\theta$ of the Hubble radius at the time of decoupling can
be written as:  
\begin{eqnarray}
\theta^{-1} &=& \frac{H(z)/H_0}{\int\limits_0^z
  {dy}/{\left(H(y)/H_0\right)}} 
\nonumber \\
 &\simeq & \frac{\sqrt{\Omega_{NR}
     \left(1+z\right)^3 }}{\int\limits_0^z 
   {dy}/{\sqrt{ \Omega_{NR} \left(1+z\right)^3 +
     \varrho^{DE}(z)/\varrho^{DE}_0 }}}  \nonumber \\
 & \equiv &  \frac{\sqrt{\Omega_{NR}
     \left(1+z\right)^3 }}{\int\limits_0^z 
   {dy}/{\sqrt{ \Omega_{NR} \left(1+z\right)^3 +
  \Omega_{de} \left(1+z\right)^{3 \left( 1 + w_{eff} \right)}}}}  .
\label{eqn:weff}
\end{eqnarray}
The second line is obtained as decoupling happens at a redshift where
dark energy is not important, and if we ignore the contribution of radiation
and relativistic matter; the last equation defines $w_{eff}$.  
Clearly, the value of the integral will be different if we change $w_0$,
$w'(z=0)$ and there will also be some dependence on the parameterized form.  
Location of peaks in the angular power spectrum of the CMB
provide a constraint, but this can only constrain $w_{eff}$ and not all of
$w_0$, $w'(z=0)$ and $p$.  
Therefore if the present value $w_0 < w_{eff}$ then it is essential that
$w'(z=0) > 0$, and similarly if $w_0 > w_{eff}$ then $w'(z=0) < 0$ is needed
to ensure that the integrals match. 
Specifically, the combination of $w_0$, $w'(z=0)$ and $p$ should give us a 
$w_{eff}$ within the allowed range.  

In our analysis, we use the angular power spectrum of the CMB temperature
anisotropies \cite{cmbrev1,cmbrev2} as observed by WMAP and these are compared
to theoretical predictions using the likelihood program provided by the WMAP
team \cite{wmap_lik}.  
We vary the amplitude of the spectrum till we get the best fit with WMAP
observations. 
Note that this is different from the commonly used approach of normalizing the
angular power spectrum at $l=10$.  
As we use the entire angular power spectrum for comparison with observations,
the impact of ISW effect on the likelihood is relatively small.  

It has been pointed out that constraints from structure formation restrict the
allowed variation of dark energy in a significant manner \cite{lss_de}.  
We use observed abundance of rich clusters
\cite{1999ApJ...517...40B,cluster2,cluster4} to apply this constraint.  
Since the mass of a typical rich cluster corresponds to the scale 
of $8$h$^{-1}$Mpc, cluster abundance observations therefore constrain
$\sigma_8$, the rms fluctuations in density contrast at $8$h$^{-1}$Mpc. 
The number density of clusters depends strongly on $\sigma_8$ and
$\Omega_{NR}$.
We use the $\sigma_8$ constraints given in  \cite{1999ApJ...517...40B} from
ROSAT deep cluster survey and are given by $\sigma_8=(0.58 \pm 0.1) \times
\Omega_{NR}^{-0.47+0.16 \Omega_{NR}}$ at $99\%$ confidence level. 
The cosmological model should predict $\sigma_8$ in the allowed range in order
to be consistent with observations. 

Recent detection of the baryon acoustic peak using luminous red galaxy
sample of the SDSS survey has provided an additional handle to
constrain the cosmological parameters \cite{2005astro.ph..1171E}.
We also used distance scale $D_V$ at redshift $z=0.35$ introduced
in the above reference to further constrain the cosmological models. 
Here $D_V(0.35)^3 \equiv D_M(z)^2 cz/H(z)$ and the observational constraint is
$D_V(0.35)=1370 \pm 64~~Mpc$ at the $1 \sigma$ level.   
This fourth observation does not add significantly to other constraints listed
here and we will not describe quantitative results from this constraint here. 

Priors used in the present study are listed in Table 1. 
Apart from these limits on the models studied here, we also assumed that
neutrinos are massless and the ratio $r$ of tensor to scalar mode is zero.  
These assumptions are consistent with the known upper bounds, and in any case
these do not make any difference to the observations used here as
constraints \cite{2005PhRvD..71j3515S}. 

\begin{table}
\label{tab:priors}
\caption{This table lists the priors used in the present work.  
Apart from the range of parameters listed in the table, we assumed
that the universe is flat.  
We assumed that the primordial power spectrum had a constant index. 
Further, we ignored the effect of tensor perturbations. 
The range of values for $w_0$ and $w'(z=0)$ is as given below, but
with the constraint that $w(z=1000) \leq -1/3$.  
Any combination of $w_0$ and $w'(z=0)$ that did not satisfy this
constraint was not considered.}
\begin{center}
\begin{tabular}{||l|l|l||}
\hline
\hline
Parameter & Lower limit & Upper limit  \\
\hline
\hline
$\Omega_B$ & $0.03$ & $0.06$  \\
\hline
$\Omega_{NR}$ & $0.1$ & $0.5$  \\
\hline
$h$ & $0.6$ & $0.8$  \\
\hline
$\tau$ & $0.0$ & $0.4$  \\
\hline
$n$ & $0.86$ & $1.10$ \\
\hline
$w_0$ & $-2.0$ & $-0.4$  \\
\hline
$w'(z=0)$ & $-5.0$ & $5.0$ \\
\hline
\hline
\end{tabular}
\end{center}
\end{table}
 

\section{Markov chain monte Carlo method} \label{sec:mcmc}

We compute $\chi^2$ using the routines provided by the WMAP team
\cite{wmap_lik}. 
The CMBFAST package \cite{1996ApJ...469..437S} is used for computing the
theoretical angular power spectrum for a given set of cosmological
parameters. 
We have combined the likelihood program with the CMBFAST code and this
required a few minor changes in the CMBFAST driver routine. 
Given the large number of parameters, the task of finding the minimum
$\chi^2$ and mapping its behavior in the entire range of values for
parameters is computationally intensive.

We adapt the Metropolis algorithm \cite{metropolis} (also known in
the context of parameter estimation as the Markov Chain Monte Carlo
(MCMC) method \cite{wmap_lik,mcmc2}) for efficiently  mapping regions with low
values of $\chi^2$. 
The algorithm used is as follows:
\begin{enumerate}
\item 
Start from a random point ${\mathbf r}_i$ in parameter space and
compute $C_l$ and $\chi^2({\mathbf r}_i)$. 
\item 
Consider a small random displacement ${\mathbf r}_{i+1} = {\mathbf r}_i +
d {\mathbf r}$ and compute $\chi^2({\mathbf r}_{i+1})$. 
\item 
If $\chi^2({\mathbf r}_{i+1}) \leq \chi^2({\mathbf r}_i)$ then $i
\rightarrow i+1$.
Go to the first step.
\item 
Else:
\begin{itemize}
\item  
Compute $\Delta \chi^2 = \chi^2({\mathbf r}_{i+1}) - \chi^2({\mathbf
  r}_i)$ and $\exp[-\alpha ~ \Delta \chi^2]$.
\item 
Compare this with a random number $0 \leq \beta \leq 1$.
\item  
If $\beta \leq \exp[-\alpha ~ \Delta \chi^2]$ then $i
\rightarrow i+1$.
Go to the first step.
\end{itemize}
\end{enumerate}

The size of the small displacement $d{\mathbf r}$ and the parameter
$\alpha$ are chosen to optimally map the regions of low $\chi^2$. 
We wish the chain to converge towards the minimum, starting from an
arbitrary point, and we also want the Markov chain to map the region
in parameter with low $\chi^2$ exhaustively without getting bogged down near
the minima.  
These two conflicting requirements are reconciled by choosing a small
but non-zero value of $\alpha$.  
Maximum displacement allowed in one step is small compared with the
range of parameters, but not small enough for the chain to get
trapped in a small region around the minimum.
The optimum values of maximum displacement and $\alpha$ are related to each
other. 
We ran several chains with a varying number of points, a typical chain has
about $10^4$ points.  
For each set, we have at least $10^5$ points (We have done calculations for
five sets: cosmological constant, constant $w$ with and without perturbations
in dark energy, time varying $w(z)$ for $p=1$ and $p=2$.  Results presented
here required an aggregate CPU time of nearly $10000$~hours on $2.4$~GHz Xeon
CPUs).  
The convergence criteria for such chains is satisfied for all the sets, and
for all the parameters in each set \cite{convergence}. 

We use the MCMC approach only for comparison of models with the CMB data.   
Observations of cluster abundance are compared with models from the Markov
chain run for CMB, after the chain has been run.  
Comparison of models with observations of high redshift supernovae is done
separately. 
This approach is more conducive to one of the questions that we wish to
address, namely, are the observational constraints consistent with each other? 


\section{Results} \label{sec:details}

We present results in the form of likelihood functions for various parameters
in sets of increasing complexity, starting with the standard $\Lambda$CDM
model. 
Before we proceed with a discussion of results in this form, we discuss a few
specific models sampling a few interesting regions of  the parameter space in
order to develop an intuitive feel for different observational constraints.
We call these fiducial or reference models.
Along with the fiducial models, we also discuss the best fit models in each
set. 
We find the best fit model for individual observations as well as for the
combination of all the observations.  


\subsection{Fiducial Models}

\subsubsection{The $\Lambda$CDM model}

The $\Lambda$CDM model is our 'standard' model and we begin our discussion
with this class of models.  
Several studies have been carried out to constrain parameters for the
$\Lambda$CDM model \cite{wmap_params,2004PhRvD..69j3501T}. 
Our results for the $\Lambda$CDM model bring out --- among other things ---
the differences from previous work which arises due to a  different method we 
use here for normalizing power spectra. (See section 2.3 for details.)
Differences introduced by priors are also apparent.
Our results are as follows: 
\begin{itemize}
\item
For $\Lambda$CDM model, if we consider SN observations alone,
we get a best fit at $\Omega_{NR}=0.28$ with a $\chi^2_S = 233.1$.  
(We will use subscript $S$ for $\chi^2$ from SN analysis and $W$
for analysis with WMAP observations.)
This model with $\Omega_{NR}=0.28$ is allowed by WMAP observations and
has a best fit $\chi^2_W = 974.3$ for $\Omega_B=0.045$, $h=0.69$, $n=0.95$
and $\tau=0.008$.   
SN observations do not fix these parameters so we varied the other
parameters to get the best fit WMAP model for $\Omega_{NR}=0.28$.
\item
The model which best fits the WMAP observations has $\Omega_B=0.05$,
$\Omega_{NR}=0.34$, $h= 0.66$, $n=0.96$ and $\tau=0.002$
with a $\chi^2_W=972.5$. 
The $\chi^2$ value corresponding to SN fit is $\chi^2_S =239.9$.    
In the context of cosmological constant models alone, this model is away from
the SN best fit by $\Delta \chi^2_S = 6.8$ and is allowed with probability
$0.009$. 
In other words, the  model most favored by WMAP observations is allowed by the
SN observations only with less than one percent probability (We define
probability $\mathcal{P}$ of a given model  to be $1 - \mathcal{C}/100$, where
$\mathcal{C}\%$ is the confidence limit at which the model is allowed.  
By using this definition we avoid dilution due to a large parameter space.
While the statement about $\chi^2$ is accurate and directly obtainable from
the analysis, the conversion of confidence intervals to probabilities has
well-known statistical caveats while dealing with multiparameter fits. 
This should be kept in mind while interpreting our statements about
probability with which a model is allowed). 
In contrast, the model most favored by SN observations is allowed by WMAP
observations with a probability $\mathcal{P} = 0.945$.  
\end{itemize} 

We now restrict some of the parameters to values favored by other
observations e.g. \cite{2001ApJ...553...47F}. 
We fix the baryon density parameter $\Omega_B = 0.05$, present day
Hubble parameter $h =0.7$ and the  spectral index $n=1$
for these models.
Allowing $\Omega_{NR}$ and $\tau$ to vary the best fit $\Lambda$CDM model
in this restricted class of models is with $\Omega_{NR} = 0.31$ and
$\tau=0.14$ and $\chi^2_W =974.8$.
This is fairly close to the best fit model found by the WMAP team using a
large set of observations \cite{wmap_params}. 
This model is allowed by the rich cluster abundance observations, and also by
SN observations ($\chi^2_S= 234.8$, the corresponding probability being
$\mathcal{P}=0.2$).

Thus convergence between the WMAP and SN observations happens \textit{only} in
a narrow window for flat $\Lambda$CDM models, with the SN constraint being the
tighter of the two. 
It is worth mentioning that in a wider class of models, (obtained by relaxing
the prior $\Omega_{\rm tot}=1$) SN data favors a closed universe with
$\Omega_{tot}=1.44 \pm 0.28$  and --- more importantly --- allows the
$\Omega_{tot}=1$ models with $\mathcal{P}=0.12$ \cite{2005A&A...429..807C}.   

\begin{table*}
\label{tab:bestfits}
\caption{This table lists best fit parameter values and $\chi^2$ for
  different models and some of the selected fiducial models. The
  abbreviation b.f. denotes the best fit model for the particular data
  set. The corresponding probabilities are given in brackets.}
\begin{center}
\begin{tabular}{||l|l|l|l|l|l|l|l|l|l|l||}
\hline
\hline
& & $\Omega_B$ & $h$ & $n$ & $\tau$ & $\Omega_{NR}$& $w$& $w'$ &
$\chi^2_W(P)$& $\chi^2_S ~~~ (\mathcal{P})$ \\
\hline
\hline
&
WMAP(b.f.)&$0.05$&$0.663$&$0.96$&$0.002$&$0.34$&$-1$&&$972.5$&$239.9 ~~~ (0.009)$\\ 
$\Lambda$CDM &SN(b.f.) &$0.045$&$0.697$&$0.95$&$0.008$&$0.28$&$-1$&&$973.7 ~~~ (0.94)$&$233.1$ \\
              &model 1& $0.05$&$0.70$&$1.0$&$0.14$&$0.31$&$-1$&&$974.8 ~~~ (0.8)$&$234.8~~~(0.21)$
\\ 
\hline
&WMAP(b.f.)&$0.06$&$0.635$&$0.99$&$0.19$&$0.34$&$-0.72$&&$971.6$&$280.0 ~~~ (<0.001)$ \\
$w=constant$&SN(b.f.)&$0.055$&$0.666$&$1.0$&$0.05$&$0.47$&$-1.99$&&$985.2 ~~~ (0.022)$&$227.46$  \\
&model 1&$0.05$&$0.70$&$1.0$&$0.04$&$0.38$&$-1.5$&&$978.8~~~(0.314)$&$229.9 ~~~ (0.299)$\\
&model
2&$0.05$&$0.70$&$1.0$&$0.18$&$0.28$&$-0.9$&&$974.5 ~~~ (0.82)$&$237.9 ~~~ (0.005)$\\ 
\hline
&WMAP(b.f.)&$0.05$&$0.664$&$0.95$&$0.0$&$0.32$&$-0.96$&&$972.8$ & $239.3 ~~~ (0.003)$ \\
$w=constant$&SN(b.f.)&$0.055$&$0.666$&$1.0$&$0.05$&$0.47$&$-1.99$&&$985.2~~~(0.022)$&$227.46$  \\
with perturbations&model 1&$0.05$&$0.70$&$1.0$&$0.06$&$0.38$&$-1.5$&& $979.2~~~(0.395)$&$229.9~~~(0.299)$\\
&model 2&$0.05$&$0.70$&$1.0$&$0.16$&$0.28$&$-0.9$&& $976.1~~~(0.77)$&$237.9~~~(0.005)$\\
\hline
&WMAP(b.f.)& $0.05$&$0.73$&$1.1$&$0.35$&$0.24$&$-1.48$&$3.86$&$970.9$&$232.1~~~(0.33)$\\
$p=2$&SN(b.f.)&$0.056$&$0.662$&$1.04$&$0.002$&$0.498$&$-1.95$&$-4.5$&$987.6~~~(0.035)$&$227.37$ \\
&model 1&$0.05$&$0.70$&$1.0$&$0.0005$&$0.37$&$-1.5$&$1.0$&$976.5~~~(0.692)$&$229.3~~~(0.75)$\\
&model 2&$0.05$&$0.70$&$1.0$&$0.0$&$0.42$&$-1.5$&$-5.0$&$983.8~~~(0.123)$&$234.9~~~(0.116)$\\
&model 3&$0.05$&$0.70$&$1.0$&$0.0013$&$0.365$&$-0.9$&$-3.0$&$977.4~~~(0.56)$&$234.4~~~(0.147)$\\
\hline
\hline
\end{tabular} 
\end{center}
\end{table*}


\subsubsection{Models with a constant $w$}

We now allow the dark energy equation of state parameter to have values
different from $w=-1$ but do not allow variation with time. 
We then find that:
\begin{itemize}
\item
SN observations favor a model with $w=-1.99$ and $\Omega_{NR} =
0.47$ and with $\chi^2_S=227.5$.  
This is the root cause of the phantom menace.  
This model is, however, ruled out at a very high probability  by WMAP data
(with a $\Delta\chi^2_W = 13.6$) and   is allowed only with
$\mathcal{P}=0.022$. 
\item
WMAP observations favor higher values for $w$ (non-phantom models)
and the best fit model has $w=-0.72$. 
The other cosmological parameters corresponding to this best fit are
$\Omega_B=0.06$, $\Omega_{NR}=0.34$, $h= 0.64$, $n=0.99$ and  $\tau=0.19$
with $\chi^2_W =971.6$.  
SN observations, on the other hand, rule out this model at a very high
significance level ($\Delta\chi^2_S \simeq 53$). 
This is the case when we do not take perturbations in dark energy into
account for computing theoretical predictions. 
\item
If we include dark energy perturbations\footnote{The CMBFAST version
  ($4.5.1$) being used here incorporates the effects of perturbations in dark
  energy for constant $w$ and non-phantom models with a variable $w(z)$.} the
model which fits best with WMAP observations is same as the best fit for the
$\Lambda$CDM models. 
This model is allowed by SN observations with $\chi^2=239.3$ which is
allowed  with probability $0.003$.   
Clearly, the discrepancy between the WMAP and SN observations is reduced when 
we take perturbations in dark energy into account, but SN observations allow
the WMAP best fit model only marginally even in this case.
\item
How well does the $\Lambda$CDM fare when we allow a range of $w$ ?
The cosmological constant is allowed with $\mathcal{P}=0.063$ 
by SN observations in the set of models with constant $w$. 
WMAP observations allow the $\Lambda$CDM models with a very high probability,
indeed the best fit model continues to be a $\Lambda$CDM model if
perturbations in dark energy are taken into account.
\end{itemize}

To illustrate these aspects, we consider two fiducial models with $w=-1.5$ and  $w=-0.9$ allowing
$\Omega_{NR}$ and $\tau$ to vary. 
Other parameters are fixed as for the $\Lambda$CDM model (see subsection 1 above). 
We do not take perturbations in dark energy into account here.
The best fit values using WMAP observations for $w=-1.5$  are  $\Omega_{NR} =
0.38$ and $\tau=0.04$. 
Generically, models with lower $w$ require higher values of $\Omega_{NR}$ which
is clear from the form of the curves in Fig.\ref{fig:w_p2} (a). 
This model is allowed by WMAP observations with $\chi^2_W =978.8$ as well as
by SN observations ($\chi^2_S= 229.9$). 
But this model is \textit{not} allowed by observations of cluster abundance. 
The value of $\sigma_8=1.1$ for this model is higher than the upper limit
$\sigma_8 = 1.02$ allowed with $99\%$ confidence level for this value of
$\Omega_{NR}$.    

With $w=-0.9$, WMAP favors a model with $\Omega_{NR} = 0.28$ and $\tau=0.18$
($\chi^2_W=974.5$). 
This model is allowed by cluster abundance observations.  
The model becomes marginal when SN observations are taken into account
with $\chi^2_S=237.9$, as compared to a minimum of $\chi^2_S=227.5$ for models
with constant $w$.  

Once again, we find that WMAP observations and SN observations favor
different regions in the parameter space.  
Generalizing to the class of models with a constant $w$ from $w=-1$, we find
that SN data has a distinct preference for $w < -1$ and the
$\Lambda$CDM model is allowed only marginally. 
WMAP data also shows a mild preference for models with $ w \neq -1$, though it 
continues to allow the $\Lambda$CDM model.  
If perturbations in dark energy are taken into account then the $\Lambda$CDM
model continues to be the most favored model for WMAP observations.

\subsubsection{Models with varying $w(z)$}

Next, we allow the dark energy equation of state parameter to vary. 
As it is not possible to take the effect of non-adiabatic perturbations into
account for these models and as we do not wish to add another parameter, we
work with models without any perturbations in dark energy \cite{perturb3}. 
It is clear that this will introduce a slight bias towards models with $w >
-1$ but will not change anything else. 
For the discussion of fiducial models, we choose the parameterization with
$p=2$ in Eq.(\ref{taylor}).
\begin{itemize}
\item
Parameter values for the best fit model with SN data are $w_0=-1.95$,
$w'(z=0)=-4.5$ and $\Omega_{NR}=0.498$ with a $\chi^2_S=227.4$. 
As with the constant $w$ models, SN data favors large negative values
of the equation of state parameter. 
{\it SN observations do not favor models with varying $w(z)$ over
 models with $w \neq -1$}, since the change in best fit $\chi^2$ is only $0.1$ when the
 additional parameters are added.  
WMAP observations allow this model with $\mathcal{P}=0.035$ ($\chi^2_W
= 987.6$) if we have $\Omega_B=0.056$, $h=0.662$, $n=1.04$ and $\tau=
0.002$.  
However, this model is ruled out by cluster abundance observations as
$\sigma_8=1.25$ for this model is higher than the allowed range at
$99\%$ confidence level.
\item 
The best fit model for WMAP data has $w_0=-1.48$, $w'(z=0)=3.86$,
$\Omega_B=0.05$, $\Omega_{NR}=0.24$, $h=0.73$, $n= 1.1$ and $\tau= 0.35$ with
$\chi^2_W=970.9$.  
This model is allowed by SN observations with $\mathcal{P}=0.33$  ($\Delta
\chi^2=4.63$).  
\end{itemize} 
The tension between the WMAP and SN observations is less serious for models
with varying $w$ than for models with a constant $w$. 
Part of the reason is that with a larger number of parameters, the same
$\Delta\chi^2_S$ gives us a larger probability for a given model. 

As in the previous cases, let us choose fiducial models by restricting some of
the parameters. 
We consider a model with dark energy parameter values $w_0 =-1.5$ and
$w'(z=0)= 1.0$.  
The WMAP best fit with these parameters  is with $\Omega_{NR}=0.37$ and
$\tau=0.0005$ with $\chi^2_W=976.5$ .
The change in acceptance level of this model is due to our restricting the
values of the spectral index and the Hubble parameter.  
This model has $\chi^2_S=229.3$ and is allowed by all the three observations
used here. 

If we move closer to the model most favored by SN observations,
$w_0=-1.5$ and $w'(z=0) =-5.0$, then we find that WMAP data favors
$\Omega_{NR}=0.42$ and $\tau=0.0$.   
Even this model is outside the range allowed by WMAP observations at $68 \%$
confidence limit and is allowed at $\mathcal{P}=0.12$.  
The model fares similarly with SN observations, i.e., it is allowed 
with $\mathcal{P}=0.12$.

We wrap up with a discussion of a model with $w_0 > -1$. 
We consider $w_0=-0.9$ and $w'(z=0)=-3.0$.
WMAP observations allow this model with $\chi^2_W=977.4$ for
$\Omega_{NR}=0.365$ and $\tau=0.0013$.
SN observations allow this model within the $95\%$ confidence limit with a
probability  $\mathcal{P} = 0.15$.
This model is also allowed by observations of cluster abundance.

Within the context of models with a variable $w(z)$, the $\Lambda$CDM model is
allowed by WMAP observations ($\Delta \chi^2_W=1.6$) as well as by supernova
observations ($\Delta \chi^2_S=5.73$). 
SN observations clearly favor models other than the $\Lambda$CDM model
in context of $\Omega_{tot}=1$, while no such preference is seen for WMAP
observations.  
However, models favored by SN observations require $\Omega_{NR}$ to be
much larger than the values favored by observations of rich clusters
\cite{crisis2}.

In summary, there is significant tension between the sets of
observations we are studying and this tension does not reduce when the
parameter space is enlarged. 
We see that best fit model for one set of observation is often ruled
out by another set  with a high level of significance. 
There is an overlap of region allowed with $95\%$ confidence limit in all
cases and within $68\%$ confidence limit in some cases; but this does not take
away the significance of the differences which the above analysis has thrown
up.
The results presented here are summarized in Table 2.

\begin{figure*}
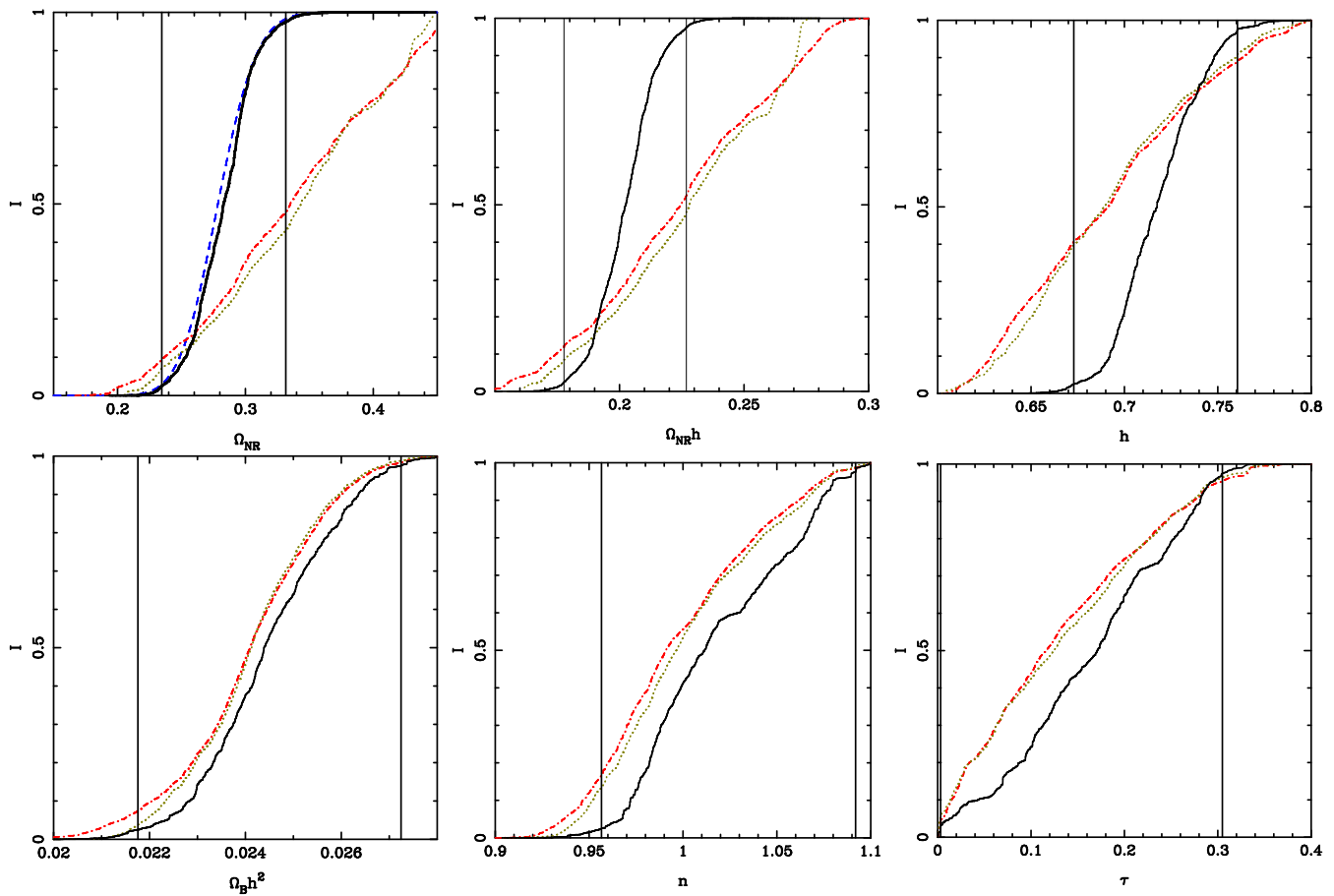

\centering
\begin{tabular}{ccc}
\begin{minipage}{2.3in}
\centering
\includegraphics[width=2.3in]{fig2a.ps}
\end{minipage}
& 
\begin{minipage}{2.3in}
\centering
\includegraphics[width=2.3in]{fig2b.ps}
\end{minipage}
&
\begin{minipage}{2.3in}
\centering
\includegraphics[width=2.3in]{fig2c.ps}
\end{minipage}
\\
\begin{minipage}{2.3in}
\centering
\includegraphics[width=2.3in]{fig2d.ps}
\end{minipage}
&
\begin{minipage}{2.3in}
\centering
\includegraphics[width=2.3in]{fig2e.ps}
\end{minipage}
& 
\begin{minipage}{2.3in}
\centering
\includegraphics[width=2.3in]{fig2f.ps}
\end{minipage}
\\
\end{tabular}
\caption{This Figure shows plots of $I$ (see eqn.\ref{eqn:i}) for various
  parameters for the $\Lambda$CDM model.  Red/dot-dashed curves show
  constraints from the WMAP observations of temperature anisotropies,
  dark-green/dotted curve shows constraints from WMAP observations and
  abundance of rich clusters, black/solid curve shows constraints from a
  combined analysis of WMAP observations, cluster abundance and high redshift
  supernovae. The blue/dashed curve shows the constraints from SN observations
  alone, this has been plotted only for the relevant parameters.  
  Vertical lines mark the $95\%$ confidence limit for each parameter.} 
\label{fig:lcdm}
\end{figure*}

\begin{figure*}
\centering
\begin{tabular}{ccc}
\begin{minipage}{2.3in}
\centering
\includegraphics[width=2.3in]{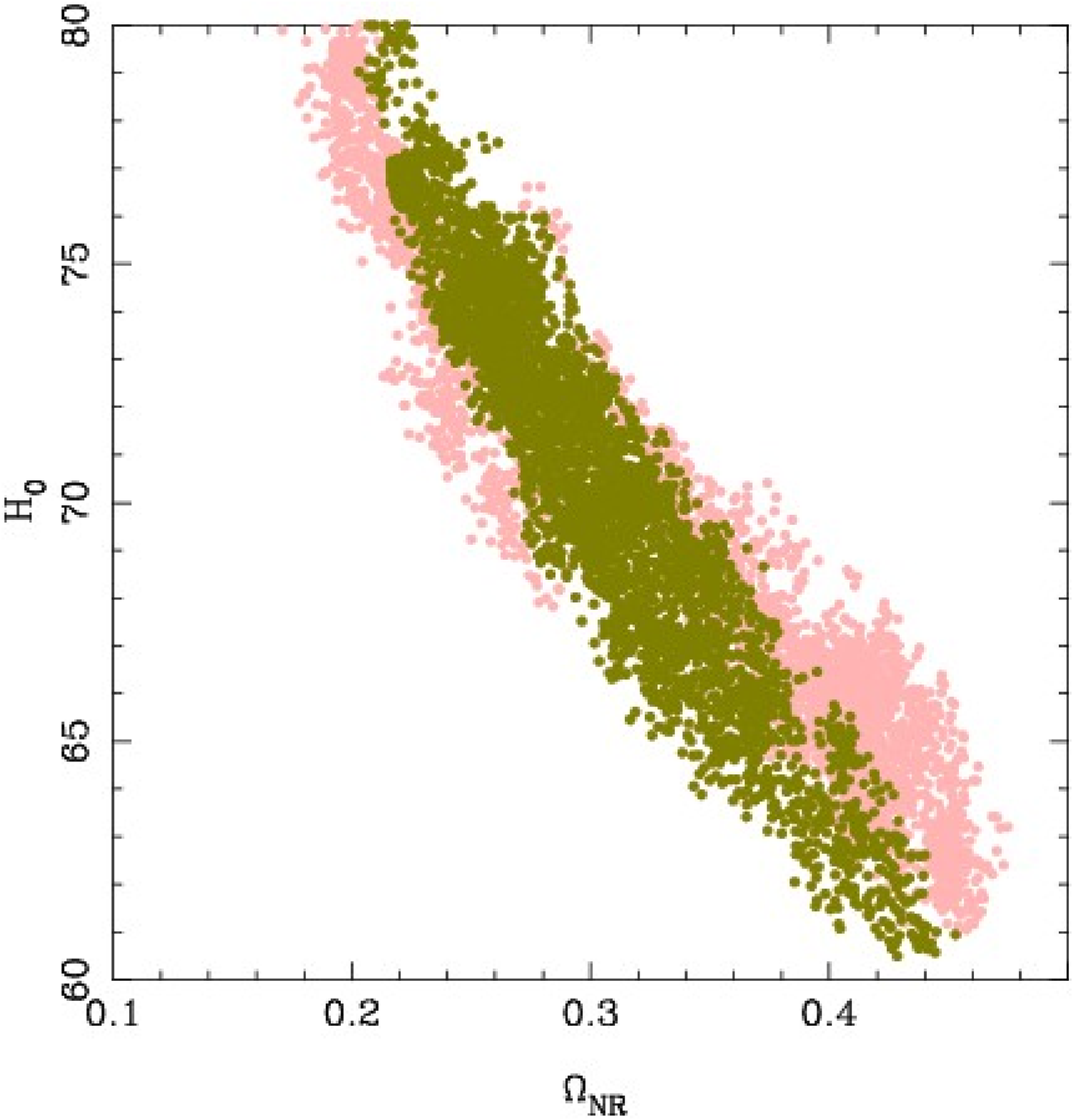}
\end{minipage}
& 
\begin{minipage}{2.3in}
\centering
\includegraphics[width=2.3in]{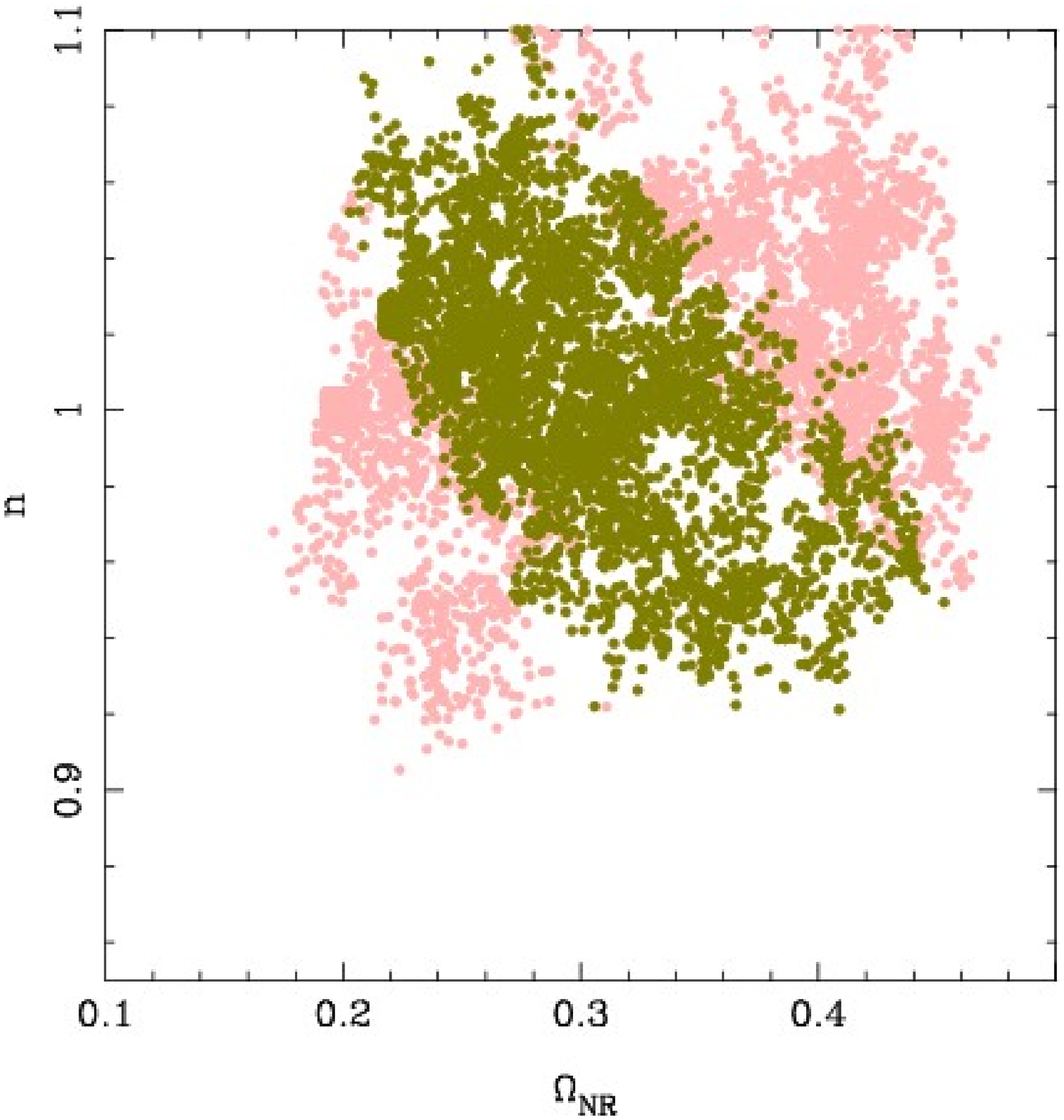}
\end{minipage}
&
\begin{minipage}{2.3in}
\centering
\includegraphics[width=2.3in]{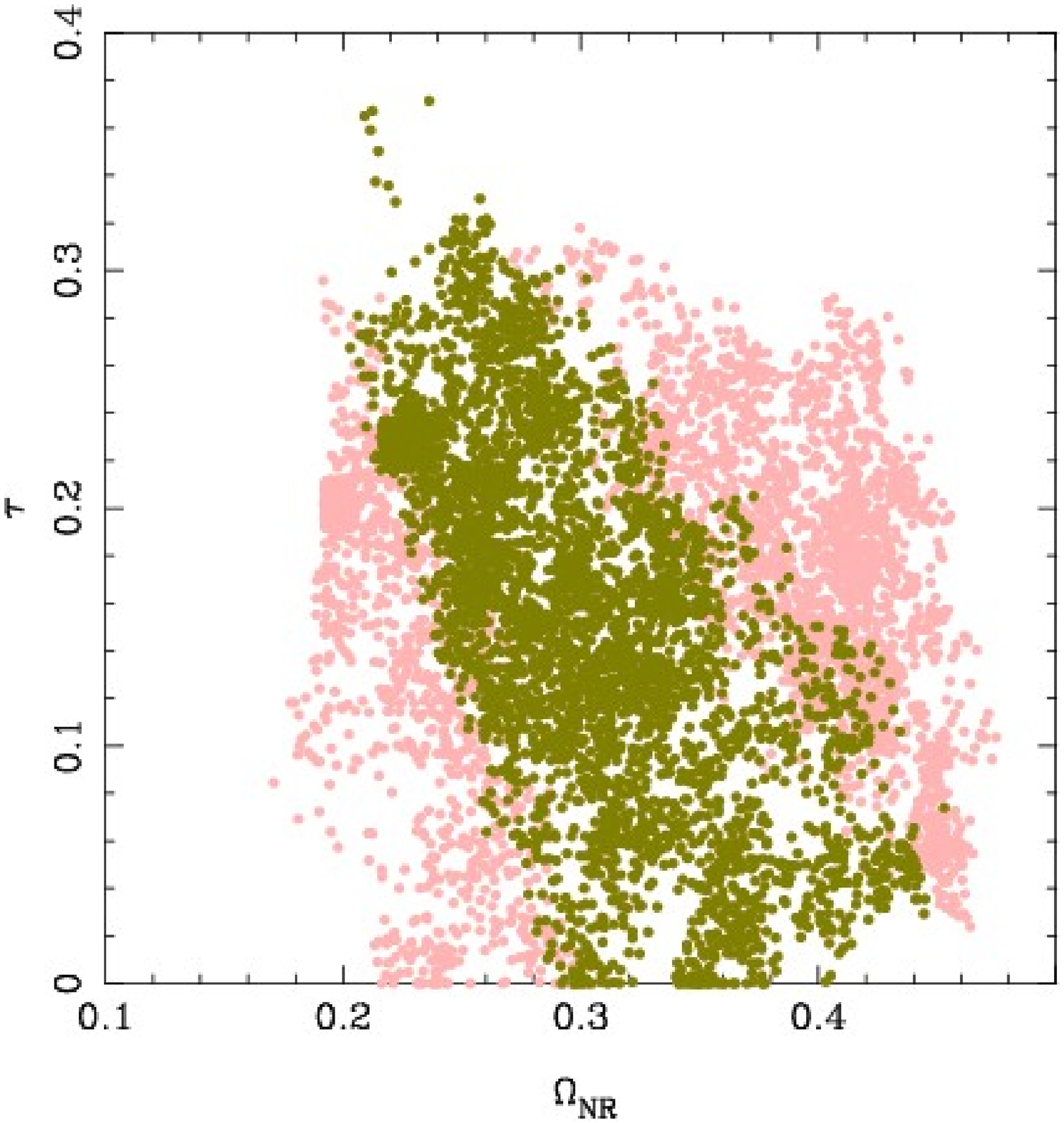}
\end{minipage}
\\
\end{tabular}
\caption{This Figure shows points in the Markov chain for $\Lambda$CDM models that are
  allowed by WMAP observations of temperature anisotropies in the CMB within
  $95\%$ confidence limit in the full parameter space (pink points).  Models
  that are also allowed within a $99\%$ confidence limit by abundance of rich
  clusters of galaxies are shown as dark green points.  The points are shown
  in a few projections to illustrate the fact that the requirement of cluster
  abundance rejects many models allowed by CMB observations.  While for some
  projections it is clear that the region in parameter space allowed by
  cluster abundance is distinctly smaller than that allowed by CMB
  observations, in other projections these seem to have an almost complete
  overlap.  This plot also highlights degeneracies in the parameter space.} 
\label{fig:lcdm_sc}
\end{figure*}

\begin{figure*}
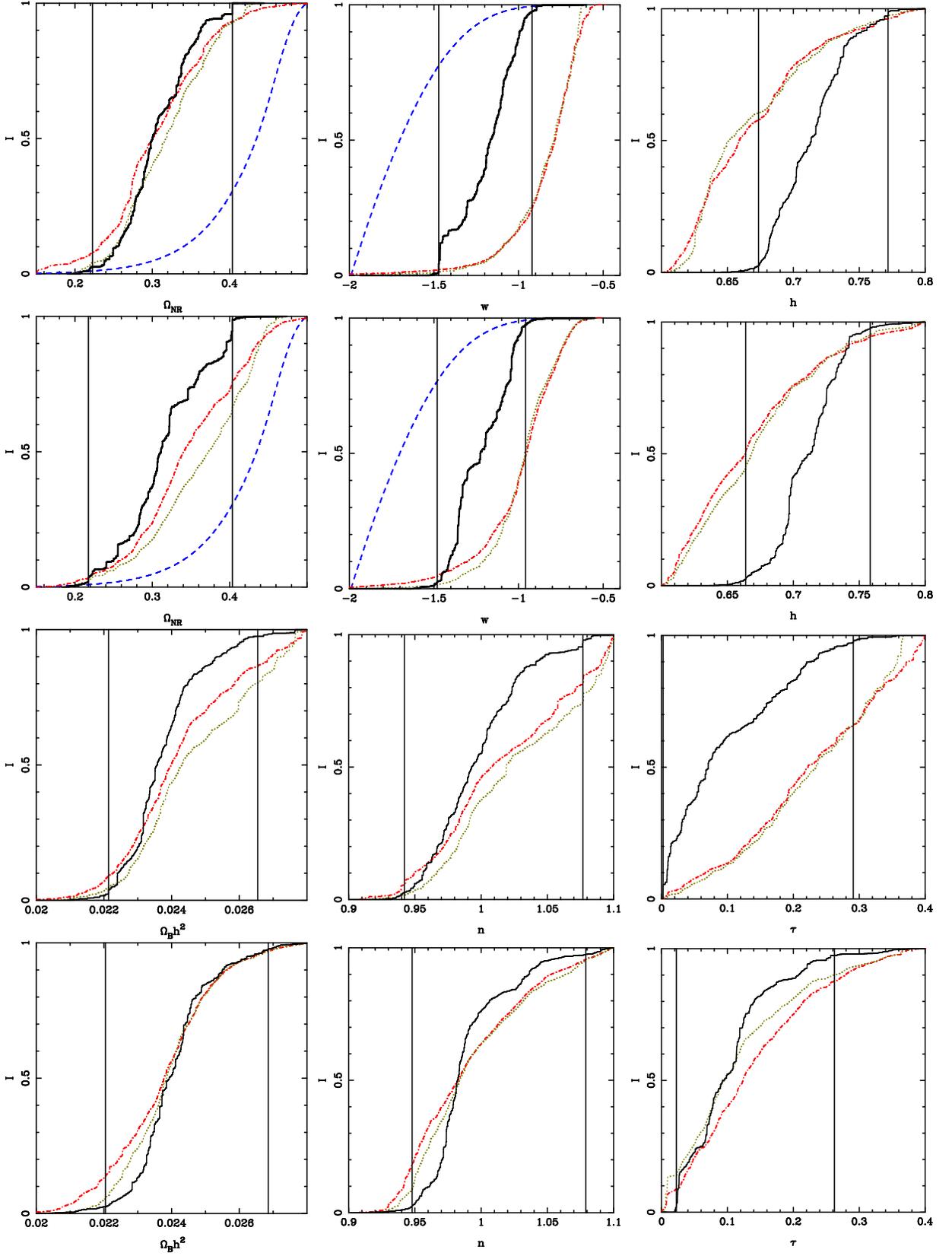

\centering
\begin{tabular}{ccc}
\begin{minipage}{2.1in}
\centering
\includegraphics[width=2.1in]{fig4a.ps}
\end{minipage}
& 
\begin{minipage}{2.1in}
\centering
\includegraphics[width=2.1in]{fig4b.ps}
\end{minipage}
&
\begin{minipage}{2.1in}
\centering
\includegraphics[width=2.1in]{fig4c.ps}
\end{minipage}
\\ 
\begin{minipage}{2.1in}
\centering
\includegraphics[width=2.1in]{figp4a.ps}
\end{minipage}
&
\begin{minipage}{2.1in}
\centering
\includegraphics[width=2.1in]{figp4b.ps}
\end{minipage}
& 
\begin{minipage}{2.1in}
\centering
\includegraphics[width=2.1in]{figp4c.ps}
\end{minipage}
\\
\begin{minipage}{2.1in}
\centering
\includegraphics[width=2.1in]{fig4d.ps}
\end{minipage}
& 
\begin{minipage}{2.1in}
\centering
\includegraphics[width=2.1in]{fig4e.ps}
\end{minipage}
&
\begin{minipage}{2.1in}
\centering
\includegraphics[width=2.1in]{fig4f.ps}
\end{minipage}
\\ 
\begin{minipage}{2.1in}
\centering
\includegraphics[width=2.1in]{figp4d.ps}
\end{minipage}
&
\begin{minipage}{2.1in}
\centering
\includegraphics[width=2.1in]{figp4e.ps}
\end{minipage}
& 
\begin{minipage}{2.1in}
\centering
\includegraphics[width=2.1in]{figp4f.ps}
\end{minipage}
\\
\end{tabular}
\caption{This figure shows $I$ for $\Omega_{NR}$, $w$ and $h$ for cosmological
  models with a constant equation of state parameter $w$.  
  Panels in the first and third row  are for models without
  perturbations in the dark energy component, whereas panels in second
  and fourth row  are for models with perturbations in dark energy.  
  Each  panel shows $I$ derived from observations of high redshift
  supernovae, WMAP   observations, WMAP observations combined with the
  constraints from abundance of rich clusters, and all three
  observations in combination (with the same color coding as in Fig. 2).   
  We have marked   $95\%$ confidence limits derived from using all the
  observations in concert   as two vertical lines. 
  The relevant parameters are written in the x-axis labels of the
  figures.} 
\label{fig:cw}
\end{figure*}


\subsection{Results in Detail}

In this section we outline the detailed results of our analysis.  
We marginalize the results in the multi-dimensional parameter space to derive 
likelihood function for each parameter of interest.  
The likelihood function is sensitive to the bin-size used for the given
parameter and tends to be somewhat noisy. 
It is customary to smooth the likelihood function with a Gaussian filter in
order to remove noise but the results are sensitive to the width of the filter
used. 
Therefore we choose to plot the cumulative likelihood as this is insensitive
to binning and smoothing is not required. 
We define the cumulative likelihood as follows:
\begin{equation}
I(x) = \frac{\int\limits_{x_{min}}^x ~ \mathcal{L}(y) ~
  dy}{\int\limits_{x_{min}}^{x_{max}} ~ \mathcal{L}(y) ~ dy} 
\label{eqn:i}
\end{equation}
where $x$ is the parameter we are interested in and it has a range $x_{min}
\leq x \leq x_{max}$, $\mathcal{L}(x)$ is the likelihood obtained by
marginalizing over other parameters.
The central value for the given variable is thus $x_c$, with $I(x_c)=0.5$.  
This can be, and in general it is different from the value of the parameter
for maximum likelihood.  
$I(x)$ is like the cumulative probability function for the parameter $x$. 
Parameter values for which $I(x)=0.025$ and $I(x)=0.975$ define the range
allowed at $95\%$ confidence limit, i.e. the probability that the variable lies
within this range is $0.95$.  
We mark this limit by two vertical lines in all the likelihood plots. 


\subsection{The $\Lambda$CDM model}

Fig.~\ref{fig:lcdm} shows marginalized cumulative likelihood $I(x)$ for the
$\Lambda$CDM model, different frames correspond to the different parameters we
have considered here.   
Curves are shown for $\Omega_{NR}$, the shape parameter $\Gamma = \Omega_{NR}
~ h$, Hubble parameter $h=H_0/100~kms^{-1}Mpc^{-1}$, $\Omega_b ~ h^2$,
spectral index $n$ and optical depth to the redshift of reionisation $\tau$. 
Red/dot-dashed curves show constraints from  WMAP observations of CMB
temperature anisotropies, dark-green/dotted curve shows constraints from WMAP
observations and abundance of rich clusters, black/solid curve shows
constraints from a combined analysis of WMAP observations, cluster abundance
and high redshift supernovae. 
The blue/dashed curve shows the constraints from SN observations alone, this
has been plotted only for $\Omega_{NR}$ as SN data does not constrain other
parameters directly.  
Combined analysis for other parameters does have an input from supernova
observations as many models allowed by WMAP and cluster abundance observations
are ruled out by SN observations. 
Vertical lines mark the $95\%$ confidence limit when all the observations are
used together. 
CMB observations allow considerable range for each of these parameters whereas
observations of high redshift supernovae provide tight constraints.  
The reason why SN observations provide a tight constraint is clear from
the discussion in the previous section, namely, SN observations favor
models with $\Omega_{tot} \geq 1$.  
Within the context of models with $\Omega_{tot} = 1$, SN observations
favor models with $w \ll -1$. 
The $\Lambda$CDM model is only marginally allowed by SN observations
within both the sets; flat $\Lambda$CDM models are allowed with
$\mathcal{P}=0.12$ \cite{2005A&A...429..807C} and within flat models the
$\Lambda$CDM model is allowed with $\mathcal{P}=0.06$.

The allowed range for all the parameters within $68\%$ confidence limit is
given in Table~3 and range allowed in $95\%$ confidence limit is given in
Table~4.
Table~3 clearly illustrates the discrepancy between SN observations and WMAP
observations.
The values are comparable with those obtained in other analyses
\cite{wmap_params,2004PhRvD..69j3501T}.  
The range of values for the shape parameter $\Gamma = \Omega_{NR} h$ favored
by these observations is consistent with values obtained from galaxy surveys
\cite{2004PhRvD..69j3501T,sdss,2005astro.ph..1174C}.
The allowed  range of values for $h$ are in agreement with direct
determination \cite{2001ApJ...553...47F}.

Constraints from abundance of rich clusters do not make a significant impact
on the likelihood function of individual parameters even though these
constraints reject a significant fraction of models allowed by WMAP
observations.  
To illustrate this, we have plotted points in the Markov chain for
$\Lambda$CDM models that are allowed by WMAP observations in a few projections
in the parameter space (Fig.~\ref{fig:lcdm_sc}).  
Also shown in the same plots are points allowed by abundance of rich clusters
of galaxies. 
This clearly shows that the region in parameter space allowed by cluster
abundance is distinctly smaller than that allowed by CMB observations.
Well known degeneracies between parameters are also highlighted by this
figure, e.g., there is a clear degeneracy between $\Omega_{NR}$ and $h$.   
An important point to note is that if the preferred range for $\sigma_8$ were
to be towards larger values than taken here
\cite{cluster4,2003ApJ...591..599S} then the cluster abundance constraint will
favor models with larger $\Omega_{NR}$.   
There is also a related shift towards $w < -1$. 

\subsection{Models with a constant $w$}

We now consider models with a constant equation of state parameter $w$.  
Introduction of this additional parameter changes the relative effectiveness 
of different observations in constraining cosmological parameters.  
Observations of high redshift supernovae of type Ia constrain the key
cosmological parameters much more strongly than CMB observations for
$\Lambda$CDM models. 
This is no longer the case once we introduce $w$ as a parameter.  
The main reason for this is the degeneracy between $w$ and $\Omega_{NR}$. 

In models with $w \neq 1$, it is necessary to take perturbations in the dark
energy component into account \cite{2004PhRvD..69h3503B,perturb3,perturb4}. 
Here, we study models with and without perturbations in dark energy in order
to illustrate the role played by these perturbations and to study how strongly
these influence determination of cosmological parameters. 
Fig.~\ref{fig:cw} shows the likelihood $I$ for parameters $\Omega_{NR}$, $w$,
$h$, $\Omega_b ~ h^2$, $n$ and $\tau$ in models with and without perturbations
in the dark energy component.   
Models with a larger $\Omega_{NR}$ and smaller $w$ are  better fits to
SN observations, whereas CMB observations prefer models
with smaller $\Omega_{NR}$ and a larger equation of state parameter $w$.  
The combination of these observations and abundance of rich clusters
constrains both the parameters to a fairly narrow range, much narrower than is
allowed by SN observations alone. 
Models with $w > -1$ fare badly with CMB observations when perturbations
in the dark energy component are taken into account.  
For models with a constant $w$, observations allow higher values of $\Omega_b
~ h^2$ than for $\Lambda$CDM models. 
The allowed range in $\tau$ is smaller then the case with no dark energy
perturbations whereas the ranges for $n$ are similar in both. 

We find that the range of $w$ allowed at the $95\%$ confidence limit is
smaller when perturbations in dark energy are allowed.  
For other parameters, the allowed range is similar. 
In other words, if we ignore perturbations in dark energy  we can still make a
reasonable estimate of the range of parameters allowed by observations.  
This fact is of immense use when we work with models that have a varying
equation of state parameter $w$.   
In order to take full effects of dark energy perturbations in these models it
is essential to know full details of the model
\cite{2004PhRvD..69h3503B,perturb3}.     
We cannot include the effect of non-adiabatic perturbations in a model
independent study of dark energy models with a varying $w$. 
Given the fact that ignoring perturbations in dark energy does not lead to an
incorrect estimate of the range of parameters (except for $w$) that is
allowed, we can safely proceed with our analysis without taking perturbations
in dark energy into account. 
As regards the equation of state parameter $w$, ignoring perturbations tends
to allow $w > -1$ models with a larger probability and we should keep this in
mind while interpreting results. 

\begin{figure}
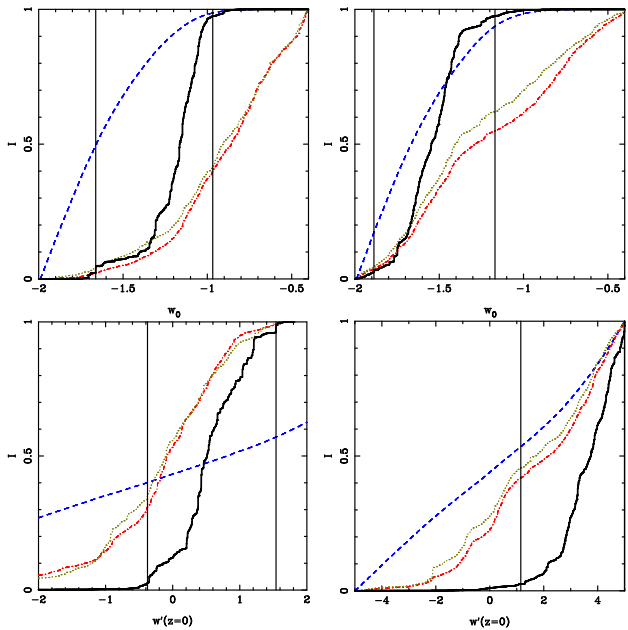

\begin{tabular}{cc}
\begin{minipage}{1.6in}
\centering
\includegraphics[width=1.6in]{fig5a.ps}
\end{minipage}
& 
\begin{minipage}{1.6in}
\centering
\includegraphics[width=1.6in]{fig5b.ps}
\end{minipage}
\\
\begin{minipage}{1.6in}
\centering
\includegraphics[width=1.6in]{fig5c.ps}
\end{minipage}
& 
\begin{minipage}{1.6in}
\centering
\includegraphics[width=1.6in]{fig5d.ps}
\end{minipage}
\\
\end{tabular}
\caption{This figure shows $I$ for parameters that describe the variation of
  the equation of state parameter: $w_0$ and $w'(z=0)$.
  Left panels are for $p=1$ and the right ones are for $p=2$.  Color
  coding used is same as in figure~2.}
\label{fig:varw}
\end{figure}

\begin{figure}
\includegraphics[width=3.3in]{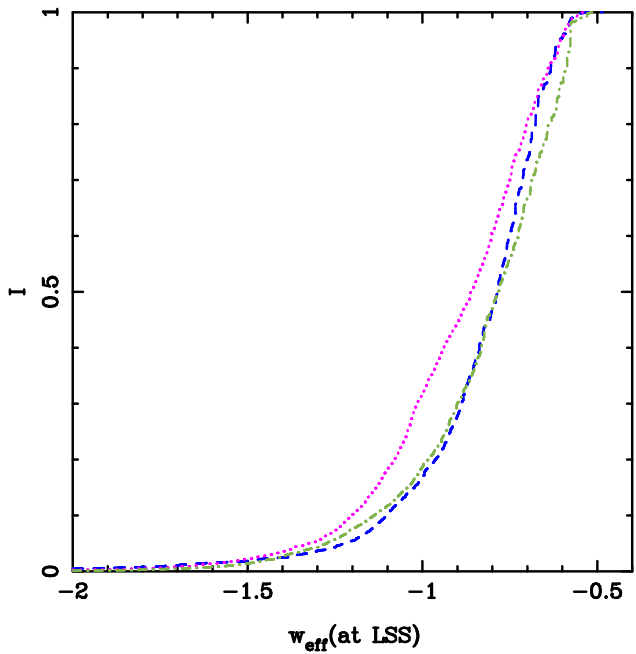}
\caption{This figure shows $I$ for $w_{eff}$ at the redshift of surface of
  last scattering for the models with constant $w$ (blue/dashed curve),
  variable $w$ with $p=1$ (green/dot-dashed curve) and with $p=2$
  (purple/dotted curve).  
  Perturbations in dark energy are not taken into account in these models.
  See text for a more detailed description of this figure.}
\label{fig:weff}
\end{figure}

\subsection{Models with varying $w(z)$}

We now proceed to the case of varying $w$, we use two parameterizations
given in \cite{2005MNRAS.356L..11J}. 
The first of these, corresponding to $p=1$, is a Taylor series expansion for
$w$ in scale factor and this is a very commonly used parameterization. 
The variation of the equation of state parameter is monotonic in this case and
rapid increase of $w$ at low redshifts cannot be allowed as it will lead to $w
\geq -1/3$ at high redshifts. 
The parameterization with $p=2$ avoids this problem to some extent as the
value of $w$ at very high redshifts is the same as the present value, but
there can be a large deviation from this at low redshifts with the deviation
peaking at $z=1$.  
Rapid variation at low red-shifts has been reported \cite{2004MNRAS.354..275A}
on the basis of SN observations \cite{nova_data1,nova_data3}, and even though
these conclusions have been contested \cite{dynamic_de6} it is useful to check
if the larger set of observations support a rapid variation of $w$ at low
redshifts.  

We have already seen in the discussion of fiducial models, supernova
observations do not distinguish between models with a constant $w$ and models
with a variable equation of state parameter.  
Also, WMAP observations do not differentiate between the three classes of
models being studied here in a statistically significant manner. 

Fig.~\ref{fig:varw} shows the allowed ranges of parameters $w_0$ and $w'(z=0)$
for $p=1$ (left panels) and $p=2$ (right panels). 
In both the parameterizations, the preferred values for $w_0$ with supernova
observations are for phantom models ($w_0 \leq -1$) and a tendency for a
larger $w$ at intermediate redshift, i.e., $w'_0 \geq 0$ though supernova
observations do not provide a clear constraint on this parameter. 
This is perhaps related to the fact that there is a strong degeneracy in
$\Omega_{NR}$ and $w_0$.  
CMB observations and abundance of rich clusters of galaxies allow models
around the $\Lambda$CDM model, which is fairly close to the centre of the
allowed region.  
A combination of these three observations rejects models with $w_0 \ll -1$ due
to CMB constraints and $w_0 > -1$ due to SN constraints. 
Even though all the observations allow $w'_0=0$, the combination of these
observations does not favor such models.
This implies that the overlap of allowed regions for the three observations is
stronger for models with $w'_0 \neq 0$, it is clear from table~3 that there
is no overlap between SN and WMAP at $68\%$ confidence limit for constant
$w$. 
The $\Lambda$CDM model, i.e., $w_0=-1$ and $w'_0=0$ is a marginally allowed 
model for both the parameterizations. 

\begin{figure*}
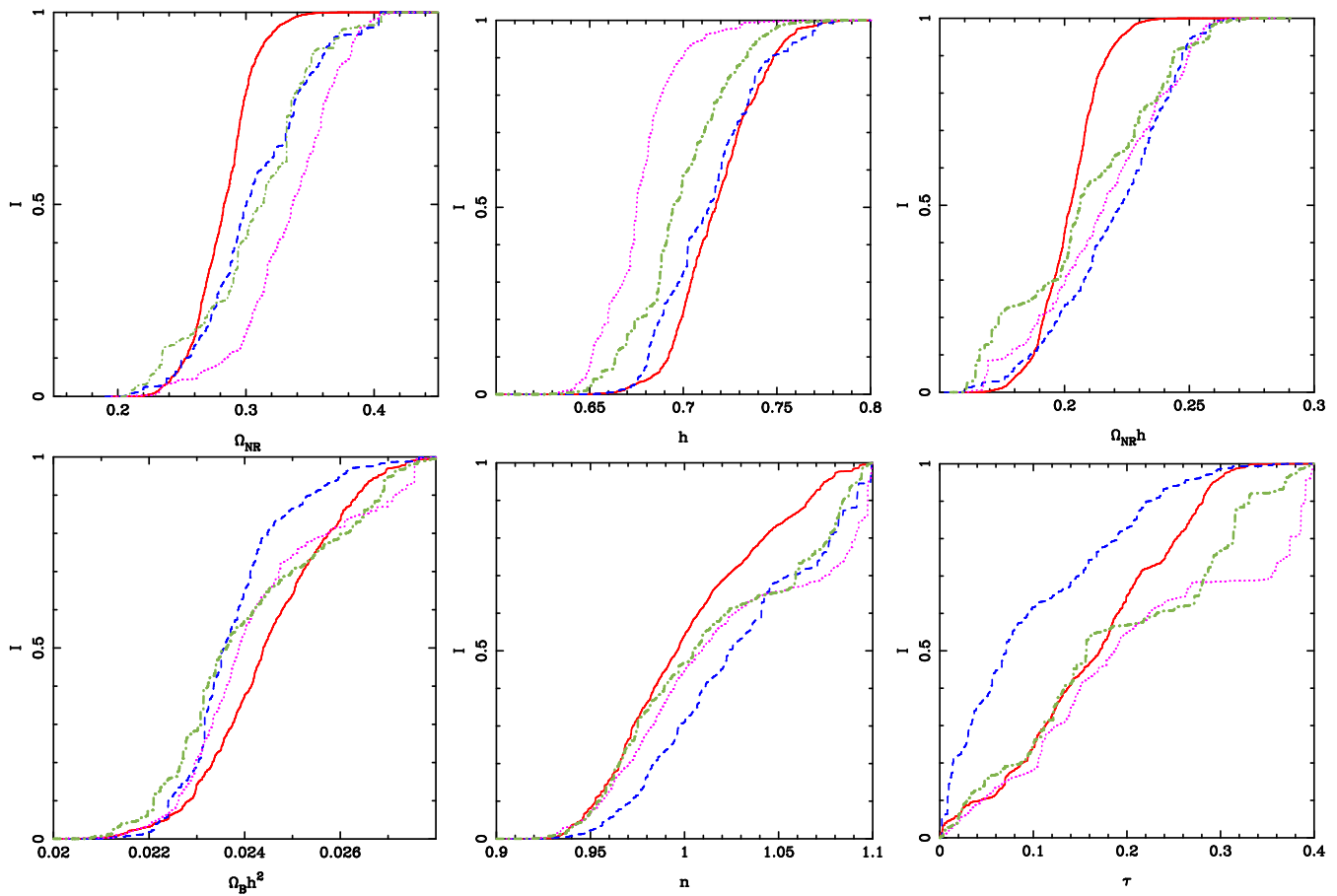

\centering
\begin{tabular}{ccc}
\begin{minipage}{2.3in}
\centering
\includegraphics[width=2.3in]{fig8a.ps}
\end{minipage}
& 
\begin{minipage}{2.3in}
\centering
\includegraphics[width=2.3in]{fig8b.ps}
\end{minipage}
&
\begin{minipage}{2.3in}
\centering
\includegraphics[width=2.3in]{fig8c.ps}
\end{minipage}
\\ 
\begin{minipage}{2.3in}
\centering
\includegraphics[width=2.3in]{fig8d.ps}
\end{minipage}
&
\begin{minipage}{2.3in}
\centering
\includegraphics[width=2.3in]{fig8e.ps}
\end{minipage}
& 
\begin{minipage}{2.3in}
\centering
\includegraphics[width=2.3in]{fig8f.ps}
\end{minipage}
\\
\end{tabular}
\caption{Panels for $\Omega_{NR}$, $h$, $\Gamma = \Omega_{NR} ~ h$, $\Omega_b ~
  h^2$, $n$ and $\tau$.  
  Red/solid curve is for $\Lambda$CDM, blue/dashed curve for models with
  constant $w$, green/dot-dashed for $p=1$ and purple/dotted curve for $p=2$.}
\label{fig:lik_all}
\end{figure*}

\begin{figure*}
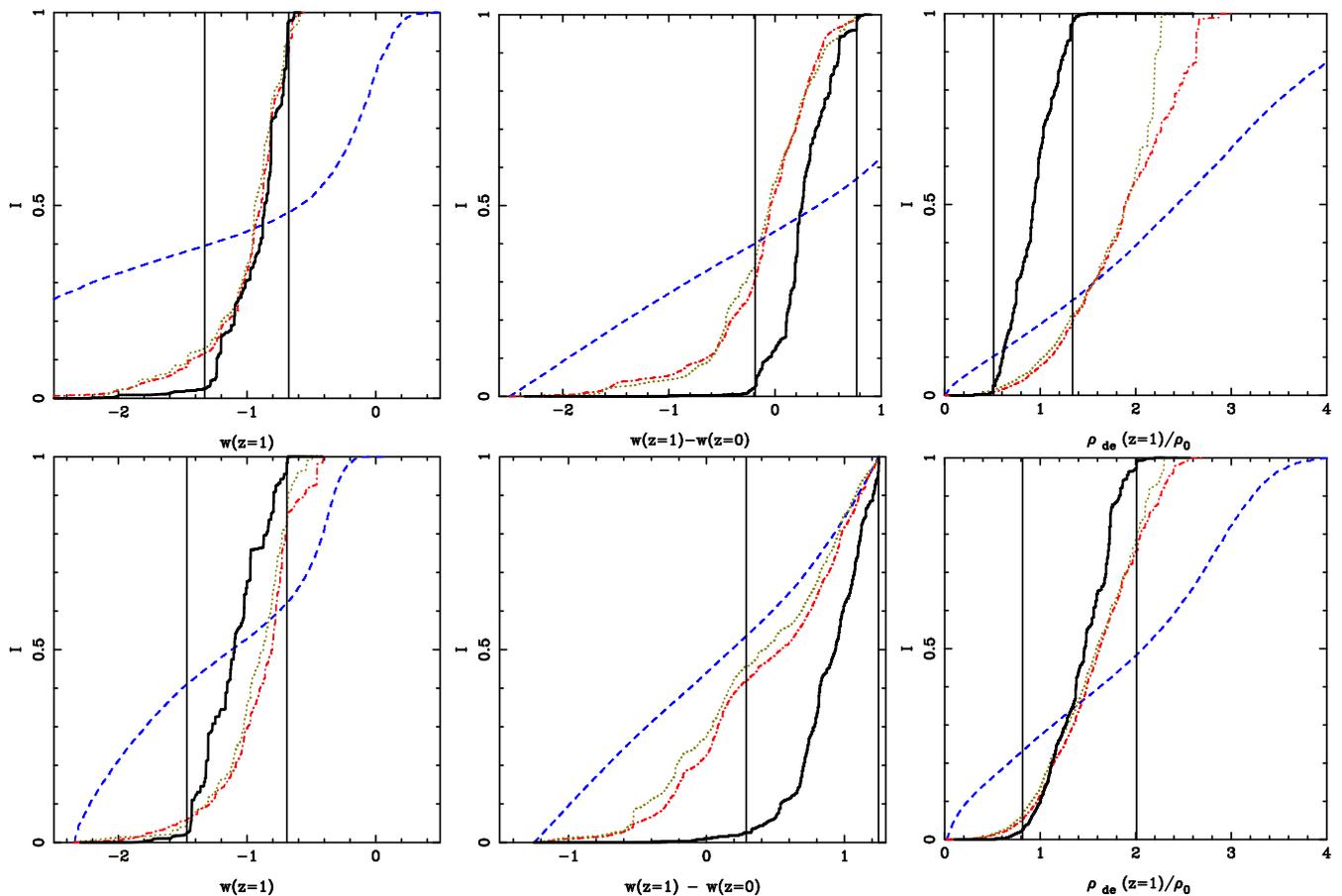

\centering
\begin{tabular}{ccc}
\begin{minipage}{2.3in}
\centering
\includegraphics[width=2.3in]{fig9a.ps}
\end{minipage}
& 
\begin{minipage}{2.3in}
\centering
\includegraphics[width=2.3in]{fig9b.ps}
\end{minipage}
&
\begin{minipage}{2.3in}
\centering
\includegraphics[width=2.3in]{fig9c.ps}
\end{minipage}
\\
\begin{minipage}{2.3in}
\centering
\includegraphics[width=2.3in]{fig9d.ps}
\end{minipage}
&
\begin{minipage}{2.3in}
\centering
\includegraphics[width=2.3in]{fig9e.ps}
\end{minipage}
& 
\begin{minipage}{2.3in}
\centering
\includegraphics[width=2.3in]{fig9f.ps}
\end{minipage}
\\
\end{tabular}
\caption{The top and bottom panels in the figure correspond to
  parameterizations with $p=1$ and $p=2$ respectively. 
  The figures on the left show likelihood for the
  equation of state parameter $w$ at redshift $z=1$ for models which
  have equation of state lying  within the range $-1.1 < w_0 < -0.9$.
In the middle panel we plot the likelihood for variation in the
equation of state parameter from the present value at redshift $z=1$. 
The third panel shows the allowed range of change in the dark energy
density upto $z=1$.
Color coding is same as in Fig. 2}
\label{fig:rhode}
\end{figure*}

To understand the nature of constraint from CMB observations
\cite{2003MNRAS.343..533D}, we computed the  likelihood of $w_{eff}$ (as
defined in eqn.(\ref{eqn:weff})) for models by comparing these with WMAP 
data. 
This is then compared with the likelihood for $w$ in models with a constant
equation of state parameter.    
We have plotted this in Fig.~\ref{fig:weff} for $p=1$ as well as $p=2$,
along with the curve for constant $w$ (with no perturbations in dark energy).
All the three curves show very similar behavior and the $95\%$ confidence
limit is identical for all three ($-1.5 \leq w_{eff} \leq -0.6$).  
This also shows that the CMB observations primarily provide a constraint for
$w_{eff}$.   
Given that adding perturbations reduces the likelihood for models with $w >
-1$, it is likely that detailed analysis of a model with perturbations in dark
energy taken into account will limit the range for $w_{eff}$ in this region. 

Lastly, we study the effect of varying dark energy on other parameters.  
The specific question we wish to address is, how the allowed ranges for
these parameters change if we allow variation of dark energy.  
Fig.~\ref{fig:lik_all} shows likelihood for the parameters studied here.   
We have plotted the likelihood using all three observations for the
$\Lambda$CDM models, constant $w$ models as well as for $p=1$ and $p=2$.  
For most parameters, the effect of $w \neq -1$ and varying dark energy is to
increase the range of allowed values.  
This increase in the allowed range is sometimes accompanied by a shift,
e.g. for $h$ where varying dark energy models fit observations better with
smaller values as compared to the $\Lambda$CDM model as well as models with
constant $w$.  
This shift is primarily due to models with $w > -1$ and this point has been
noted in other analyses as well \cite{wmap_params}.
If this is the case then including perturbations in dark energy may well
remove this shift.  

Similarly, larger values of spectral index $n$ and optical depth to the epoch
of reionisation $\tau$ fit observations better.  
As these parameters can be constrained using other observations, we may be
able to restrict models with varying $w$ by constraining the values of these
parameters. 
For example, polarization anisotropies in the CMB can be used to constrain
$\tau$ \cite{cmbrev1,2003ApJ...583...24K}.  
We find that the presently available information from WMAP about polarization
anisotropies does not lead to a significantly improved constraint on
the parameter $\tau$. 

\begin{figure*}
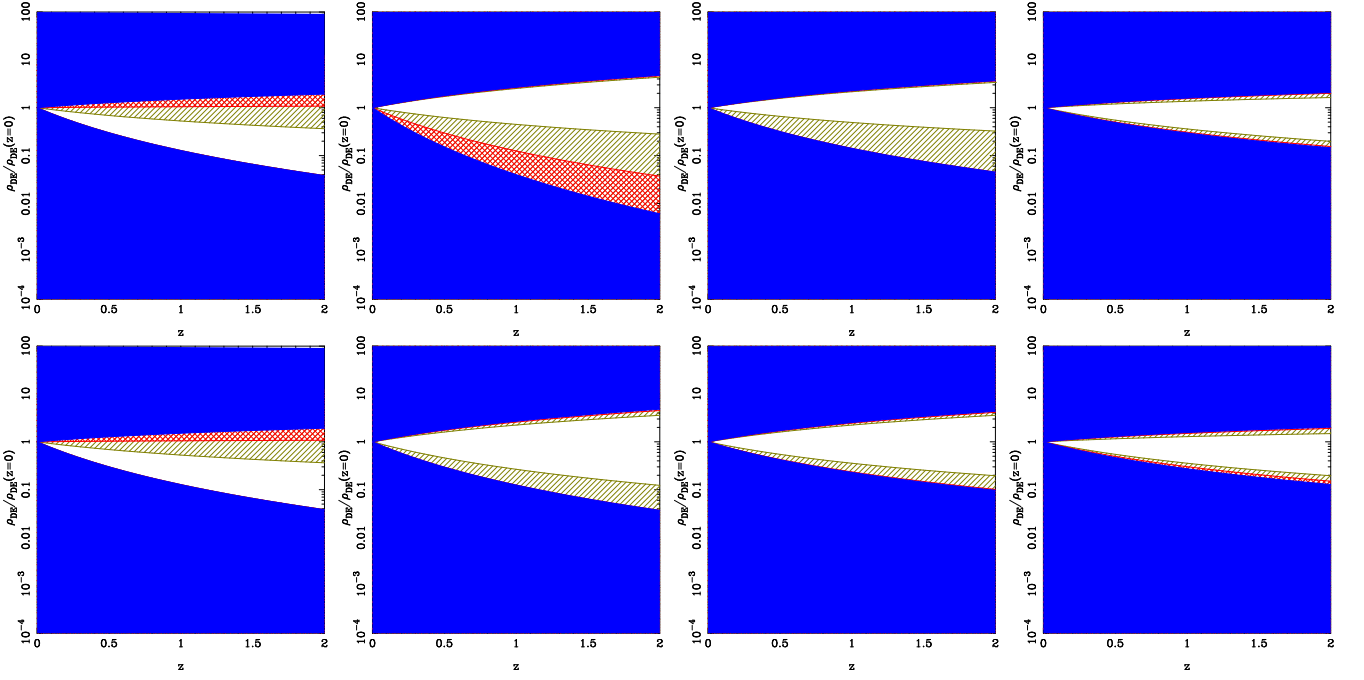

\centering
\begin{tabular}{cccc}
\begin{minipage}{1.7in}
\centering
\includegraphics[width=1.7in]{fig10a.ps}
\end{minipage}
&
\begin{minipage}{1.7in}
\centering
\includegraphics[width=1.7in]{fig10b.ps}
\end{minipage}
&
\begin{minipage}{1.7in}
\centering
\includegraphics[width=1.7in]{fig10c.ps}
\end{minipage}
&
\begin{minipage}{1.7in}
\centering
\includegraphics[width=1.7in]{fig10d.ps}
\end{minipage}
\\
\begin{minipage}{1.7in}
\centering
\includegraphics[width=1.7in]{fig10a.ps}
\end{minipage}
&
\begin{minipage}{1.7in}
\centering
\includegraphics[width=1.7in]{fig10e.ps}
\end{minipage}
&
\begin{minipage}{1.7in}
\centering
\includegraphics[width=1.7in]{fig10f.ps}
\end{minipage}
&
\begin{minipage}{1.7in}
\centering
\includegraphics[width=1.7in]{fig10g.ps}
\end{minipage}
\\
\end{tabular}
\caption{In this figure we plot evolution of dark energy density as a function
of redshift. The top panel is for models without dark energy perturbations and
the lower panel is with dark energy perturbations included. 
The left plot in both the rows shows  the variation in energy density
allowed by Supernova observations. The green/hatched region is excluded at $68\%$
confidence level, red/cross-hatched at $95\%$ confidence level and blue/solid
at $99\%$ confidence level. The white region is the allowed region in variation at $68\%$
significance level. The plot which is second from left (in both the
cases) displays the allowed range by WMAP data alone, color scheme being the
same as  in left figure. The next figure shows allowed range by WMAP and
cluster abundance observations. The last column  shows contribution from
combined analysis of the three observations.}  
\label{fig:rhode1}
\end{figure*}
\begin{figure*}
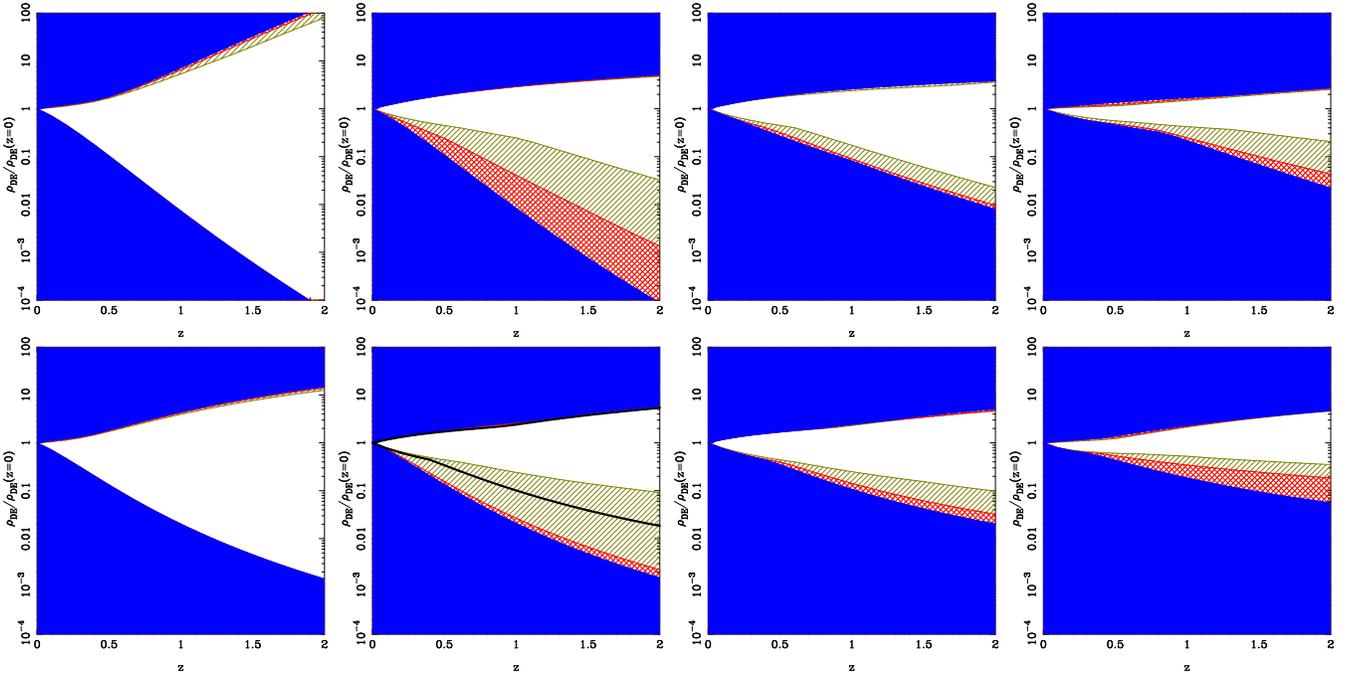

\begin{tabular}{cccc}
\begin{minipage}{1.7in}
\centering
\includegraphics[width=1.7in]{fig11a.ps}
\end{minipage}
&
\begin{minipage}{1.7in}
\centering
\includegraphics[width=1.7in]{fig11b.ps}
\end{minipage}
&
\begin{minipage}{1.7in}
\centering
\includegraphics[width=1.7in]{fig11c.ps}
\end{minipage}
&
\begin{minipage}{1.7in}
\centering
\includegraphics[width=1.7in]{fig11d.ps}
\end{minipage}
\\ 
\begin{minipage}{1.7in}
\centering
\includegraphics[width=1.7in]{fig11e.ps}
\end{minipage}
&
\begin{minipage}{1.7in}
\centering
\includegraphics[width=1.7in]{fig11f.ps}
\end{minipage}
&
\begin{minipage}{1.7in}
\centering
\includegraphics[width=1.7in]{fig11g.ps}
\end{minipage}
&
\begin{minipage}{1.7in}
\centering
\includegraphics[width=1.7in]{fig11h.ps}
\end{minipage}
\\
\end{tabular}
\caption{Same as figure~9 but 
  for varying dark energy models. The top panel is for $p=1$ and the lower
  panel is for $p=2$. Starting from left, the figures show allowed dark energy
  density variation from SN observations, WMAP observations, WMAP observations
  combined with cluster abundance requirements and by  combined analysis.  The
  region enclosed by black solid lines (lower panel, second frame from left)
  is obtained from the derived constraints on $w_{eff}$, see text for
    details.}  
\label{fig:rhode2}
\end{figure*}

\subsection{Evolution of dark energy}

We now summarize our results for the allowed variation of dark energy, once
all three observational constraints have been taken into account.  
In the left panel of  Fig. \ref{fig:rhode}, we have plotted the
cumulative likelihood for the equation of state parameter $w$ at redshift
$z=1$. 
Here we have used models which lie within the range $-1.1 < w_0 <
-0.9$.
The upper and lower panels correspond to parameterizations with $p=1$
and $p=2$ respectively.
The allowed range of variation in the equation of state by supernova
observations is much larger than that allowed by WMAP results. 
This, again, is a reflection of the strong preference of supernova
observations for $w \ll -1$ and of the large parameter space allowed by
SN data. 
(Our result that SN data prefers  $w \ll -1$ with large variation is consistent
with previous published analysis e.g., in
\cite{2004ApJ...617L...1B}.)  
The range of values in both the parameterizations are similar for this subset
of models. 
In the middle panel we have shown the likelihood for variation in the
equation of state parameter from the present to its value at redshift $z=1$.   
The allowed ranges of variation in dark energy equations of state are
different for these two parameterizations. 
In fact, the constant dark energy equation of state is ruled out at
$95\%$ confidence level for $p=2$ (that is, the probability of occurrence is
less than $0.05$) when all the constraints are taken into account even though
each observational constraint allows such models individually. 
Clearly, the models with constant $w$ allowed by each of these observations
are ruled out by other observations. 
The $\Lambda$CDM model is allowed for $p=2$ at $77\%$ C.L. (with probability
of $23\%$) by SN observations and by $0.9\%$ C.L. ($\mathcal{P}=0.991$) by
WMAP observations.  

In the right panel, we have shown the ratio of dark energy density at $z=1$
and the present value.  
The variation allowed by SN observations is very large, whereas WMAP
limits the variation to within a factor $2.5$ at $95\%$ confidence limits. 
This drives the joint analysis to restrict variation even further.  
That WMAP observations provide a much tighter constraint on the equation
of state as compared to SN observations was earlier shown
in \cite{2005MNRAS.356L..11J}.

In Fig.~\ref{fig:rhode1} we show the allowed range of variation of dark energy
as function of redshift for $w=constant$ models, with and without
perturbations at $68\%$, $95\%$ and $99\%$ confidence levels.   
The figure shows the disparity in allowed range by SN observations and
WMAP observations at $68\%$ confidence level. 
Allowing perturbations in dark energy gives a similar range as compared to the
case where dark energy perturbations are absent.
In Fig.~\ref{fig:rhode2} we plot this range for varying dark energy models,
top panel for models with $p=1$ and lower panel for $p=2$. 
As mentioned earlier (see also \cite{2005MNRAS.356L..11J}), SN observations
 allow a much wider range in change of dark energy density with redshift.  
The variation allowed by WMAP is smaller in all cases except constant
 $w$.  
The combination of the three constraints allows very little variation, with 
maximum allowed variation in dark energy density being by a factor $5$ up to 
$z=2$ at $68\%$ confidence limit.
The allowed variation in dark energy density is similar in 
both the cases, indicating that the constraints on this quantity are 
parameterization independent to a large extent. 

Finally, we would like to make some comments regarding the fact that WMAP
constrains the evolution of dark energy more effectively than SN. 
This arises essentially from the constraint on the angular diameter distance
to the last  scattering surface, or --- equivalently --- the effective
equation of state parameter $w_{eff}$.
(The Integrated Sachs-Wolfe effect and the contribution of other
parameters turns out to be less important.)
To illustrate this point, we have compared the constraints on dark
energy density for $p=2$ with those implied by constraints on $w_{eff}$. 
The second figure in the lower panel in Fig.~\ref{fig:rhode2} shows the
allowed range in evolution of dark energy density allowed by WMAP data alone
for $p=2$. 
We have also plotted the allowed range for dark energy density as a function
of redshift if $-1.6 \leq w_{eff} \leq -0.6$, by thick black lines.  
This is derived by using all $w_0$ and $w'_0$ that lead to $w_{eff}$ in the
range given above, and computing the highest and lowest dark energy density
amongst this set of models at each redshift. 
We allow $w_0$ and $w'_0$ to vary in the range specified in the priors. 
The region allowed by the range in $w_{eff}$ and that directly obtained from
all allowed models is similar, with the latter allowing larger variation for
phantom models. 
This reiterates our claim that the main constraint from WMAP data on
dark energy parameters is on the value of $w_{eff}$ at the last scattering
surface. 
We believe that the larger range allowed at $95\%$ confidence limit is due
mainly to the ISW effect. 

\begin{table*}
\label{tab:range1}
\caption{This table lists the range of relevant parameters allowed within 68\%
  confidence limit from SN,  WMAP, WMAP + cluster   abundance (CA)
  requirements and after combining all observations.}  
\begin{center}
\begin{tabular}{||l|c|c|c|c|c|c||}
\hline
\hline
Parameter         &$\Lambda$CDM&w=const. & w=const. &p=1         & p=2 & \\
                  &            &         & with perturbations  &     &  & \\
\hline
\hline
                  &$0.256$---$0.3$  & $0.37$---$0.47$ &$0.37$---$0.47$&$0.19$---$0.47$ &$0.18$---$0.47$ & SN\\
                  &$0.26$---$0.42$  & $0.25$---$0.37$ &$0.28$---$0.42$&$0.22$---$0.39$ &$0.21$---$0.37$& WMAP \\
$\Omega_{NR}$     &$0.265$---$0.424$& $0.26$---$0.38$ &$0.29$---$0.43$&$0.24$---$0.42$ &$0.24$---$0.4$ &  WMAP+CA\\
                  &$0.26$---$0.3$   & $0.27$---$0.35$ &$0.26$---$0.38$&$0.26$---$0.35$ &$0.3$---$0.37$ & SN+WMAP+CA\\
\hline
                  &           &$-1.9$ --- $-1.39$ &$-1.9$ --- $-1.39$&$-1.89$ --- $-1.31$&$-1.9$ --- $-1.33$&SN\\
                  &           &$-1.0$ --- $-0.67$ &$-1.2$ --- $-0.77$&$-1.19$ --- $-0.56$&$-1.67$ --- $-0.7$&WMAP\\
$w_0$             &           &$-1.02$ --- $-0.67$&$-1.13$ --- $-0.78$ &$-1.28$ --- $-0.56$&$-1.71$ --- $-0.75$ &WMAP+CA\\
                  &           &$-1.41$ --- $-1.02$&$-1.37$ --- $-1.04$&$-1.3$ --- $-1.05$&$-1.69$ --- $-1.4$&SN+WMAP+CA\\
\hline
                  &           &              && $-3.2$ --- $3.46$ &$-3.3$ --- $3.99$ &SN\\
                  &           &              && $-0.94$ --- $0.67$ &$-0.74$ --- $4.15$ &WMAP\\
$w'(z=0)$         &           &              && $-1.0$ ---  $0.75$ &$-1.03$ --- $3.94$ &WMAP+CA\\
                  &           &              && $0.21$ --- $1.06$ &$2.74$ --- $4.58$ &SN+WMAP+CA\\

\hline
\hline
\end{tabular}
\end{center}
\end{table*}
\begin{table*}
\label{tab:range2}
\caption{This table lists the range of parameters allowed within 95\%
  confidence limit from SN,  WMAP, WMAP + cluster
  abundance (CA) requirements and after combining all observations.} 
\begin{center}
\begin{tabular}{||l|c|c|c|c|c|c||}
\hline
\hline
Parameter         &$\Lambda$CDM&w=const. & w=const. &p=1         & p=2 &\\
                  &            &         & with perturbations  &     & &\\
\hline
\hline
                  &$0.23$ --- $0.33$  & $0.27$ --- $0.49$   &$0.27$ ---
$0.495$&$0.11$ --- $0.49$ &$0.11$ --- $0.49$ & SN\\
                  &$0.20$ --- $0.45$  & $0.16$ --- $0.43$   &$0.20$ ---
$0.47$&$0.17$ --- $0.45$ &$0.18$ --- $0.44$& WMAP \\
$\Omega_{NR}$     &$0.22$ --- $0.44$  & $0.21$ --- $0.42$   &$0.22$ ---
$0.46$&$0.21$ --- $0.46$  &$0.2$ --- $0.44$ &  WMAP+CA\\
                  &$0.23$ --- $0.33$  & $0.22$ --- $0.4$   &$0.22$ --- $0.4$&
$0.21$ --- $0.403$ &$0.23$ --- $0.39$ & SN+WMAP+CA\\
\hline
                  &           &$-1.97$ --- $-1.1$ &$-1.97$ --- $-1.1$&$-1.97$ --- $-1.03$&$-1.97$ --- $-1.05$&SN\\
                  &           &$-1.39$ --- $-0.58$&$-1.6$ --- $-0.63$&$-1.64$ --- $-0.42$&$-1.93$ --- $-0.43$&WMAP\\
$w_0$             &           &$-1.34$ --- $-0.63$&$-1.4$ --- $-0.66$ &$-1.73$
--- $-0.42$&$-1.95$ --- $-0.47$ &WMAP+CA\\
                  &           &$-1.47$ --- $-0.91$&$-1.48$ --- $-0.96$&$-1.66$
--- $-0.97$&$-1.89$ --- $-1.17$&SN+WMAP+CA\\
\hline
                  &           &              && $-4.72$ --- $4.6$ &$-4.73$ --- $4.85$ &SN\\
                  &           &              && $-3.09$ --- $1.32$ &$-2.5$ --- $4.87$ &WMAP\\
$w'(z=0)$         &           &              && $-3$ --- $1.35$ &$-2.7$ ---    $4.8$ &WMAP+CA\\
                  &           &              && $-0.38$ --- $1.54$ &$1.15$ --- $4.99$ &SN+WMAP+CA\\
\hline
$w_{eff}(at LSS)$   &           &$-1.4$ --- $-0.5$&& $-1.48$ --- $-0.59$& $-1.41$ --- $-0.58$& \\
\hline
                  &$0.02$ --- $0.027$&$0.021$ --- $0.028$ &$0.02$ ---
$0.027$&$0.02$ --- $0.027$ &$0.02$ --- $0.027$& WMAP\\
$\Omega_b ~ h^2$  &$0.021$ --- $0.027$& $0.022$ --- $0.028$ &$0.022$ ---
$0.027$&$0.021$ --- $0.027$ &$0.022$ --- $0.028$&WMAP+CA\\
                  &$0.022$ --- $0.027$& $0.022$ --- $0.027$  &$0.022$ ---
$0.027$ &$0.021$ --- $0.027$ &$0.022$ --- $0.028$&SN+WMAP+CA\\
\hline
                  &$0.61$ --- $0.79$  & $0.61$ --- $0.78$  &$0.6$ ---
$0.79$&$0.6$ --- $0.78$  & $0.61$ --- $0.78$ & WMAP\\
$h$               &$0.61$ -- $0.78$  & $0.61$ --- $0.77$  &$0.61$ ---
$0.78$&$0.61$ --- $0.7$8  & $0.6$ --- $0.78$ &WMAP+CA\\
                  &$0.67$ --- $0.76$  & $0.67$ --- $0.77$  &$0.66$ --- $0.76$
&$0.65$ --- $0.75$  & $0.64$ --- $0.72$ &SN+WMAP+CA\\
\hline
                  &$0.002$ --- $0.33$  & $0.011$ --- $0.39$  &$0.007$ ---
$0.35$ &$0.13$ --- $0.4$  &$0.016$ --- $0.39$ &WMAP\\
$\tau$            &$0.001$ --- $0.32$  & $0.028$ --- $0.36$  &$0.006$ ---
$0.35$&$0.01$ --- $0.37$  &$0.015$ --- $0.387$ & WMAP+CA\\
                  &$0.003$ --- $0.3$  & $0.002$ --- $0.29$  &$0.022$ ---
$0.26$&$0.006$ --- $0.38$  &$0.01$ --- $0.39$ &SN+WMAP+CA\\
\hline
                  &$0.93$ --- $1.08$  & $0.93$ --- $1.1$   &$0.93$ ---
$1.09$&$0.93$ --- $1.1$   &$0.93$ --- $1.09$&WMAP\\
$n$               &$0.94$ --- $1.09$  & $0.94$ --- $1.1$   &$0.94$ ---
$1.09$&$0.94$ --- $1.09$   &$0.94$ --- $1.09$ &WMAP+CA\\
                  &$0.96$ --- $1.09$  & $0.94$ --- $1.08$  &$0.95$ --- $1.08$
&$0.94$ --- $1.09$   &0$.93$ --- $1.09$ &SN+WMAP+CA\\
\hline
\hline
\end{tabular}
\end{center}
\end{table*}

\section{Conclusions} \label{sec:conclude}

In this paper we presented a detailed analysis  of constraints on
cosmological parameters from different observations. 
In particular we focussed on constraints on dark energy equation of
state, its present value and the allowed range of variation in it.

It is demonstrated that the allowed range for the equation of state
parameter $w$ is smaller if dark energy is allowed to cluster.  
Including perturbations mainly affects models with $w > -1$.

We find that WMAP observations do not distinguish between the $\Lambda$CDM
model, models with a constant equations of state parameter $w$ and models with
a variable $w$, the change in $\chi^2_W$ for best fit models is less than $3$
even as the number of parameters is increased by three. 
WMAP allows only a modest variation in energy density of dark energy, with
maximum variation being less than a factor of three in $99\%$ confidence
limit up to $z=1$.  
We infer that the main constraint from WMAP observations is for the
derived quantity $w_{eff}$, essentially representing the distance to the last
scattering surface. 

SN observations favour models with $w < -1$ and $\Omega_{nr} > 0.4$.  
A corollory is that if we restrict to models with $w \geq -1$ then the
$\Lambda$CDM model is the most favoured model.  
Without this restriction the $\Lambda$CDM model is allowed only marginally by
the combination of observations used here, this is driven mainly by SN
observations.  

Allowing variation in dark energy has an impact on other cosmological
parameters as the allowed range for many of these parameters becomes larger.  
Conversely, better measurements of these parameters will allow us to constrain
models of dark energy. 

We find significant tension between different observations. 
Our key conclusions in this regard may be summarised as follows:
\begin{itemize}
\item
SN observations favor models with large $\Omega_{NR}$ and $w \ll -1$.
Indeed, the best fit model is at the edge of our priors. 
\item
Enlarging priors to $0.1 \leq \Omega_{NR} \leq 0.6$, $-0.3 \geq w \geq -3.0$
does not lead to a better fit model for SN observations, indicating
that our default priors are sufficiently wide for joint estimation of
parameters. 
This is  because $w \ll -1$ is rejected by WMAP observations. 
\item
WMAP observations favor models with $w \sim -1$ with a marginal preference
for $w > -1$.  
Including perturbations in dark energy removes this marginal preference as
well. 
\item
For constant $w$ models and models with variable $w$, the best fit model of
each observation is ruled out by the other observations at a high
significance level.  
As an example, the model that best fits the WMAP observations is completely
ruled out by SN observations ($\Delta\chi_S^2 = 53$).  
The problem is slightly less serious if perturbations in dark energy are taken
into account ($\Delta\chi_S^2 = 12.5$). 
\item
There is overlap of allowed regions  at $95\%$ (or better) by these
observations, though there is little overlap of allowed regions  at $68\%$
confidence limit (see table~3).  It can, of course, be argued that situation
is not alarming given that there is an overlap of allowed regions in parameter
space at $95\%$.  But we find this offset noteworthy.
\item
Using larger values for $\sigma_8$, as indicated by some recent analyses
\cite{cluster4,2003ApJ...591..599S} favors models with a slightly larger
$\Omega_{NR}$ and slightly lower $w_{eff}$.   
\item
Given that the preference of individual observations for different types of
models is not understood, and the fact that the best fit model of one is ruled
out by the other, it is necessary to use a combination of observations for
reliable constraints on models of dark energy.  
Use of either one of the observations is likely to mislead. 
\item
Our conclusions are not sensitive to priors used for parameters other than
$\Omega_{NR}$.  
Limiting priors for matter density to $0.1 \leq \Omega_{NR} \leq 0.3$ enhances
the overlap between SN and WMAP observations and removes the tension
between SN and WMAP observations for constant $w$ models. 
\item
If we repeat the analysis of models with variable $w$ with this restricted
priors then we find that SN observations strongly favor a variable $w$ as
compared to constant $w$.
There is no significant tension between SN and WMAP models in this
class of models even for the wider priors, but this is mainly due to a larger
number of parameters.  
\item
Using only the {\sl Gold} data set for supernovae instead of the {\sl Gold $+$
  Silver} used here reduces the tension between the WMAP and supernova
observations by a marginal amount. 
\end{itemize}

Given the points noted here regarding tension between different observations,
it is important that some effort is made to look for systematic effects in
observations as well as in analysis of observations. 
We have tested our analysis for systematic effects by varying priors and our
findings appear to be independent of the chosen priors, the only instance of
change in results is mentioned above. 
Since the SN data set which is used by most people (including in this work)
arises from different sources, one needs to be careful regarding hidden
systematics (see e.g., the discussion in \cite{saul}).
When larger, homogeneous SN datasets are available in future (like for
example, from SNLS), it is likely that the tension between the SN observations
and WMAP results disappear. If it does not, and the agreement continues to
exist only at 3-sigma level, there is some cause for concern.


\section*{Acknowledgements}

JSB thanks Stefano Borgani and U. Seljak and TP thanks S.Perlmutter for 
useful comments.
The numerical work in this paper was done using cluster computing
facilities at the Harish-Chandra Research Institute
(http://cluster.mri.ernet.in/). 
This research has made use of NASA's Astrophysics Data System.


\begin{thebibliography}{158}
\expandafter\ifx\csname natexlab\endcsname\relax\def\natexlab#1{#1}\fi
\expandafter\ifx\csname bibnamefont\endcsname\relax
  \def\bibnamefont#1{#1}\fi
\expandafter\ifx\csname bibfnamefont\endcsname\relax
  \def\bibfnamefont#1{#1}\fi
\expandafter\ifx\csname citenamefont\endcsname\relax
  \def\citenamefont#1{#1}\fi
\expandafter\ifx\csname url\endcsname\relax
  \def\url#1{\texttt{#1}}\fi
\expandafter\ifx\csname urlprefix\endcsname\relax\def\urlprefix{URL }\fi
\providecommand{\bibinfo}[2]{#2}
\providecommand{\eprint}[2][]{\url{#2}}

\bibitem{crisis2}
\bibinfo{author}{\bibfnamefont{J.~P.} \bibnamefont{{Ostriker}}}
  \bibnamefont{and} \bibinfo{author}{\bibfnamefont{P.~J.}
  \bibnamefont{{Steinhardt}}}, \bibinfo{journal}{\nat}
  \textbf{\bibinfo{volume}{377}}, \bibinfo{pages}{600} (\bibinfo{year}{1995}).
\bibinfo{author}{\bibfnamefont{S.~D.~M.} \bibnamefont{{White}}},
  \bibinfo{author}{\bibfnamefont{J.~F.} \bibnamefont{{Navarro}}},
  \bibinfo{author}{\bibfnamefont{A.~E.} \bibnamefont{{Evrard}}},
  \bibnamefont{and} \bibinfo{author}{\bibfnamefont{C.~S.}
  \bibnamefont{{Frenk}}}, \bibinfo{journal}{\nat}
  \textbf{\bibinfo{volume}{366}}, \bibinfo{pages}{429} (\bibinfo{year}{1993}).
\bibinfo{author}{\bibfnamefont{J.~S.} \bibnamefont{{Bagla}}},
  \bibinfo{author}{\bibfnamefont{T.}~\bibnamefont{{Padmanabhan}}},
  \bibnamefont{and} \bibinfo{author}{\bibfnamefont{J.~V.}
  \bibnamefont{{Narlikar}}}, \bibinfo{journal}{Comments on Astrophysics}
  \textbf{\bibinfo{volume}{18}}, \bibinfo{pages}{275} (\bibinfo{year}{1996}),
  \eprint{arXiv:astro-ph/9511102} ;  
\bibinfo{author}{\bibfnamefont{G.}~\bibnamefont{{Efstathiou}}},
  \bibinfo{author}{\bibfnamefont{W.~J.} \bibnamefont{{Sutherland}}},
  \bibnamefont{and} \bibinfo{author}{\bibfnamefont{S.~J.}
  \bibnamefont{{Maddox}}}, \bibinfo{journal}{\nat}
  \textbf{\bibinfo{volume}{348}}, \bibinfo{pages}{705} (\bibinfo{year}{1990}).



\bibitem{nova_data1}
\bibinfo{author}{\bibfnamefont{J.~L.} \bibnamefont{{Tonry}}},
  \bibinfo{author}{\bibfnamefont{B.~P.} \bibnamefont{{Schmidt}}},
  \bibinfo{author}{\bibfnamefont{B.}~\bibnamefont{{Barris}}},
  \bibinfo{author}{\bibfnamefont{P.}~\bibnamefont{{Candia}}},
  \bibinfo{author}{\bibfnamefont{P.}~\bibnamefont{{Challis}}},
  \bibinfo{author}{\bibfnamefont{A.}~\bibnamefont{{Clocchiatti}}},
  \bibinfo{author}{\bibfnamefont{A.~L.} \bibnamefont{{Coil}}},
  \bibinfo{author}{\bibfnamefont{A.~V.} \bibnamefont{{Filippenko}}},
  \bibinfo{author}{\bibfnamefont{P.}~\bibnamefont{{Garnavich}}},
  \bibinfo{author}{\bibfnamefont{C.}~\bibnamefont{{Hogan}}},
  \bibnamefont{et~al.}, \bibinfo{journal}{\apj} \textbf{\bibinfo{volume}{594}},
  \bibinfo{pages}{1} (\bibinfo{year}{2003}); 
\bibinfo{author}{\bibfnamefont{B.~J.} \bibnamefont{{Barris}}},
  \bibinfo{author}{\bibfnamefont{J.~L.} \bibnamefont{{Tonry}}},
  \bibinfo{author}{\bibfnamefont{S.}~\bibnamefont{{Blondin}}},
  \bibinfo{author}{\bibfnamefont{P.}~\bibnamefont{{Challis}}},
  \bibinfo{author}{\bibfnamefont{R.}~\bibnamefont{{Chornock}}},
  \bibinfo{author}{\bibfnamefont{A.}~\bibnamefont{{Clocchiatti}}},
  \bibinfo{author}{\bibfnamefont{A.~V.} \bibnamefont{{Filippenko}}},
  \bibinfo{author}{\bibfnamefont{P.}~\bibnamefont{{Garnavich}}},
  \bibinfo{author}{\bibfnamefont{S.~T.} \bibnamefont{{Holland}}},
  \bibinfo{author}{\bibfnamefont{S.}~\bibnamefont{{Jha}}},
  \bibnamefont{et~al.}, \bibinfo{journal}{\apj} \textbf{\bibinfo{volume}{602}},
  \bibinfo{pages}{571} (\bibinfo{year}{2004}).

\bibitem{nova_data3}
\bibinfo{author}{\bibfnamefont{A.~G.} \bibnamefont{{Riess}}},
  \bibinfo{author}{\bibfnamefont{L.}~\bibnamefont{{Strolger}}},
  \bibinfo{author}{\bibfnamefont{J.}~\bibnamefont{{Tonry}}},
  \bibinfo{author}{\bibfnamefont{S.}~\bibnamefont{{Casertano}}},
  \bibinfo{author}{\bibfnamefont{H.~C.} \bibnamefont{{Ferguson}}},
  \bibinfo{author}{\bibfnamefont{B.}~\bibnamefont{{Mobasher}}},
  \bibinfo{author}{\bibfnamefont{P.}~\bibnamefont{{Challis}}},
  \bibinfo{author}{\bibfnamefont{A.~V.} \bibnamefont{{Filippenko}}},
  \bibinfo{author}{\bibfnamefont{S.}~\bibnamefont{{Jha}}},
  \bibinfo{author}{\bibfnamefont{W.}~\bibnamefont{{Li}}}, \bibnamefont{et~al.},
  \bibinfo{journal}{\apj} \textbf{\bibinfo{volume}{607}}, \bibinfo{pages}{665}
  (\bibinfo{year}{2004}).

\bibitem{boomerang}
\bibinfo{author}{\bibfnamefont{A.}~\bibnamefont{{Melchiorri}}},
  \bibinfo{author}{\bibfnamefont{P.~A.~R.} \bibnamefont{{Ade}}},
  \bibinfo{author}{\bibfnamefont{P.}~\bibnamefont{{de Bernardis}}},
  \bibinfo{author}{\bibfnamefont{J.~J.} \bibnamefont{{Bock}}},
  \bibinfo{author}{\bibfnamefont{J.}~\bibnamefont{{Borrill}}},
  \bibinfo{author}{\bibfnamefont{A.}~\bibnamefont{{Boscaleri}}},
  \bibinfo{author}{\bibfnamefont{B.~P.} \bibnamefont{{Crill}}},
  \bibinfo{author}{\bibfnamefont{G.}~\bibnamefont{{De Troia}}},
  \bibinfo{author}{\bibfnamefont{P.}~\bibnamefont{{Farese}}},
  \bibinfo{author}{\bibfnamefont{P.~G.} \bibnamefont{{Ferreira}}},
  \bibnamefont{et~al.}, \bibinfo{journal}{\apjl}
  \textbf{\bibinfo{volume}{536}}, \bibinfo{pages}{L63} (\bibinfo{year}{2000}).
\bibitem{wmap_params}
\bibinfo{author}{\bibfnamefont{D.~N.} \bibnamefont{{Spergel}}},
  \bibinfo{author}{\bibfnamefont{L.}~\bibnamefont{{Verde}}},
  \bibinfo{author}{\bibfnamefont{H.~V.} \bibnamefont{{Peiris}}},
  \bibinfo{author}{\bibfnamefont{E.}~\bibnamefont{{Komatsu}}},
  \bibinfo{author}{\bibfnamefont{M.~R.} \bibnamefont{{Nolta}}},
  \bibinfo{author}{\bibfnamefont{C.~L.} \bibnamefont{{Bennett}}},
  \bibinfo{author}{\bibfnamefont{M.}~\bibnamefont{{Halpern}}},
  \bibinfo{author}{\bibfnamefont{G.}~\bibnamefont{{Hinshaw}}},
  \bibinfo{author}{\bibfnamefont{N.}~\bibnamefont{{Jarosik}}},
  \bibinfo{author}{\bibfnamefont{A.}~\bibnamefont{{Kogut}}},
  \bibnamefont{et~al.}, \bibinfo{journal}{\apjs}
  \textbf{\bibinfo{volume}{148}}, \bibinfo{pages}{175} (\bibinfo{year}{2003}).
\bibitem{2df}
\bibinfo{author}{\bibfnamefont{E.}~\bibnamefont{{Hawkins}}},
  \bibinfo{author}{\bibfnamefont{S.}~\bibnamefont{{Maddox}}},
  \bibinfo{author}{\bibfnamefont{S.}~\bibnamefont{{Cole}}},
  \bibinfo{author}{\bibfnamefont{O.}~\bibnamefont{{Lahav}}},
  \bibinfo{author}{\bibfnamefont{D.~S.} \bibnamefont{{Madgwick}}},
  \bibinfo{author}{\bibfnamefont{P.}~\bibnamefont{{Norberg}}},
  \bibinfo{author}{\bibfnamefont{J.~A.} \bibnamefont{{Peacock}}},
  \bibinfo{author}{\bibfnamefont{I.~K.} \bibnamefont{{Baldry}}},
  \bibinfo{author}{\bibfnamefont{C.~M.} \bibnamefont{{Baugh}}},
  \bibinfo{author}{\bibfnamefont{J.}~\bibnamefont{{Bland-Hawthorn}}},
  \bibnamefont{et~al.}, \bibinfo{journal}{\mnras}
  \textbf{\bibinfo{volume}{346}}, \bibinfo{pages}{78} (\bibinfo{year}{2003}).
\bibitem{sdss}
\bibinfo{author}{\bibfnamefont{A.~C.} \bibnamefont{{Pope}}},
  \bibinfo{author}{\bibfnamefont{T.}~\bibnamefont{{Matsubara}}},
  \bibinfo{author}{\bibfnamefont{A.~S.} \bibnamefont{{Szalay}}},
  \bibinfo{author}{\bibfnamefont{M.~R.} \bibnamefont{{Blanton}}},
  \bibinfo{author}{\bibfnamefont{D.~J.} \bibnamefont{{Eisenstein}}},
  \bibinfo{author}{\bibfnamefont{J.}~\bibnamefont{{Gray}}},
  \bibinfo{author}{\bibfnamefont{B.}~\bibnamefont{{Jain}}},
  \bibinfo{author}{\bibfnamefont{N.~A.} \bibnamefont{{Bahcall}}},
  \bibinfo{author}{\bibfnamefont{J.}~\bibnamefont{{Brinkmann}}},
  \bibinfo{author}{\bibfnamefont{T.}~\bibnamefont{{Budavari}}},
  \bibnamefont{et~al.}, \bibinfo{journal}{\apj} \textbf{\bibinfo{volume}{607}},
  \bibinfo{pages}{655} (\bibinfo{year}{2004}).
\bibitem{2003Sci...299.1532B}
\bibinfo{author}{\bibfnamefont{S.~L.} \bibnamefont{{Bridle}}},
  \bibinfo{author}{\bibfnamefont{O.}~\bibnamefont{{Lahav}}},
  \bibinfo{author}{\bibfnamefont{J.~P.} \bibnamefont{{Ostriker}}},
  \bibnamefont{and} \bibinfo{author}{\bibfnamefont{P.~J.}
  \bibnamefont{{Steinhardt}}}, \bibinfo{journal}{Science}
  \textbf{\bibinfo{volume}{299}}, \bibinfo{pages}{1532} (\bibinfo{year}{2003}).
\bibitem{2004PhRvD..69j3501T}
\bibinfo{author}{\bibfnamefont{M.}~\bibnamefont{{Tegmark}}},
  \bibinfo{author}{\bibfnamefont{M.~A.} \bibnamefont{{Strauss}}},
  \bibinfo{author}{\bibfnamefont{M.~R.} \bibnamefont{{Blanton}}},
  \bibinfo{author}{\bibfnamefont{K.}~\bibnamefont{{Abazajian}}},
  \bibinfo{author}{\bibfnamefont{S.}~\bibnamefont{{Dodelson}}},
  \bibinfo{author}{\bibfnamefont{H.}~\bibnamefont{{Sandvik}}},
  \bibinfo{author}{\bibfnamefont{X.}~\bibnamefont{{Wang}}},
  \bibinfo{author}{\bibfnamefont{D.~H.} \bibnamefont{{Weinberg}}},
  \bibinfo{author}{\bibfnamefont{I.}~\bibnamefont{{Zehavi}}},
  \bibinfo{author}{\bibfnamefont{N.~A.} \bibnamefont{{Bahcall}}},
  \bibnamefont{et~al.}, \bibinfo{journal}{\prd} \textbf{\bibinfo{volume}{69}},
  \bibinfo{pages}{103501} (\bibinfo{year}{2004}).
\bibitem{TPabhay}
\bibinfo{author}{\bibfnamefont{T.}~\bibnamefont{{Padmanabhan}}},
  \bibinfo{journal}{ }
  (\bibinfo{year}{2005}), \eprint{arXiv:gr-qc/0503107}.
\bibitem{ccprob}
\bibinfo{author}{\bibfnamefont{S.}~\bibnamefont{{Weinberg}}},
  \bibinfo{journal}{ Rev. Mod. Phys.} \textbf{\bibinfo{volume}{61}},
  \bibinfo{pages}{1} (\bibinfo{year}{1989}).
\bibitem{review3}
\bibinfo{author}{\bibfnamefont{T.}~\bibnamefont{{Padmanabhan}}},
  \bibinfo{journal}{\physrep} \textbf{\bibinfo{volume}{380}},
  \bibinfo{pages}{235} (\bibinfo{year}{2003}),\eprint{arXiv:hep-th/0212290}; 
\bibinfo{author}{\bibfnamefont{P.~J.} \bibnamefont{{Peebles}}}
  \bibnamefont{and} \bibinfo{author}{\bibfnamefont{B.}~\bibnamefont{{Ratra}}},
  \bibinfo{journal}{ Rev. Mod. Phys.} \textbf{\bibinfo{volume}{75}},
  \bibinfo{pages}{559} (\bibinfo{year}{2003}); 
\bibinfo{author}{\bibfnamefont{V.}~\bibnamefont{{Sahni}}} \bibnamefont{and}
  \bibinfo{author}{\bibfnamefont{A.}~\bibnamefont{{Starobinsky}}},
  \bibinfo{journal}{Int.Jour.Mod.Phys.D}
  \textbf{\bibinfo{volume}{9}}, \bibinfo{pages}{373} (\bibinfo{year}{2000}); 
\bibinfo{author}{\bibfnamefont{J.}~\bibnamefont{{Ellis}}},
  \bibinfo{journal}{Royal Society of London Philosophical Transactions Series
  A} \textbf{\bibinfo{volume}{361}}, \bibinfo{pages}{2607}
  (\bibinfo{year}{2003}); 
\bibinfo{author}{\bibfnamefont{T.}~\bibnamefont{{Padmanabhan}}},
  \bibinfo{journal}{ }
  (\bibinfo{year}{2004}{\natexlab{a}}), \eprint{arXiv:astro-ph/0411044}.

\bibitem{quint1}
\bibinfo{author}{\bibfnamefont{P.~J.} \bibnamefont{{Steinhardt}}},
  \bibinfo{journal}{Royal Society of London Philosophical Transactions Series
  A} \textbf{\bibinfo{volume}{361}}, \bibinfo{pages}{2497}
  (\bibinfo{year}{2003}); 
\bibinfo{author}{\bibfnamefont{A.~D.} \bibnamefont{{Macorra}}}
  \bibnamefont{and}
  \bibinfo{author}{\bibfnamefont{G.}~\bibnamefont{{Piccinelli}}},
  \bibinfo{journal}{\prd} \textbf{\bibinfo{volume}{61}},
  \bibinfo{pages}{123503} (\bibinfo{year}{2000}); 
\bibinfo{author}{\bibfnamefont{L.~A.} \bibnamefont{{Ure{\~ n}a-L{\' o}pez}}}
  \bibnamefont{and} \bibinfo{author}{\bibfnamefont{T.}~\bibnamefont{{Matos}}},
  \bibinfo{journal}{\prd} \textbf{\bibinfo{volume}{62}},
  \bibinfo{pages}{081302} (\bibinfo{year}{2000}); 
\bibinfo{author}{\bibfnamefont{P.~F.} \bibnamefont{{Gonz{\'
  a}lez-D{\'{\i}}az}}}, \bibinfo{journal}{\prd} \textbf{\bibinfo{volume}{62}},
  \bibinfo{pages}{023513} (\bibinfo{year}{2000}{\natexlab{a}}); 
\bibinfo{author}{\bibfnamefont{R.}~\bibnamefont{{de Ritis}}} \bibnamefont{and}
  \bibinfo{author}{\bibfnamefont{A.~A.} \bibnamefont{{Marino}}},
  \bibinfo{journal}{\prd} \textbf{\bibinfo{volume}{64}},
  \bibinfo{pages}{083509} (\bibinfo{year}{2001}); 
\bibinfo{author}{\bibfnamefont{S.}~\bibnamefont{{Sen}}} \bibnamefont{and}
  \bibinfo{author}{\bibfnamefont{T.~R.} \bibnamefont{{Seshadri}}},
  \bibinfo{journal}{Int.Jour.Mod.Phys.D}
  \textbf{\bibinfo{volume}{12}}, \bibinfo{pages}{445} (\bibinfo{year}{2003});
\bibinfo{author}{\bibfnamefont{C.}~\bibnamefont{Rubano}} \bibnamefont{and}
  \bibinfo{author}{\bibfnamefont{P.}~\bibnamefont{Scudellaro}},
  \bibinfo{journal}{Gen. Rel. Grav.} \textbf{\bibinfo{volume}{34}},
  \bibinfo{pages}{307} (\bibinfo{year}{2002}), \eprint{astro-ph/0103335}; 
\bibinfo{author}{\bibfnamefont{S.~A.} \bibnamefont{{Bludman}}}
  \bibnamefont{and} \bibinfo{author}{\bibfnamefont{M.}~\bibnamefont{{Roos}}},
  \bibinfo{journal}{\prd} \textbf{\bibinfo{volume}{65}},
  \bibinfo{pages}{043503} (\bibinfo{year}{2002}).

\bibitem{2001PhRvD..63j3510A}
\bibinfo{author}{\bibfnamefont{C.}~\bibnamefont{{Armendariz-Picon}}},
  \bibinfo{author}{\bibfnamefont{V.}~\bibnamefont{{Mukhanov}}},
  \bibnamefont{and} \bibinfo{author}{\bibfnamefont{P.~J.}
  \bibnamefont{{Steinhardt}}}, \bibinfo{journal}{\prd}
  \textbf{\bibinfo{volume}{63}}, \bibinfo{pages}{103510}
  (\bibinfo{year}{2001}); 
\bibinfo{author}{\bibfnamefont{T.}~\bibnamefont{{Chiba}}},
  \bibinfo{journal}{\prd} \textbf{\bibinfo{volume}{66}},
  \bibinfo{pages}{063514} (\bibinfo{year}{2002}); 
\bibinfo{author}{\bibfnamefont{M.}~\bibnamefont{{Malquarti}}},
  \bibinfo{author}{\bibfnamefont{E.~J.} \bibnamefont{{Copeland}}},
  \bibinfo{author}{\bibfnamefont{A.~R.} \bibnamefont{{Liddle}}},
  \bibnamefont{and}
  \bibinfo{author}{\bibfnamefont{M.}~\bibnamefont{{Trodden}}},
  \bibinfo{journal}{\prd} \textbf{\bibinfo{volume}{67}},
  \bibinfo{pages}{123503} (\bibinfo{year}{2003}); 
\bibinfo{author}{\bibfnamefont{L.~P.} \bibnamefont{{Chimento}}}
  \bibnamefont{and}
  \bibinfo{author}{\bibfnamefont{A.}~\bibnamefont{{Feinstein}}},
  \bibinfo{journal}{Mod. Phys. Letts. A} \textbf{\bibinfo{volume}{19}},
  \bibinfo{pages}{761} (\bibinfo{year}{2004}); 
\bibinfo{author}{\bibfnamefont{R.~J.} \bibnamefont{{Scherrer}}},
  \bibinfo{journal}{Phys. Rev.  Letts. } \textbf{\bibinfo{volume}{93}},
  \bibinfo{pages}{011301} (\bibinfo{year}{2004}).
\bibitem{tachyon1}
\bibinfo{author}{\bibfnamefont{T.}~\bibnamefont{{Padmanabhan}}},
  \bibinfo{journal}{\prd} \textbf{\bibinfo{volume}{66}},
  \bibinfo{pages}{021301} (\bibinfo{year}{2002}).
\bibitem{2003PhRvD..67f3504B}
\bibinfo{author}{\bibfnamefont{J.~S.} \bibnamefont{{Bagla}}},
  \bibinfo{author}{\bibfnamefont{H.~K.} \bibnamefont{{Jassal}}},
  \bibnamefont{and}
  \bibinfo{author}{\bibfnamefont{T.}~\bibnamefont{{Padmanabhan}}},
  \bibinfo{journal}{\prd} \textbf{\bibinfo{volume}{67}},
  \bibinfo{pages}{063504} (\bibinfo{year}{2003}), \eprint{arXiv:astro-ph/0212198}; 
\bibinfo{author}{\bibfnamefont{H.~K.} \bibnamefont{{Jassal}}},
  \bibinfo{journal}{ }
  (\bibinfo{year}{2003}{\natexlab{a}}), \eprint{arXiv:astro-ph/0303406};
\bibinfo{author}{\bibfnamefont{J.~M.} \bibnamefont{{Aguirregabiria}}}
  \bibnamefont{and} \bibinfo{author}{\bibfnamefont{R.}~\bibnamefont{{Lazkoz}}},
  \bibinfo{journal}{\prd} \textbf{\bibinfo{volume}{69}},
  \bibinfo{pages}{123502} (\bibinfo{year}{2004}); 
\bibinfo{author}{\bibfnamefont{A.}~\bibnamefont{{Sen}}},
  \bibinfo{journal}{ }
  (\bibinfo{year}{2003}), \eprint{arXiv:hep-th/0312153}; 
\bibinfo{author}{\bibfnamefont{V.}~\bibnamefont{{Gorini}}},
  \bibinfo{author}{\bibfnamefont{A.}~\bibnamefont{{Kamenshchik}}},
  \bibinfo{author}{\bibfnamefont{U.}~\bibnamefont{{Moschella}}},
  \bibnamefont{and}
  \bibinfo{author}{\bibfnamefont{V.}~\bibnamefont{{Pasquier}}},
  \bibinfo{journal}{\prd} \textbf{\bibinfo{volume}{69}},
  \bibinfo{pages}{123512} (\bibinfo{year}{2004}); 
\bibinfo{author}{\bibfnamefont{G.~W.} \bibnamefont{{Gibbons}}},
  \bibinfo{journal}{Class. Quan. Grav.}
  \textbf{\bibinfo{volume}{20}}, \bibinfo{pages}{321}
  (\bibinfo{year}{2003}{\natexlab{a}}); 
\bibinfo{author}{\bibfnamefont{C.}~\bibnamefont{{Kim}}},
  \bibinfo{author}{\bibfnamefont{H.~B.} \bibnamefont{{Kim}}}, \bibnamefont{and}
  \bibinfo{author}{\bibfnamefont{Y.}~\bibnamefont{{Kim}}},
  \bibinfo{journal}{Phys. Letts.  B} \textbf{\bibinfo{volume}{552}},
  \bibinfo{pages}{111} (\bibinfo{year}{2003}); 
\bibinfo{author}{\bibfnamefont{G.}~\bibnamefont{{Shiu}}} \bibnamefont{and}
  \bibinfo{author}{\bibfnamefont{I.}~\bibnamefont{{Wasserman}}},
  \bibinfo{journal}{Phys. Letts.  B} \textbf{\bibinfo{volume}{541}},
  \bibinfo{pages}{6} (\bibinfo{year}{2002}); 
\bibinfo{author}{\bibfnamefont{D.}~\bibnamefont{{Choudhury}}},
  \bibinfo{author}{\bibfnamefont{D.}~\bibnamefont{{Ghoshal}}},
  \bibinfo{author}{\bibfnamefont{D.~P.} \bibnamefont{{Jatkar}}},
  \bibnamefont{and} \bibinfo{author}{\bibfnamefont{S.}~\bibnamefont{{Panda}}},
  \bibinfo{journal}{Phys. Letts.  B} \textbf{\bibinfo{volume}{544}},
  \bibinfo{pages}{231} (\bibinfo{year}{2002}); 
\bibinfo{author}{\bibfnamefont{A.}~\bibnamefont{{Frolov}}},
  \bibinfo{author}{\bibfnamefont{L.}~\bibnamefont{{Kofman}}}, \bibnamefont{and}
  \bibinfo{author}{\bibfnamefont{A.}~\bibnamefont{{Starobinsky}}},
  \bibinfo{journal}{Phys. Letts.  B} \textbf{\bibinfo{volume}{545}},
  \bibinfo{pages}{8} (\bibinfo{year}{2002}); 
\bibinfo{author}{\bibfnamefont{G.~W.} \bibnamefont{{Gibbons}}},
  \bibinfo{journal}{Phys. Letts.  B} \textbf{\bibinfo{volume}{537}},
  \bibinfo{pages}{1} (\bibinfo{year}{2002}); 
\bibinfo{author}{\bibfnamefont{A.}~\bibnamefont{{Das}}},
  \bibinfo{author}{\bibfnamefont{S.}~\bibnamefont{{Gupta}}},
  \bibinfo{author}{\bibfnamefont{T.}~\bibnamefont{{Deep Saini}}},
  \bibnamefont{and} \bibinfo{author}{\bibfnamefont{S.}~\bibnamefont{{Kar}}},
  \bibinfo{journal}{ }  (\bibinfo{year}{2005}),
  \eprint{arXiv:astro-ph/0505509};
\bibinfo{author}{\bibfnamefont{I.~Y.}~\bibnamefont{{Aref'eva}}},
  \bibinfo{journal}{ }  (\bibinfo{year}{2004}),
  \eprint{arXiv:astro-ph/0410443}.


\bibitem{2002PhLB..545...23C}
\bibinfo{author}{\bibfnamefont{R.~R.} \bibnamefont{{Caldwell}}},
  \bibinfo{journal}{Phys. Letts.  B} \textbf{\bibinfo{volume}{545}},
  \bibinfo{pages}{23} (\bibinfo{year}{2002}); 
\bibinfo{author}{\bibfnamefont{J.}~\bibnamefont{{Hao}}} \bibnamefont{and}
  \bibinfo{author}{\bibfnamefont{X.}~\bibnamefont{{Li}}},
  \bibinfo{journal}{\prd} \textbf{\bibinfo{volume}{68}},
  \bibinfo{pages}{043501} (\bibinfo{year}{2003}{\natexlab{a}}); 
\bibinfo{author}{\bibfnamefont{G.~W.} \bibnamefont{{Gibbons}}},
  \bibinfo{journal}{ }
  (\bibinfo{year}{2003}{\natexlab{b}}), \eprint{arXiv:hep-th/0302199}; 
\bibinfo{author}{\bibfnamefont{V.~K.}~\bibnamefont{{Onemli}}} \bibnamefont{and}
  \bibinfo{author}{\bibfnamefont{R.~P.}~\bibnamefont{{Woodard}}},
  \bibinfo{journal}{\prd} \textbf{\bibinfo{volume}{70}},
  \bibinfo{pages}{107301} (\bibinfo{year}{2004}{\natexlab{a}});
\bibinfo{author}{\bibfnamefont{S.}~\bibnamefont{{Nojiri}}} \bibnamefont{and}
  \bibinfo{author}{\bibfnamefont{S.~D.} \bibnamefont{{Odintsov}}},
  \bibinfo{journal}{Phys. Letts.  B} \textbf{\bibinfo{volume}{562}},
  \bibinfo{pages}{147} (\bibinfo{year}{2003}); 
\bibinfo{author}{\bibfnamefont{S.~M.} \bibnamefont{{Carroll}}},
  \bibinfo{author}{\bibfnamefont{M.}~\bibnamefont{{Hoffman}}},
  \bibnamefont{and}
  \bibinfo{author}{\bibfnamefont{M.}~\bibnamefont{{Trodden}}},
  \bibinfo{journal}{\prd} \textbf{\bibinfo{volume}{68}},
  \bibinfo{pages}{023509} (\bibinfo{year}{2003}); 
\bibinfo{author}{\bibfnamefont{P.}~\bibnamefont{{Singh}}},
  \bibinfo{author}{\bibfnamefont{M.}~\bibnamefont{{Sami}}}, \bibnamefont{and}
  \bibinfo{author}{\bibfnamefont{N.}~\bibnamefont{{Dadhich}}},
  \bibinfo{journal}{\prd} \textbf{\bibinfo{volume}{68}},
  \bibinfo{pages}{023522} (\bibinfo{year}{2003}); 
\bibinfo{author}{\bibfnamefont{P.~H.} \bibnamefont{{Frampton}}},
  \bibinfo{journal}{Mod. Phys. Letts. A} \textbf{\bibinfo{volume}{19}},
  \bibinfo{pages}{801} (\bibinfo{year}{2004}); 
\bibinfo{author}{\bibfnamefont{J.}~\bibnamefont{{Hao}}} \bibnamefont{and}
  \bibinfo{author}{\bibfnamefont{X.}~\bibnamefont{{Li}}},
  \bibinfo{journal}{\prd} \textbf{\bibinfo{volume}{67}},
  \bibinfo{pages}{107303} (\bibinfo{year}{2003}{\natexlab{b}}); 
\bibinfo{author}{\bibfnamefont{P.}~\bibnamefont{{Gonz{\' a}lez-D{\'{\i}}az}}},
  \bibinfo{journal}{\prd} \textbf{\bibinfo{volume}{68}},
  \bibinfo{pages}{021303} (\bibinfo{year}{2003}{\natexlab{a}}); 
\bibinfo{author}{\bibfnamefont{M.~P.} \bibnamefont{{Dabrowski}}},
  \bibinfo{author}{\bibfnamefont{T.}~\bibnamefont{{Stachowiak}}},
  \bibnamefont{and}
  \bibinfo{author}{\bibfnamefont{M.}~\bibnamefont{{Szyd{\l}owski}}},
  \bibinfo{journal}{\prd} \textbf{\bibinfo{volume}{68}},
  \bibinfo{pages}{103519} (\bibinfo{year}{2003}); 
\bibinfo{author}{\bibfnamefont{J.~M.} \bibnamefont{{Cline}}},
  \bibinfo{author}{\bibfnamefont{S.}~\bibnamefont{{Jeon}}}, \bibnamefont{and}
  \bibinfo{author}{\bibfnamefont{G.~D.} \bibnamefont{{Moore}}},
  \bibinfo{journal}{\prd} \textbf{\bibinfo{volume}{70}},
  \bibinfo{pages}{043543} (\bibinfo{year}{2004}); 
\bibinfo{author}{\bibfnamefont{W.}~\bibnamefont{{Fang}}},
  \bibinfo{author}{\bibfnamefont{H.~Q.} \bibnamefont{{Lu}}},
  \bibinfo{author}{\bibfnamefont{Z.~G.} \bibnamefont{{Huang}}},
  \bibnamefont{and} \bibinfo{author}{\bibfnamefont{K.~F.}
  \bibnamefont{{Zhang}}}, \bibinfo{journal}{ArXiv High Energy Physics - Theory
  e-prints}  (\bibinfo{year}{2004}), \eprint{arXiv:hep-th/0409080};
\bibinfo{author}{\bibfnamefont{S.}~\bibnamefont{{Nojiri}}}
\bibnamefont{and}
\bibinfo{author}{\bibfnamefont{S.~D.}~\bibnamefont{{Odinstov}}},
\bibinfo{journal}{ArXiv High Energy Physics - Theory
  e-prints}  (\bibinfo{year}{2005}), \eprint{arXiv:hep-th/0505215};
  \bibinfo{author}{\bibfnamefont{S.}~\bibnamefont{{Nesseris}}}
  \bibnamefont{and} 
  \bibinfo{author}{\bibfnamefont{L.} \bibnamefont{{Perivolaropoulos}}},
  \bibinfo{journal}{\prd} \textbf{\bibinfo{volume}{70}},
  \bibinfo{pages}{123529} (\bibinfo{year}{2004});
\bibinfo{author}{\bibfnamefont{S.}~\bibnamefont{{Nojiri}}}
\bibnamefont{and}
\bibinfo{author}{\bibfnamefont{S.~D.}~\bibnamefont{{Odinstov}}},
  \bibinfo{journal}{\prd} \textbf{\bibinfo{volume}{70}},
  \bibinfo{pages}{103522} (\bibinfo{year}{2004});
\bibinfo{author}{\bibfnamefont{E.}~\bibnamefont{{Elizalde}}},
\bibinfo{author}{\bibfnamefont{S.}~\bibnamefont{{Nojiri}}}
\bibnamefont{and}
\bibinfo{author}{\bibfnamefont{S.~D.}~\bibnamefont{{Odinstov}}},
  \bibinfo{journal}{\prd} \textbf{\bibinfo{volume}{70}},
  \bibinfo{pages}{043539} (\bibinfo{year}{2004});
\bibinfo{author}{\bibfnamefont{S.}~\bibnamefont{{Nojiri}}},
\bibinfo{author}{\bibfnamefont{S.~D.}~\bibnamefont{{Odinstov}}},
\bibnamefont{and}
\bibinfo{author}{\bibfnamefont{S.}~\bibnamefont{{Tsujikawa}}},
  \bibinfo{journal}{\prd} \textbf{\bibinfo{volume}{71}},
  \bibinfo{pages}{063004} (\bibinfo{year}{2005}).

\bibitem{brane1}
\bibinfo{author}{\bibfnamefont{K.}~\bibnamefont{{Uzawa}}} \bibnamefont{and}
  \bibinfo{author}{\bibfnamefont{J.}~\bibnamefont{{Soda}}},
  \bibinfo{journal}{Mod. Phys. Letts. A} \textbf{\bibinfo{volume}{16}},
  \bibinfo{pages}{1089} (\bibinfo{year}{2001}); 
\bibinfo{author}{\bibfnamefont{H.~K.} \bibnamefont{{Jassal}}},
  \bibinfo{journal}{ }
  (\bibinfo{year}{2003}{\natexlab{b}}), \eprint{arXiv:hep-th/0312253}; 
\bibinfo{author}{\bibfnamefont{C.~P.} \bibnamefont{{Burgess}}},
  \bibinfo{journal}{Int.Jour.Mod.Phys.D}
  \textbf{\bibinfo{volume}{12}}, \bibinfo{pages}{1737} (\bibinfo{year}{2003}); 
\bibinfo{author}{\bibfnamefont{K.~A.} \bibnamefont{Milton}},
  \bibinfo{journal}{Grav. Cosmol.} \textbf{\bibinfo{volume}{9}},
  \bibinfo{pages}{66} (\bibinfo{year}{2003}), \eprint{hep-ph/0210170}; 
\bibinfo{author}{\bibfnamefont{P.~F.} \bibnamefont{{Gonz{\'
  a}lez-D{\'{\i}}az}}}, \bibinfo{journal}{Phys. Letts.  B}
  \textbf{\bibinfo{volume}{481}}, \bibinfo{pages}{353}
  (\bibinfo{year}{2000}{\natexlab{b}}).

\bibitem{water}
\bibinfo{author}{\bibfnamefont{R.}~\bibnamefont{{Holman}}} \bibnamefont{and}
  \bibinfo{author}{\bibfnamefont{S.}~\bibnamefont{{Naidu}}},
  \bibinfo{journal}{ }  (\bibinfo{year}{2004}),
  \eprint{arXiv:astro-ph/0408102}.

\bibitem{tp173}
\bibinfo{author}{\bibfnamefont{T.}~\bibnamefont{{Padmanabhan}}},
  \bibinfo{journal}{ }
  (\bibinfo{year}{2004}{\natexlab{b}}), \eprint{arXiv:hep-th/0406060};
  \bibinfo{author}{\bibfnamefont{I.}~\bibnamefont{{Shapiro}}}\bibnamefont{ and}
 \bibinfo{author}{\bibfnamefont{J.}~\bibnamefont{{Sola}}} 
  \bibinfo{journal}{JHEP } \bibinfo{pages}{006}  (\bibinfo{year}{2002}),
  \eprint{arXiv:hep-th/0012227}; 
 \bibinfo{author}{\bibfnamefont{J.}~\bibnamefont{{Sola}}} \bibnamefont{and}
\bibinfo{author}{\bibfnamefont{H.}~\bibnamefont{{Stefancic}}},
  \bibinfo{journal}{ }  (\bibinfo{year}{2005}),
  \eprint{arXiv:astro-ph/0505133}.
\bibitem{unified_dedm1}
\bibinfo{author}{\bibfnamefont{T.}~\bibnamefont{{Padmanabhan}}}
  \bibnamefont{and} \bibinfo{author}{\bibfnamefont{T.~R.}
  \bibnamefont{{Choudhury}}}, \bibinfo{journal}{\prd}
  \textbf{\bibinfo{volume}{66}}, \bibinfo{pages}{081301}
  (\bibinfo{year}{2002}); 
\bibinfo{author}{\bibfnamefont{V.~F.} \bibnamefont{{Cardone}}},
  \bibinfo{author}{\bibfnamefont{A.}~\bibnamefont{{Troisi}}}, \bibnamefont{and}
  \bibinfo{author}{\bibfnamefont{S.}~\bibnamefont{{Capozziello}}},
  \bibinfo{journal}{\prd} \textbf{\bibinfo{volume}{69}},
  \bibinfo{pages}{083517} (\bibinfo{year}{2004}); 
\bibinfo{author}{\bibfnamefont{P.~F.} \bibnamefont{{Gonz{\'
  a}lez-D{\'{\i}}az}}}, \bibinfo{journal}{Phys. Letts.  B}
  \textbf{\bibinfo{volume}{562}}, \bibinfo{pages}{1}
  (\bibinfo{year}{2003}{\natexlab{b}}); 
\bibinfo{author}{\bibfnamefont{M.~A.~M.~C.} \bibnamefont{{Calik}}},
  \bibinfo{journal}{ }
  (\bibinfo{year}{2005}), \eprint{arXiv:gr-qc/0505035};
\bibinfo{author}{\bibfnamefont{S.} \bibnamefont{{Capozziello}}},
\bibinfo{author}{\bibfnamefont{S.} \bibnamefont{{Nojiri}}}
\bibnamefont{and}
\bibinfo{author}{\bibfnamefont{S.~D.} \bibnamefont{{Odintsov}}},
\bibinfo{journal}{ }  (\bibinfo{year}{2005}), \eprint{arXiv:hep-th/0507182}.

\bibitem{2005astro.ph..5133S}
\bibinfo{author}{\bibfnamefont{V.~K.}~\bibnamefont{{Onemli}}} \bibnamefont{and}
  \bibinfo{author}{\bibfnamefont{R.~P.}~\bibnamefont{{Woodard}}},
  \bibinfo{journal}{Class. Quant. Grav.}   \textbf{\bibinfo{volume}{19}},
 \bibinfo{pages}{4607} (\bibinfo{year}{2002}),
 \bibinfo{author}{\bibfnamefont{T.}~\bibnamefont{Padmanabhan}},
  \bibinfo{journal}{Phys. Reports,} \textbf{\bibinfo{volume}{49}},
  \bibinfo{pages}{406},\bibinfo{year}{2005}
     \eprint{arXiv:gr-qc/0311036};
  \bibinfo{author}{\bibfnamefont{T.}~\bibnamefont{{Padmanabhan}}},
  \bibinfo{journal}{Class.Quant.Grav,} \textbf{\bibinfo{volume}{19}},
  \bibinfo{pages}{5387},\bibinfo{year}{2002}
     \eprint{arXiv:gr-qc/0204019};
 \bibinfo{author}{\bibfnamefont{A.~A.} \bibnamefont{{Andrianov}}},
  \bibinfo{author}{\bibfnamefont{F.}~\bibnamefont{{Cannata}}},
  \bibnamefont{and} \bibinfo{author}{\bibfnamefont{A.~Y.}
  \bibnamefont{{Kamenshchik}}}, \bibinfo{journal}{ }  (\bibinfo{year}{2005}),
  \eprint{arXiv:gr-qc/0505087};
\bibinfo{author}{\bibfnamefont{R.}~\bibnamefont{{Lazkoz}}},
\bibinfo{author}{\bibfnamefont{S.}~\bibnamefont{{Nesseris}}} \bibnamefont{and}
\bibinfo{author}{\bibfnamefont{L.}~\bibnamefont{{Perivolaropoulos}}},
  \bibinfo{journal}{ }  (\bibinfo{year}{2005}),
  \eprint{arXiv:astro-ph/0503230};
\bibinfo{author}{\bibfnamefont{M.}~\bibnamefont{{Szydlowski}}},
\bibinfo{author}{\bibfnamefont{W.}~\bibnamefont{{Godlowski}}}
\bibnamefont{and} 
\bibinfo{author}{\bibfnamefont{R.}~\bibnamefont{{Wojtak}}},
  \bibinfo{journal}{ }  (\bibinfo{year}{2005}),
  \eprint{arXiv:astro-ph/0503202}.


\bibitem{2005MNRAS.356L..11J}
\bibinfo{author}{\bibfnamefont{H.~K.} \bibnamefont{{Jassal}}},
  \bibinfo{author}{\bibfnamefont{J.~S.} \bibnamefont{{Bagla}}},
  \bibnamefont{and}
  \bibinfo{author}{\bibfnamefont{T.}~\bibnamefont{{Padmanabhan}}},
  \bibinfo{journal}{\mnrasl} \textbf{\bibinfo{volume}{356}},
  \bibinfo{pages}{L11} (\bibinfo{year}{2005}),\eprint{arXiv:astro-ph/0404378}.
\bibitem{2005A&A...429..807C}
\bibinfo{author}{\bibfnamefont{T.~R.} \bibnamefont{{Choudhury}}}
  \bibnamefont{and}
  \bibinfo{author}{\bibfnamefont{T.}~\bibnamefont{{Padmanabhan}}},
  \bibinfo{journal}{\aap} \textbf{\bibinfo{volume}{429}}, \bibinfo{pages}{807}
  (\bibinfo{year}{2005})
  \eprint{arXiv:astro-ph/0311622}.
\bibitem{2005PhRvD..71j3515S}
\bibinfo{author}{\bibfnamefont{U.}~\bibnamefont{{Seljak}}},
  \bibinfo{author}{\bibfnamefont{A.}~\bibnamefont{{Makarov}}},
  \bibinfo{author}{\bibfnamefont{P.}~\bibnamefont{{McDonald}}},
  \bibinfo{author}{\bibfnamefont{S.~F.} \bibnamefont{{Anderson}}},
  \bibinfo{author}{\bibfnamefont{N.~A.} \bibnamefont{{Bahcall}}},
  \bibinfo{author}{\bibfnamefont{J.}~\bibnamefont{{Brinkmann}}},
  \bibinfo{author}{\bibfnamefont{S.}~\bibnamefont{{Burles}}},
  \bibinfo{author}{\bibfnamefont{R.}~\bibnamefont{{Cen}}},
  \bibinfo{author}{\bibfnamefont{M.}~\bibnamefont{{Doi}}},
  \bibinfo{author}{\bibfnamefont{J.~E.} \bibnamefont{{Gunn}}},
  \bibnamefont{et~al.}, \bibinfo{journal}{\prd} \textbf{\bibinfo{volume}{71}},
  \bibinfo{pages}{103515} (\bibinfo{year}{2005}).
\bibitem{constraints_2}
\bibinfo{author}{\bibfnamefont{Y.}~\bibnamefont{{Wang}}},
  \bibinfo{author}{\bibfnamefont{V.}~\bibnamefont{{Kostov}}},
  \bibinfo{author}{\bibfnamefont{K.}~\bibnamefont{{Freese}}},
  \bibinfo{author}{\bibfnamefont{J.~A.} \bibnamefont{{Frieman}}},
  \bibnamefont{and}
  \bibinfo{author}{\bibfnamefont{P.}~\bibnamefont{{Gondolo}}},
  \bibinfo{journal}{ }  (\bibinfo{year}{2004}),
  \eprint{arXiv:astro-ph/0402080}.
\bibitem{2004ApJ...617L...1B}
\bibinfo{author}{\bibfnamefont{B.~A.} \bibnamefont{{Bassett}}},
  \bibinfo{author}{\bibfnamefont{P.~S.} \bibnamefont{{Corasaniti}}},
  \bibnamefont{and} \bibinfo{author}{\bibfnamefont{M.}~\bibnamefont{{Kunz}}},
  \bibinfo{journal}{\apjl} \textbf{\bibinfo{volume}{617}}, \bibinfo{pages}{L1}
  (\bibinfo{year}{2004}).
\bibitem{constraints_10}
\bibinfo{author}{\bibfnamefont{S.}~\bibnamefont{{Lee}}},
  \bibinfo{journal}{ }  (\bibinfo{year}{2005}),
  \eprint{arXiv:astro-ph/0504650}.
\bibitem{holographic}
\bibinfo{author}{\bibfnamefont{M.}~\bibnamefont{{Li}}},
  \bibinfo{journal}{Phys. Letts.  B} \textbf{\bibinfo{volume}{603}},
  \bibinfo{pages}{1} (\bibinfo{year}{2004}).
\bibitem{Hannestad:2004cb}
\bibinfo{author}{\bibfnamefont{S.}~\bibnamefont{{Hannestad}}} \bibnamefont{and}
  \bibinfo{author}{\bibfnamefont{E.}~\bibnamefont{{M{\" o}rtsell}}},
  \bibinfo{journal}{JCAP}
  \textbf{\bibinfo{volume}{9}}, \bibinfo{pages}{1} (\bibinfo{year}{2004}).

\bibitem{p1}
\bibinfo{author}{\bibfnamefont{M.}~\bibnamefont{{Chevallier}}}
  \bibnamefont{and} \bibinfo{author}{\bibfnamefont{D.}
  \bibnamefont{{Polarski}}}, \bibinfo{journal}{Int. J. Mod. Phys.}
  \textbf{\bibinfo{volume}{D10}}, \bibinfo{pages}{213} (\bibinfo{year}{2001}); 
\bibinfo{author}{\bibfnamefont{E.~V.}~\bibnamefont{{Linder}}},
 \bibinfo{journal}{\prl}, \textbf{\bibinfo{volume}{90}},
 \bibinfo{pages}{091301} (\bibinfo{year}{2003}) 


\bibitem{1998ApJ...509...74G}
\bibinfo{author}{\bibfnamefont{M.} \bibnamefont{{Hamuy}}},
\bibinfo{author}{\bibfnamefont{M.~M.} \bibnamefont{{Phillips}}},
\bibinfo{author}{\bibfnamefont{N.~B.} \bibnamefont{{Suntzeff}}},
\bibinfo{author}{\bibfnamefont{R.~A.} \bibnamefont{{Schommer}}},
\bibinfo{author}{\bibfnamefont{J.} \bibnamefont{{Maza}}},
\bibinfo{author}{\bibfnamefont{A.~R.} \bibnamefont{{Antezan}}},
\bibinfo{author}{\bibfnamefont{M.} \bibnamefont{{Wischnjewsky}}},
\bibinfo{author}{\bibfnamefont{G.} \bibnamefont{{Valladares}}},
\bibinfo{author}{\bibfnamefont{C.} \bibnamefont{{Muena}}},
\bibinfo{author}{\bibfnamefont{L.~E.} \bibnamefont{{Gonzales}}},
\bibnamefont{et~al.}, \bibinfo{journal}{\apj} \textbf{\bibinfo{volume}{112}},
\bibinfo{pages}{2408} (\bibinfo{year}{1996});
\bibinfo{author}{\bibfnamefont{A.~G.} \bibnamefont{{Riess}}},
 \bibinfo{author}{\bibfnamefont{A.~V.} \bibnamefont{{Filippenko}}},
 \bibinfo{author}{\bibfnamefont{P.}~\bibnamefont{{Challis}}},
\bibinfo{author}{\bibfnamefont{A.}~\bibnamefont{{Clocchiatti}}},
\bibinfo{author}{\bibfnamefont{A.}~\bibnamefont{{Diercks}}},
\bibinfo{author}{\bibfnamefont{P.~M.}~\bibnamefont{{Garnavich}}},
\bibinfo{author}{\bibfnamefont{R.~L.}~\bibnamefont{{Gilliland}}},
\bibinfo{author}{\bibfnamefont{C.~J.}~\bibnamefont{{Hogan}}},
\bibinfo{author}{\bibfnamefont{S.}~\bibnamefont{{Jha}}},
\bibinfo{author}{\bibfnamefont{R.~P.}~\bibnamefont{{Kirshner}}},
 \bibnamefont{et~al.}, \bibinfo{journal}{\aj} \textbf{\bibinfo{volume}{116}},
  \bibinfo{pages}{1009} (\bibinfo{year}{1998});
\bibinfo{author}{\bibfnamefont{S.}~\bibnamefont{{Perlmutter}}},
  \bibinfo{author}{\bibfnamefont{G.}~\bibnamefont{{Aldering}}},
  \bibinfo{author}{\bibfnamefont{G.}~\bibnamefont{{Goldhaber}}},
  \bibinfo{author}{\bibfnamefont{R.~A.} \bibnamefont{{Knop}}},
  \bibinfo{author}{\bibfnamefont{P.}~\bibnamefont{{Nugent}}},
  \bibinfo{author}{\bibfnamefont{P.~G.} \bibnamefont{{Castro}}},
  \bibinfo{author}{\bibfnamefont{S.}~\bibnamefont{{Deustua}}},
  \bibinfo{author}{\bibfnamefont{S.}~\bibnamefont{{Fabbro}}},
  \bibinfo{author}{\bibfnamefont{A.}~\bibnamefont{{Goobar}}},
  \bibinfo{author}{\bibfnamefont{D.~E.} \bibnamefont{{Groom}}},
  \bibnamefont{et~al.}, \bibinfo{journal}{\apj} \textbf{\bibinfo{volume}{517}},
  \bibinfo{pages}{565} (\bibinfo{year}{1999}{\natexlab{a}}); 
\bibinfo{author}{\bibfnamefont{A.~G.} \bibnamefont{{Riess}}},
 \bibinfo{author}{\bibfnamefont{R.~P.}~\bibnamefont{{Kirshner}}},
\bibinfo{author}{\bibfnamefont{B.~P.}~\bibnamefont{{Schmidt}}},
 \bibinfo{author}{\bibfnamefont{S.}~\bibnamefont{{Jha}}},
 \bibinfo{author}{\bibfnamefont{P.}~\bibnamefont{{Challis}}},
\bibinfo{author}{\bibfnamefont{P.~M.}~\bibnamefont{{Garnavich}}},
\bibinfo{author}{\bibfnamefont{A.~A.}~\bibnamefont{{Esin}}},
\bibinfo{author}{\bibfnamefont{C.}~\bibnamefont{{Carpenter}}},
\bibinfo{author}{\bibfnamefont{R.}~\bibnamefont{{Grashius}}},
\bibinfo{author}{\bibfnamefont{R.~E.}~\bibnamefont{{Schild}}},
 \bibnamefont{et~al.}, \bibinfo{journal}{\aj} \textbf{\bibinfo{volume}{117}},
  \bibinfo{pages}{1009} (\bibinfo{year}{1999});
\bibinfo{author}{\bibfnamefont{A.~G.} \bibnamefont{{Riess}}},
 \bibinfo{author}{\bibfnamefont{P.~E.}~\bibnamefont{{Nugent}}},
\bibinfo{author}{\bibfnamefont{R.~L.}~\bibnamefont{{Gilliland}}},
\bibinfo{author}{\bibfnamefont{B.~P.}~\bibnamefont{{Schmidt}}},
 \bibinfo{author}{\bibfnamefont{J.}~\bibnamefont{{Tonry}}},
 \bibinfo{author}{\bibfnamefont{M.}~\bibnamefont{{Dickinson}}},
 \bibinfo{author}{\bibfnamefont{R.~I}~\bibnamefont{{Thompson}}},
 \bibinfo{author}{\bibfnamefont{T.}~\bibnamefont{{Budav{\' a}ri}}},
 \bibinfo{author}{\bibfnamefont{S.}~\bibnamefont{{Casertano}}},
 \bibinfo{author}{\bibfnamefont{A.~S.}~\bibnamefont{{Evans}}},
 \bibnamefont{et~al.}, \bibinfo{journal}{\apj} \textbf{\bibinfo{volume}{560}},
  \bibinfo{pages}{49} (\bibinfo{year}{2001}{\natexlab{a}}); 
\bibinfo{author}{\bibfnamefont{R.~A.} \bibnamefont{{Knop}}},
  \bibinfo{author}{\bibfnamefont{G.}~\bibnamefont{{Aldering}}},
  \bibinfo{author}{\bibfnamefont{R.}~\bibnamefont{{Amanullah}}},
  \bibinfo{author}{\bibfnamefont{P.}~\bibnamefont{{Astier}}},
  \bibinfo{author}{\bibfnamefont{G.}~\bibnamefont{{Blanc}}},
  \bibinfo{author}{\bibfnamefont{M.~S.} \bibnamefont{{Burns}}},
  \bibinfo{author}{\bibfnamefont{A.}~\bibnamefont{{Conley}}},
  \bibinfo{author}{\bibfnamefont{S.~E.} \bibnamefont{{Deustua}}},
  \bibinfo{author}{\bibfnamefont{M.}~\bibnamefont{{Doi}}},
  \bibinfo{author}{\bibfnamefont{R.}~\bibnamefont{{Ellis}}},
  \bibnamefont{et~al.}, \bibinfo{journal}{\apj} \textbf{\bibinfo{volume}{598}},
  \bibinfo{pages}{102} (\bibinfo{year}{2003});
\bibinfo{author}{\bibfnamefont{P.~M.} \bibnamefont{{Garnavich}}},
  \bibinfo{author}{\bibfnamefont{S.}~\bibnamefont{{Jha}}},
  \bibinfo{author}{\bibfnamefont{P.}~\bibnamefont{{Challis}}},
  \bibinfo{author}{\bibfnamefont{A.}~\bibnamefont{{Clocchiatti}}},
  \bibinfo{author}{\bibfnamefont{A.}~\bibnamefont{{Diercks}}},
  \bibinfo{author}{\bibfnamefont{A.~V.} \bibnamefont{{Filippenko}}},
  \bibinfo{author}{\bibfnamefont{R.~L.} \bibnamefont{{Gilliland}}},
  \bibinfo{author}{\bibfnamefont{C.~J.} \bibnamefont{{Hogan}}},
  \bibinfo{author}{\bibfnamefont{R.~P.} \bibnamefont{{Kirshner}}},
  \bibinfo{author}{\bibfnamefont{B.}~\bibnamefont{{Leibundgut}}},
  \bibnamefont{et~al.}, \bibinfo{journal}{\apj} \textbf{\bibinfo{volume}{509}},
  \bibinfo{pages}{74} (\bibinfo{year}{1998}).




\bibitem{dynamic_de1}
\bibinfo{author}{\bibfnamefont{T.}~\bibnamefont{{Padmanabhan}}}
  \bibnamefont{and} \bibinfo{author}{\bibfnamefont{T.~R.}
  \bibnamefont{{Choudhury}}}, \bibinfo{journal}{\mnras}
  \textbf{\bibinfo{volume}{344}}, \bibinfo{pages}{823} (\bibinfo{year}{2003})
  \eprint{arXiv:astro-ph/0212573}.
\bibitem{2004MNRAS.354..275A}
\bibinfo{author}{\bibfnamefont{U.}~\bibnamefont{{Alam}}},
  \bibinfo{author}{\bibfnamefont{V.}~\bibnamefont{{Sahni}}},
  \bibinfo{author}{\bibfnamefont{T.}~\bibnamefont{{Deep Saini}}},
  \bibnamefont{and} \bibinfo{author}{\bibfnamefont{A.~A.}
  \bibnamefont{{Starobinsky}}}, \bibinfo{journal}{\mnras}
  \textbf{\bibinfo{volume}{344}}, \bibinfo{pages}{1057}
  (\bibinfo{year}{2003}); 
\bibinfo{author}{\bibfnamefont{U.}~\bibnamefont{{Alam}}},
  \bibinfo{author}{\bibfnamefont{V.}~\bibnamefont{{Sahni}}}, \bibnamefont{and}
  \bibinfo{author}{\bibfnamefont{A.~A.} \bibnamefont{{Starobinsky}}},
  \bibinfo{journal}{JCAP}
  \textbf{\bibinfo{volume}{6}}, \bibinfo{pages}{8} (\bibinfo{year}{2004}).
\bibitem{dynamic_de5}
\bibinfo{author}{\bibfnamefont{Y.}~\bibnamefont{{Wang}}} \bibnamefont{and}
  \bibinfo{author}{\bibfnamefont{M.}~\bibnamefont{{Tegmark}}},
  \bibinfo{journal}{Phys. Rev.  Letts. } \textbf{\bibinfo{volume}{92}},
  \bibinfo{pages}{241302} (\bibinfo{year}{2004}); 
\bibinfo{author}{\bibfnamefont{Y.}~\bibnamefont{{Wang}}}, \bibinfo{journal}{New
  Astronomy Review} \textbf{\bibinfo{volume}{49}}, \bibinfo{pages}{97}
  (\bibinfo{year}{2005}).
\bibitem{2004PhRvD..70j3523E}
\bibinfo{author}{\bibfnamefont{D.}~\bibnamefont{{Eisenstein}}}
  \bibnamefont{and} \bibinfo{author}{\bibfnamefont{M.}~\bibnamefont{{White}}},
  \bibinfo{journal}{\prd} \textbf{\bibinfo{volume}{70}},
  \bibinfo{pages}{103523} (\bibinfo{year}{2004}).
\bibitem{2005astro.ph..1171E}
\bibinfo{author}{\bibfnamefont{D.~J.} \bibnamefont{{Eisenstein}}},
  \bibinfo{author}{\bibfnamefont{I.}~\bibnamefont{{Zehavi}}},
  \bibinfo{author}{\bibfnamefont{D.~W.} \bibnamefont{{Hogg}}},
  \bibinfo{author}{\bibfnamefont{R.}~\bibnamefont{{Scoccimarro}}},
  \bibinfo{author}{\bibfnamefont{M.~R.} \bibnamefont{{Blanton}}},
  \bibinfo{author}{\bibfnamefont{R.~C.} \bibnamefont{{Nichol}}},
  \bibinfo{author}{\bibfnamefont{R.}~\bibnamefont{{Scranton}}},
  \bibinfo{author}{\bibfnamefont{H.}~\bibnamefont{{Seo}}},
  \bibinfo{author}{\bibfnamefont{M.}~\bibnamefont{{Tegmark}}},
  \bibinfo{author}{\bibfnamefont{Z.}~\bibnamefont{{Zheng}}},
  \bibnamefont{et~al.}, \bibinfo{journal}{ }
  (\bibinfo{year}{2005}), \eprint{arXiv:astro-ph/0501171}.
\bibitem{2002PhRvD..65f3001H}
\bibinfo{author}{\bibfnamefont{D.}~\bibnamefont{{Huterer}}},
  \bibinfo{journal}{\prd} \textbf{\bibinfo{volume}{65}},
  \bibinfo{pages}{063001} (\bibinfo{year}{2002}); 
\bibinfo{author}{\bibfnamefont{M.}~\bibnamefont{{Bartelmann}}},
  \bibinfo{author}{\bibfnamefont{F.}~\bibnamefont{{Perrotta}}},
  \bibnamefont{and}
  \bibinfo{author}{\bibfnamefont{C.}~\bibnamefont{{Baccigalupi}}},
  \bibinfo{journal}{\aap} \textbf{\bibinfo{volume}{396}}, \bibinfo{pages}{21}
  (\bibinfo{year}{2002}).
\bibitem{2003ApJ...583..566M}
\bibinfo{author}{\bibfnamefont{D.}~\bibnamefont{{Munshi}}} \bibnamefont{and}
  \bibinfo{author}{\bibfnamefont{Y.}~\bibnamefont{{Wang}}},
  \bibinfo{journal}{\apj} \textbf{\bibinfo{volume}{583}}, \bibinfo{pages}{566}
  (\bibinfo{year}{2003}).
\bibitem{2003MNRAS.341..251W}
\bibinfo{author}{\bibfnamefont{N.~N.} \bibnamefont{{Weinberg}}}
  \bibnamefont{and}
  \bibinfo{author}{\bibfnamefont{M.}~\bibnamefont{{Kamionkowski}}},
  \bibinfo{journal}{\mnras} \textbf{\bibinfo{volume}{341}},
  \bibinfo{pages}{251} (\bibinfo{year}{2003}).
\bibitem{2003PhRvL..91d1301A}
\bibinfo{author}{\bibfnamefont{K.}~\bibnamefont{{Abazajian}}} \bibnamefont{and}
  \bibinfo{author}{\bibfnamefont{S.}~\bibnamefont{{Dodelson}}},
  \bibinfo{journal}{Phys. Rev.  Letts. } \textbf{\bibinfo{volume}{91}},
  \bibinfo{pages}{041301} (\bibinfo{year}{2003}).
\bibitem{2003PhRvL..91n1302J}
\bibinfo{author}{\bibfnamefont{B.}~\bibnamefont{{Jain}}} \bibnamefont{and}
  \bibinfo{author}{\bibfnamefont{A.}~\bibnamefont{{Taylor}}},
  \bibinfo{journal}{Phys. Rev.  Letts. } \textbf{\bibinfo{volume}{91}},
  \bibinfo{pages}{141302} (\bibinfo{year}{2003}); 
\bibinfo{author}{\bibfnamefont{G.}~\bibnamefont{{Bernstein}}} \bibnamefont{and}
  \bibinfo{author}{\bibfnamefont{B.}~\bibnamefont{{Jain}}},
  \bibinfo{journal}{\apj} \textbf{\bibinfo{volume}{600}}, \bibinfo{pages}{17}
  (\bibinfo{year}{2004}); 
\bibinfo{author}{\bibfnamefont{R.}~\bibnamefont{{Massey}}},
  \bibinfo{author}{\bibfnamefont{A.}~\bibnamefont{{Refregier}}},
  \bibnamefont{and} \bibinfo{author}{\bibfnamefont{J.}~\bibnamefont{{Rhodes}}},
  \bibinfo{journal}{ }  (\bibinfo{year}{2004}),
  \eprint{arXiv:astro-ph/0403229}; 
\bibinfo{author}{\bibfnamefont{L.}~\bibnamefont{{Knox}}},
  \bibinfo{author}{\bibfnamefont{A.}~\bibnamefont{{Albrecht}}},
  \bibnamefont{and} \bibinfo{author}{\bibfnamefont{Y.~S.}
  \bibnamefont{{Song}}}, \bibinfo{journal}{ }
  (\bibinfo{year}{2004}), \eprint{arXiv:astro-ph/0408141}.
\bibitem{2001PhRvD..64h3501B}
\bibinfo{author}{\bibfnamefont{K.}~\bibnamefont{{Benabed}}} \bibnamefont{and}
  \bibinfo{author}{\bibfnamefont{F.}~\bibnamefont{{Bernardeau}}},
  \bibinfo{journal}{\prd} \textbf{\bibinfo{volume}{64}},
  \bibinfo{pages}{083501} (\bibinfo{year}{2001}); 
\bibinfo{author}{\bibfnamefont{L.}~\bibnamefont{{Amendola}}},
  \bibinfo{journal}{\prd} \textbf{\bibinfo{volume}{69}},
  \bibinfo{pages}{103524} (\bibinfo{year}{2004}); 
\bibinfo{author}{\bibfnamefont{S.}~\bibnamefont{{Dedeo}}},
  \bibinfo{author}{\bibfnamefont{R.~R.} \bibnamefont{{Caldwell}}},
  \bibnamefont{and} \bibinfo{author}{\bibfnamefont{P.~J.}
  \bibnamefont{{Steinhardt}}}, \bibinfo{journal}{\prd}
  \textbf{\bibinfo{volume}{67}}, \bibinfo{pages}{103509}
  (\bibinfo{year}{2003}).
\bibitem{1999ApJ...517...40B}
\bibinfo{author}{\bibfnamefont{S.}~\bibnamefont{{Borgani}}},
  \bibinfo{author}{\bibfnamefont{P.}~\bibnamefont{{Rosati}}},
  \bibinfo{author}{\bibfnamefont{P.}~\bibnamefont{{Tozzi}}}, \bibnamefont{and}
  \bibinfo{author}{\bibfnamefont{C.}~\bibnamefont{{Norman}}},
  \bibinfo{journal}{\apj} \textbf{\bibinfo{volume}{517}}, \bibinfo{pages}{40}
  (\bibinfo{year}{1999}).
\bibitem{cluster2}
\bibinfo{author}{\bibfnamefont{P.~T.~P.} \bibnamefont{{Viana}}}
  \bibnamefont{and} \bibinfo{author}{\bibfnamefont{A.~R.}
  \bibnamefont{{Liddle}}}, \bibinfo{journal}{\mnras}
  \textbf{\bibinfo{volume}{281}}, \bibinfo{pages}{323} (\bibinfo{year}{1996}); 
\bibinfo{author}{\bibfnamefont{T.}~\bibnamefont{{Kitayama}}} \bibnamefont{and}
  \bibinfo{author}{\bibfnamefont{Y.}~\bibnamefont{{Suto}}},
  \bibinfo{journal}{\apj} \textbf{\bibinfo{volume}{469}}, \bibinfo{pages}{480}
  (\bibinfo{year}{1996}).
\bibitem{cluster4}
\bibinfo{author}{\bibfnamefont{E.}~\bibnamefont{{Rasia}}},
  \bibinfo{author}{\bibfnamefont{P.}~\bibnamefont{{Mazzotta}}},
  \bibinfo{author}{\bibfnamefont{S.}~\bibnamefont{{Borgani}}},
  \bibinfo{author}{\bibfnamefont{L.}~\bibnamefont{{Moscardini}}},
  \bibinfo{author}{\bibfnamefont{K.}~\bibnamefont{{Dolag}}},
  \bibinfo{author}{\bibfnamefont{G.}~\bibnamefont{{Tormen}}},
  \bibinfo{author}{\bibfnamefont{A.}~\bibnamefont{{Diaferio}}},
  \bibnamefont{and}
  \bibinfo{author}{\bibfnamefont{G.}~\bibnamefont{{Murante}}},
  \bibinfo{journal}{\apjl} \textbf{\bibinfo{volume}{618}}, \bibinfo{pages}{L1}
  (\bibinfo{year}{2005}).
\bibitem{2004astro.ph..8252V}
\bibinfo{author}{\bibfnamefont{P.}~\bibnamefont{{Vielva}}},
  \bibinfo{author}{\bibfnamefont{E.}~\bibnamefont{{Martinez-Gonzalez}}},
  \bibnamefont{and} \bibinfo{author}{\bibfnamefont{M.}~\bibnamefont{{Tucci}}},
  \bibinfo{journal}{ }  (\bibinfo{year}{2004}),
  \eprint{arXiv:astro-ph/0408252}; 
\bibinfo{author}{\bibfnamefont{P.~S.} \bibnamefont{{Corasaniti}}},
  \bibinfo{author}{\bibfnamefont{B.~A.} \bibnamefont{{Bassett}}},
  \bibinfo{author}{\bibfnamefont{C.}~\bibnamefont{{Ungarelli}}},
  \bibnamefont{and} \bibinfo{author}{\bibfnamefont{E.~J.}
  \bibnamefont{{Copeland}}}, \bibinfo{journal}{Phys. Rev.  Letts. }
  \textbf{\bibinfo{volume}{90}}, \bibinfo{pages}{091303}
  (\bibinfo{year}{2003}); 
\bibinfo{author}{\bibfnamefont{P.~S.} \bibnamefont{{Corasaniti}}},
  \bibinfo{author}{\bibfnamefont{M.}~\bibnamefont{{Kunz}}},
  \bibinfo{author}{\bibfnamefont{D.}~\bibnamefont{{Parkinson}}},
  \bibinfo{author}{\bibfnamefont{E.~J.} \bibnamefont{{Copeland}}},
  \bibnamefont{and} \bibinfo{author}{\bibfnamefont{B.~A.}
  \bibnamefont{{Bassett}}}, \bibinfo{journal}{\prd}
  \textbf{\bibinfo{volume}{70}}, \bibinfo{pages}{083006}
  (\bibinfo{year}{2004}); 
\bibinfo{author}{\bibfnamefont{B.}~\bibnamefont{{Gold}}},
  \bibinfo{journal}{\prd} \textbf{\bibinfo{volume}{71}},
  \bibinfo{pages}{063522} (\bibinfo{year}{2005}).
\bibitem{2004astro.ph..9207Y}
\bibinfo{author}{\bibfnamefont{K.}~\bibnamefont{{Yamamoto}}},
  \bibinfo{author}{\bibfnamefont{B.~A.} \bibnamefont{{Bassett}}},
  \bibnamefont{and}
  \bibinfo{author}{\bibfnamefont{H.}~\bibnamefont{{Nishioka}}},
  \bibinfo{journal}{Phys. Rev.  Letts. } \textbf{\bibinfo{volume}{94}},
  \bibinfo{pages}{051301} (\bibinfo{year}{2005}); 
\bibinfo{author}{\bibfnamefont{T.}~\bibnamefont{{Matsubara}}} \bibnamefont{and}
  \bibinfo{author}{\bibfnamefont{A.~S.} \bibnamefont{{Szalay}}},
  \bibinfo{journal}{Phys. Rev.  Letts. } \textbf{\bibinfo{volume}{90}},
  \bibinfo{pages}{021302} (\bibinfo{year}{2003}).
\bibitem{Calvao:2002zr}
\bibinfo{author}{\bibfnamefont{M.~O.} \bibnamefont{{Calv{\~ a}o}}},
  \bibinfo{author}{\bibfnamefont{J.~R.} \bibnamefont{{de Mello Neto}}},
  \bibnamefont{and} \bibinfo{author}{\bibfnamefont{I.}~\bibnamefont{{Waga}}},
  \bibinfo{journal}{Phys. Rev.  Letts. } \textbf{\bibinfo{volume}{88}},
  \bibinfo{pages}{091302} (\bibinfo{year}{2002}).
\bibitem{lss_de}
\bibinfo{author}{\bibfnamefont{S.}~\bibnamefont{{Perlmutter}}},
  \bibinfo{author}{\bibfnamefont{M.~S.} \bibnamefont{{Turner}}},
  \bibnamefont{and} \bibinfo{author}{\bibfnamefont{M.}~\bibnamefont{{White}}},
  \bibinfo{journal}{Phys. Rev.  Letts. } \textbf{\bibinfo{volume}{83}},
  \bibinfo{pages}{670} (\bibinfo{year}{1999}{\natexlab{b}}).
\bibitem{1979Natur.281..358A}
\bibinfo{author}{\bibfnamefont{C.}~\bibnamefont{{Alcock}}} \bibnamefont{and}
  \bibinfo{author}{\bibfnamefont{B.}~\bibnamefont{{Paczynski}}},
  \bibinfo{journal}{\nat} \textbf{\bibinfo{volume}{281}}, \bibinfo{pages}{358}
  (\bibinfo{year}{1979}).
\bibitem{2003NewAR..47..775W}
\bibinfo{author}{\bibfnamefont{J.}~\bibnamefont{{Weller}}} \bibnamefont{and}
  \bibinfo{author}{\bibfnamefont{R.~A.} \bibnamefont{{Battye}}},
  \bibinfo{journal}{New Astronomy Review} \textbf{\bibinfo{volume}{47}},
  \bibinfo{pages}{775} (\bibinfo{year}{2003}); 
\bibinfo{author}{\bibfnamefont{J.~J.} \bibnamefont{{Mohr}}},
  \bibinfo{journal}{ }  (\bibinfo{year}{2004}),
  \eprint{arXiv:astro-ph/0408484}; 
\bibinfo{author}{\bibfnamefont{J.}~\bibnamefont{{Weller}}},
  \bibinfo{author}{\bibfnamefont{R.~A.} \bibnamefont{{Battye}}},
  \bibnamefont{and}
  \bibinfo{author}{\bibfnamefont{R.}~\bibnamefont{{Kneissl}}},
  \bibinfo{journal}{Phys. Rev.  Letts. } \textbf{\bibinfo{volume}{88}},
  \bibinfo{pages}{231301} (\bibinfo{year}{2002}).
\bibitem{isw0}
\bibinfo{author}{\bibfnamefont{H.~V.} \bibnamefont{{Peiris}}} \bibnamefont{and}
  \bibinfo{author}{\bibfnamefont{D.~N.} \bibnamefont{{Spergel}}},
  \bibinfo{journal}{\apj} \textbf{\bibinfo{volume}{540}}, \bibinfo{pages}{605}
  (\bibinfo{year}{2000}); 
\bibinfo{author}{\bibfnamefont{P.}~\bibnamefont{{Fosalba}}},
  \bibinfo{author}{\bibfnamefont{E.}~\bibnamefont{{Gazta{\~ n}aga}}},
  \bibnamefont{and} \bibinfo{author}{\bibfnamefont{F.~J.}
  \bibnamefont{{Castander}}}, \bibinfo{journal}{\apjl}
  \textbf{\bibinfo{volume}{597}}, \bibinfo{pages}{L89} (\bibinfo{year}{2003}); 
\bibinfo{author}{\bibfnamefont{R.}~\bibnamefont{{Scranton}}},
  \bibinfo{author}{\bibfnamefont{A.~J.} \bibnamefont{{Connolly}}},
  \bibinfo{author}{\bibfnamefont{R.~C.} \bibnamefont{{Nichol}}},
  \bibinfo{author}{\bibfnamefont{A.}~\bibnamefont{{Stebbins}}},
  \bibinfo{author}{\bibfnamefont{I.}~\bibnamefont{{Szapudi}}},
  \bibinfo{author}{\bibfnamefont{D.~J.} \bibnamefont{{Eisenstein}}},
  \bibinfo{author}{\bibfnamefont{N.}~\bibnamefont{{Afshordi}}},
  \bibinfo{author}{\bibfnamefont{T.}~\bibnamefont{{Budavari}}},
  \bibinfo{author}{\bibfnamefont{I.}~\bibnamefont{{Csabai}}},
  \bibinfo{author}{\bibfnamefont{J.~A.} \bibnamefont{{Frieman}}},
  \bibnamefont{et~al.}, \bibinfo{journal}{ }
  (\bibinfo{year}{2003}), \eprint{arXiv:astro-ph/0307335}; 
\bibinfo{author}{\bibfnamefont{N.}~\bibnamefont{{Afshordi}}},
  \bibinfo{journal}{\prd} \textbf{\bibinfo{volume}{70}},
  \bibinfo{pages}{083536} (\bibinfo{year}{2004}); 
\bibinfo{author}{\bibfnamefont{N.}~\bibnamefont{{Padmanabhan}}},
  \bibinfo{author}{\bibfnamefont{C.~M.} \bibnamefont{{Hirata}}},
  \bibinfo{author}{\bibfnamefont{U.}~\bibnamefont{{Seljak}}},
  \bibinfo{author}{\bibfnamefont{D.}~\bibnamefont{{Schlegel}}},
  \bibinfo{author}{\bibfnamefont{J.}~\bibnamefont{{Brinkmann}}},
  \bibnamefont{and} \bibinfo{author}{\bibfnamefont{D.~P.}
  \bibnamefont{{Schneider}}}, \bibinfo{journal}{ }
  (\bibinfo{year}{2004}), \eprint{arXiv:astro-ph/0410360}; 
\bibinfo{author}{\bibfnamefont{L.}~\bibnamefont{{Pogosian}}}, 
  \bibinfo{author}{\bibfnamefont{P.~S.~} \bibnamefont{{Corasaniti}}}, 
  \bibinfo{author}{\bibfnamefont{C.}~\bibnamefont{{Stephan-Otto}}}, 
  \bibinfo{author}{\bibfnamefont{R.}~\bibnamefont{{Crittenden}}}
  \bibnamefont{and} \bibinfo{author}{\bibfnamefont{R.}~\bibnamefont{{Nichol}}},
 \bibinfo{journal}{ }
  (\bibinfo{year}{2005}), \eprint{arXiv:astro-ph/0506396}.
\bibitem{1992MNRAS.258P...1E}
\bibinfo{author}{\bibfnamefont{G.}~\bibnamefont{{Efstathiou}}},
  \bibinfo{author}{\bibfnamefont{J.~R.} \bibnamefont{{Bond}}},
  \bibnamefont{and} \bibinfo{author}{\bibfnamefont{S.~D.~M.}
  \bibnamefont{{White}}}, \bibinfo{journal}{\mnras}
  \textbf{\bibinfo{volume}{258}}, \bibinfo{pages}{1P} (\bibinfo{year}{1992}).
\bibitem{constraints_3}
\bibinfo{author}{\bibfnamefont{D.}~\bibnamefont{{Huterer}}} \bibnamefont{and}
  \bibinfo{author}{\bibfnamefont{A.}~\bibnamefont{{Cooray}}},
  \bibinfo{journal}{\prd} \textbf{\bibinfo{volume}{71}},
  \bibinfo{pages}{023506} (\bibinfo{year}{2005}); 
\bibinfo{author}{\bibfnamefont{T.}~\bibnamefont{{Multam{\" a}ki}}},
  \bibinfo{author}{\bibfnamefont{M.}~\bibnamefont{{Manera}}}, \bibnamefont{and}
  \bibinfo{author}{\bibfnamefont{E.}~\bibnamefont{{Gazta{\~ n}aga}}},
  \bibinfo{journal}{\prd} \textbf{\bibinfo{volume}{69}},
  \bibinfo{pages}{023004} (\bibinfo{year}{2004}); 
\bibinfo{author}{\bibfnamefont{A.~V.} \bibnamefont{{Macci{\` o}}}},
  \bibinfo{author}{\bibfnamefont{S.~A.} \bibnamefont{{Bonometto}}},
  \bibinfo{author}{\bibfnamefont{R.}~\bibnamefont{{Mainini}}},
  \bibnamefont{and} \bibinfo{author}{\bibfnamefont{A.}~\bibnamefont{{Klypin}}},
  in \emph{\bibinfo{booktitle}{Multiwavelength Cosmology. Ed. Manolis Plionis.; 
   Kluwer Academic Publishers, Dordrecht, The Netherlands, 2004, p.199}}
  (\bibinfo{year}{2004}), pp. \bibinfo{pages}{199--+}; 
\bibinfo{author}{\bibfnamefont{E.~V.} \bibnamefont{{Linder}}} \bibnamefont{and}
  \bibinfo{author}{\bibfnamefont{A.}~\bibnamefont{{Jenkins}}},
  \bibinfo{journal}{\mnras} \textbf{\bibinfo{volume}{346}},
  \bibinfo{pages}{573} (\bibinfo{year}{2003}); 
\bibinfo{author}{\bibfnamefont{D.}~\bibnamefont{{Pogosyan}}},
  \bibinfo{author}{\bibfnamefont{J.~R.} \bibnamefont{{Bond}}},
  \bibnamefont{and} \bibinfo{author}{\bibfnamefont{C.~R.}
  \bibnamefont{{Contaldi}}}, \bibinfo{journal}{ }
  (\bibinfo{year}{2003}), \eprint{arXiv:astro-ph/0301310}; 
\bibinfo{author}{\bibfnamefont{W.}~\bibnamefont{{Lee}}} \bibnamefont{and}
  \bibinfo{author}{\bibfnamefont{K.}~\bibnamefont{{Ng}}},
  \bibinfo{journal}{\prd} \textbf{\bibinfo{volume}{67}},
  \bibinfo{pages}{107302} (\bibinfo{year}{2003}); 
\bibinfo{author}{\bibfnamefont{D.}~\bibnamefont{{Lee}}},
  \bibinfo{author}{\bibfnamefont{W.}~\bibnamefont{{Lee}}}, \bibnamefont{and}
  \bibinfo{author}{\bibfnamefont{K.}~\bibnamefont{{Ng}}},
  \bibinfo{journal}{Int.Jour.Mod.Phys.D}
  \textbf{\bibinfo{volume}{14}}, \bibinfo{pages}{335} (\bibinfo{year}{2005}); 
\bibinfo{author}{\bibfnamefont{R.}~\bibnamefont{{Mainini}}},
  \bibinfo{author}{\bibfnamefont{L.~P.~L.} \bibnamefont{{Colombo}}},
  \bibnamefont{and} \bibinfo{author}{\bibfnamefont{S.~A.}
  \bibnamefont{{Bonometto}}}, \bibinfo{journal}{ }
  (\bibinfo{year}{2005}), \eprint{arXiv:astro-ph/0503036}; 
\bibinfo{author}{\bibfnamefont{L.}~\bibnamefont{{Pogosian}}},
  \bibinfo{journal}{JCAP}
  \textbf{\bibinfo{volume}{4}}, \bibinfo{pages}{15} (\bibinfo{year}{2005}); 
\bibinfo{author}{\bibfnamefont{D.}~\bibnamefont{{Rapetti}}},
  \bibinfo{author}{\bibfnamefont{S.~W.} \bibnamefont{{Allen}}},
  \bibnamefont{and} \bibinfo{author}{\bibfnamefont{J.}~\bibnamefont{{Weller}}},
  \bibinfo{journal}{\mnras} \textbf{\bibinfo{volume}{360}},
  \bibinfo{pages}{555} (\bibinfo{year}{2005}); 
\bibinfo{author}{\bibfnamefont{J.}~\bibnamefont{{Shen}}},
  \bibinfo{author}{\bibfnamefont{B.}~\bibnamefont{{Wang}}},
  \bibinfo{author}{\bibfnamefont{E.}~\bibnamefont{{Abdalla}}},
  \bibnamefont{and} \bibinfo{author}{\bibfnamefont{R.}~\bibnamefont{{Su}}},
  \bibinfo{journal}{Phys. Letts.  B} \textbf{\bibinfo{volume}{609}},
  \bibinfo{pages}{200} (\bibinfo{year}{2005}); 
\bibinfo{author}{\bibfnamefont{F.}~\bibnamefont{{Giovi}}},
  \bibinfo{author}{\bibfnamefont{C.}~\bibnamefont{{Baccigalupi}}},
  \bibnamefont{and}
  \bibinfo{author}{\bibfnamefont{F.}~\bibnamefont{{Perrotta}}},
  \bibinfo{journal}{\prd} \textbf{\bibinfo{volume}{71}},
  \bibinfo{pages}{103009} (\bibinfo{year}{2005}); 
\bibinfo{author}{\bibfnamefont{G.}~\bibnamefont{{Chen}}} \bibnamefont{and}
  \bibinfo{author}{\bibfnamefont{B.}~\bibnamefont{{Ratra}}},
  \bibinfo{journal}{\apjl} \textbf{\bibinfo{volume}{612}}, \bibinfo{pages}{L1}
  (\bibinfo{year}{2004}); 
\bibinfo{author}{\bibfnamefont{P.~S.} \bibnamefont{{Corasaniti}}},
  \bibinfo{author}{\bibfnamefont{T.}~\bibnamefont{{Giannantonio}}},
  \bibnamefont{and}
  \bibinfo{author}{\bibfnamefont{A.}~\bibnamefont{{Melchiorri}}},
  \bibinfo{journal}{ }  (\bibinfo{year}{2005}),
  \eprint{arXiv:astro-ph/0504115}; 
\bibinfo{author}{\bibfnamefont{R.}~\bibnamefont{{Jimenez}}},
  \bibinfo{author}{\bibfnamefont{L.}~\bibnamefont{{Verde}}},
  \bibinfo{author}{\bibfnamefont{T.}~\bibnamefont{{Treu}}} \bibnamefont{and}
  \bibinfo{author}{\bibfnamefont{D.}~\bibnamefont{{Stern}}},
  \bibinfo{journal}{\apj} \textbf{\bibinfo{volume}{593}}, \bibinfo{pages}{622}
  (\bibinfo{year}{2003});
\bibinfo{author}{\bibfnamefont{L.}~\bibnamefont{{Perivolaropoulos}}},
  \bibinfo{journal}{ }  (\bibinfo{year}{2005}),
  \eprint{arXiv:astro-ph/0504582};
\bibinfo{author}{\bibfnamefont{L.}~\bibnamefont{{Perivolaropoulos}}},
  \bibinfo{journal}{\prd} \textbf{\bibinfo{volume}{71}},
  \bibinfo{pages}{063503}(\bibinfo{year}{2005});
\bibinfo{author}{\bibfnamefont{L.}~\bibnamefont{{Perivolaropoulos}}},
  \bibinfo{journal}{\prd} \textbf{\bibinfo{volume}{70}},
  \bibinfo{pages}{043531}(\bibinfo{year}{2004});
\bibinfo{author}{\bibfnamefont{W.}~\bibnamefont{{Godlowski}}}
\bibnamefont{and}
\bibinfo{author}{\bibfnamefont{M.}~\bibnamefont{{Szydlowski}}},
  \bibinfo{journal}{ }  (\bibinfo{year}{2005}),
  \eprint{arXiv:astro-ph/0507322};
\bibinfo{author}{\bibfnamefont{A.}~\bibnamefont{{Krawiec}}},
\bibinfo{author}{\bibfnamefont{M.}~\bibnamefont{{Szydlowski}}},
\bibnamefont{and}
\bibinfo{author}{\bibfnamefont{W.}~\bibnamefont{{Godlowski}}},
  \bibinfo{journal}{Phys. Lett. B} \textbf{\bibinfo{volume}{619}},
  \bibinfo{pages}{219}(\bibinfo{year}{2005}).




\bibitem{cmbrev1}
\bibinfo{author}{\bibfnamefont{W.}~\bibnamefont{{Hu}}} \bibnamefont{and}
  \bibinfo{author}{\bibfnamefont{S.}~\bibnamefont{{Dodelson}}},
  \bibinfo{journal}{\araa} \textbf{\bibinfo{volume}{40}}, \bibinfo{pages}{171}
  (\bibinfo{year}{2002}).
\bibitem[{\citenamefont{{White} and {Cohn}}(2002)}]{cmbrev2}
\bibinfo{author}{\bibfnamefont{M.}~\bibnamefont{{White}}} \bibnamefont{and}
  \bibinfo{author}{\bibfnamefont{J.~D.} \bibnamefont{{Cohn}}},
  \bibinfo{journal}{ }  (\bibinfo{year}{2002}),
  \eprint{arXiv:astro-ph/0203120}; 
\bibinfo{author}{\bibfnamefont{K.}~\bibnamefont{{Subramanian}}},
  \bibinfo{journal}{ }  (\bibinfo{year}{2004}),
  \eprint{arXiv:astro-ph/0411049}; 
\bibinfo{author}{\bibfnamefont{M.}~\bibnamefont{{Giovannini}}},
  \bibinfo{journal}{Int.Jour.Mod.Phys.D}
  \textbf{\bibinfo{volume}{14}}, \bibinfo{pages}{363} (\bibinfo{year}{2005}).
\bibitem{wmap_lik}
\bibinfo{author}{\bibfnamefont{L.}~\bibnamefont{{Verde}}},
  \bibinfo{author}{\bibfnamefont{H.~V.} \bibnamefont{{Peiris}}},
  \bibinfo{author}{\bibfnamefont{D.~N.} \bibnamefont{{Spergel}}},
  \bibinfo{author}{\bibfnamefont{M.~R.} \bibnamefont{{Nolta}}},
  \bibinfo{author}{\bibfnamefont{C.~L.} \bibnamefont{{Bennett}}},
  \bibinfo{author}{\bibfnamefont{M.}~\bibnamefont{{Halpern}}},
  \bibinfo{author}{\bibfnamefont{G.}~\bibnamefont{{Hinshaw}}},
  \bibinfo{author}{\bibfnamefont{N.}~\bibnamefont{{Jarosik}}},
  \bibinfo{author}{\bibfnamefont{A.}~\bibnamefont{{Kogut}}},
  \bibinfo{author}{\bibfnamefont{M.}~\bibnamefont{{Limon}}},
  \bibnamefont{et~al.}, \bibinfo{journal}{\apjs}
  \textbf{\bibinfo{volume}{148}}, \bibinfo{pages}{195} (\bibinfo{year}{2003}).
\bibitem{1996ApJ...469..437S}
\bibinfo{author}{\bibfnamefont{U.}~\bibnamefont{{Seljak}}} \bibnamefont{and}
  \bibinfo{author}{\bibfnamefont{M.}~\bibnamefont{{Zaldarriaga}}},
  \bibinfo{journal}{\apj} \textbf{\bibinfo{volume}{469}}, \bibinfo{pages}{437}
  (\bibinfo{year}{1996}); http://www.cmbfast.org/
\bibitem{metropolis}
\bibinfo{author}{\bibfnamefont{N.}~\bibnamefont{{Metropolis}}},
  \bibinfo{author}{\bibfnamefont{A.~W.} \bibnamefont{{Rosenbluth}}},
  \bibinfo{author}{\bibfnamefont{M.~N.} \bibnamefont{{Rosenbluth}}},
  \bibinfo{author}{\bibfnamefont{A.~H.} \bibnamefont{{Teller}}},
  \bibnamefont{and} \bibinfo{author}{\bibfnamefont{E.}~\bibnamefont{{Teller}}},
  \bibinfo{journal}{J. Chem. Phys} \textbf{\bibinfo{volume}{21}},
  \bibinfo{pages}{1087} (\bibinfo{year}{1953}).
\bibitem{mcmc2}
\bibinfo{author}{\bibfnamefont{A.}~\bibnamefont{{Lewis}}} \bibnamefont{and}
  \bibinfo{author}{\bibfnamefont{S.}~\bibnamefont{{Bridle}}},
  \bibinfo{journal}{\prd} \textbf{\bibinfo{volume}{66}},
  \bibinfo{pages}{103511} (\bibinfo{year}{2002}).
\bibitem{convergence}
\bibinfo{author}{\bibfnamefont{A.}~\bibnamefont{{Gelman}}} \bibnamefont{and}
  \bibinfo{author}{\bibfnamefont{D.~B.} \bibnamefont{{Rubin}}},
  \bibinfo{journal}{Statistical Science} \textbf{\bibinfo{volume}{7}},
  \bibinfo{pages}{457} (\bibinfo{year}{1992}).
\bibitem{2001ApJ...553...47F}
\bibinfo{author}{\bibfnamefont{W.~L.} \bibnamefont{{Freedman}}},
  \bibinfo{author}{\bibfnamefont{B.~F.} \bibnamefont{{Madore}}},
  \bibinfo{author}{\bibfnamefont{B.~K.} \bibnamefont{{Gibson}}},
  \bibinfo{author}{\bibfnamefont{L.}~\bibnamefont{{Ferrarese}}},
  \bibinfo{author}{\bibfnamefont{D.~D.} \bibnamefont{{Kelson}}},
  \bibinfo{author}{\bibfnamefont{S.}~\bibnamefont{{Sakai}}},
  \bibinfo{author}{\bibfnamefont{J.~R.} \bibnamefont{{Mould}}},
  \bibinfo{author}{\bibfnamefont{R.~C.} \bibnamefont{{Kennicutt}}},
  \bibinfo{author}{\bibfnamefont{H.~C.} \bibnamefont{{Ford}}},
  \bibinfo{author}{\bibfnamefont{J.~A.} \bibnamefont{{Graham}}},
  \bibnamefont{et~al.}, \bibinfo{journal}{\apj} \textbf{\bibinfo{volume}{553}},
  \bibinfo{pages}{47} (\bibinfo{year}{2001}).
\bibitem{perturb3} \bibinfo{author}{\bibfnamefont{J.}~\bibnamefont{{Weller}}}
  \bibnamefont{and} 
  \bibinfo{author}{\bibfnamefont{A.~M.} \bibnamefont{{Lewis}}},
  \bibinfo{journal}{\mnras} \textbf{\bibinfo{volume}{346}},
  \bibinfo{pages}{987} (\bibinfo{year}{2003}).
\bibitem{2005astro.ph..1174C}
\bibinfo{author}{\bibfnamefont{S.}~\bibnamefont{{Cole}}},
  \bibinfo{author}{\bibfnamefont{W.~J.} \bibnamefont{{Percival}}},
  \bibinfo{author}{\bibfnamefont{J.~A.} \bibnamefont{{Peacock}}},
  \bibinfo{author}{\bibfnamefont{P.}~\bibnamefont{{Norberg}}},
  \bibinfo{author}{\bibfnamefont{C.~M.} \bibnamefont{{Baugh}}},
  \bibinfo{author}{\bibfnamefont{C.~S.} \bibnamefont{{Frenk}}},
  \bibinfo{author}{\bibfnamefont{I.}~\bibnamefont{{Baldry}}},
  \bibinfo{author}{\bibfnamefont{J.}~\bibnamefont{{Bland-Hawthorn}}},
  \bibinfo{author}{\bibfnamefont{T.}~\bibnamefont{{Bridges}}},
  \bibinfo{author}{\bibfnamefont{R.}~\bibnamefont{{Cannon}}},
  \bibnamefont{et~al.}, \bibinfo{journal}{ }
  (\bibinfo{year}{2005}), \eprint{arXiv:astro-ph/0501174}.
\bibitem{2003ApJ...591..599S}
\bibinfo{author}{\bibfnamefont{J.~L.} \bibnamefont{{Sievers}}},
  \bibinfo{author}{\bibfnamefont{J.~R.} \bibnamefont{{Bond}}},
  \bibinfo{author}{\bibfnamefont{J.~K.} \bibnamefont{{Cartwright}}},
  \bibinfo{author}{\bibfnamefont{C.~R.} \bibnamefont{{Contaldi}}},
  \bibinfo{author}{\bibfnamefont{B.~S.} \bibnamefont{{Mason}}},
  \bibinfo{author}{\bibfnamefont{S.~T.} \bibnamefont{{Myers}}},
  \bibinfo{author}{\bibfnamefont{S.}~\bibnamefont{{Padin}}},
  \bibinfo{author}{\bibfnamefont{T.~J.} \bibnamefont{{Pearson}}},
  \bibinfo{author}{\bibfnamefont{U.-L.} \bibnamefont{{Pen}}},
  \bibinfo{author}{\bibfnamefont{D.}~\bibnamefont{{Pogosyan}}},
  \bibnamefont{et~al.}, \bibinfo{journal}{\apj} \textbf{\bibinfo{volume}{591}},
  \bibinfo{pages}{599} (\bibinfo{year}{2003}); 
\bibinfo{author}{\bibfnamefont{J.~R.} \bibnamefont{{Bond}}},
  \bibinfo{author}{\bibfnamefont{C.~R.} \bibnamefont{{Contaldi}}},
  \bibinfo{author}{\bibfnamefont{U.-L.} \bibnamefont{{Pen}}},
  \bibinfo{author}{\bibfnamefont{D.}~\bibnamefont{{Pogosyan}}},
  \bibinfo{author}{\bibfnamefont{S.}~\bibnamefont{{Prunet}}},
  \bibinfo{author}{\bibfnamefont{M.~I.} \bibnamefont{{Ruetalo}}},
  \bibinfo{author}{\bibfnamefont{J.~W.} \bibnamefont{{Wadsley}}},
  \bibinfo{author}{\bibfnamefont{P.}~\bibnamefont{{Zhang}}},
  \bibinfo{author}{\bibfnamefont{B.~S.} \bibnamefont{{Mason}}},
  \bibinfo{author}{\bibfnamefont{S.~T.} \bibnamefont{{Myers}}},
  \bibnamefont{et~al.}, \bibinfo{journal}{\apj} \textbf{\bibinfo{volume}{626}},
  \bibinfo{pages}{12} (\bibinfo{year}{2005}).
\bibitem{2004PhRvD..69h3503B}
\bibinfo{author}{\bibfnamefont{R.}~\bibnamefont{{Bean}}} \bibnamefont{and}
  \bibinfo{author}{\bibfnamefont{O.}~\bibnamefont{{Dor{\' e}}}},
  \bibinfo{journal}{\prd} \textbf{\bibinfo{volume}{69}},
  \bibinfo{pages}{083503} (\bibinfo{year}{2004}); 
\bibinfo{author}{\bibfnamefont{R.~R.} \bibnamefont{{Caldwell}}},
  \bibinfo{author}{\bibfnamefont{R.}~\bibnamefont{{Dave}}}, \bibnamefont{and}
  \bibinfo{author}{\bibfnamefont{P.~J.} \bibnamefont{{Steinhardt}}},
  \bibinfo{journal}{Phys. Rev.  Letts. } \textbf{\bibinfo{volume}{80}},
  \bibinfo{pages}{1582} (\bibinfo{year}{1998}).
\bibitem{perturb4}
\bibinfo{author}{\bibfnamefont{S.}~\bibnamefont{{Hannestad}}},
  \bibinfo{journal}{\prd} \textbf{\bibinfo{volume}{71}},
  \bibinfo{pages}{103519} (\bibinfo{year}{2005}).
\bibitem{dynamic_de6}
\bibinfo{author}{\bibfnamefont{J.}~\bibnamefont{{J{\" o}nsson}}},
  \bibinfo{author}{\bibfnamefont{A.}~\bibnamefont{{Goobar}}},
  \bibinfo{author}{\bibfnamefont{R.}~\bibnamefont{{Amanullah}}},
  \bibnamefont{and} \bibinfo{author}{\bibfnamefont{L.}~\bibnamefont{{Bergstr{\"
  o}m}}}, \bibinfo{journal}{JCAP}
  \textbf{\bibinfo{volume}{9}}, \bibinfo{pages}{7} (\bibinfo{year}{2004}).
\bibitem{2003MNRAS.343..533D} 
\bibinfo{author}{\bibfnamefont{T.}~\bibnamefont{{Deep Saini}}},
  \bibinfo{author}{\bibfnamefont{T.}~\bibnamefont{{Padmanabhan}}}
 \bibnamefont{and} \bibinfo{author}{\bibfnamefont{S.}~\bibnamefont{{Bridle}}},
  \bibinfo{journal}{\mnras} \textbf{\bibinfo{volume}{343}},
  \bibinfo{pages}{533} (\bibinfo{year}{2003}).
\bibitem{2003ApJ...583...24K}
\bibinfo{author}{\bibfnamefont{M.}~\bibnamefont{{Kaplinghat}}},
  \bibinfo{author}{\bibfnamefont{M.}~\bibnamefont{{Chu}}},
  \bibinfo{author}{\bibfnamefont{Z.}~\bibnamefont{{Haiman}}},
  \bibinfo{author}{\bibfnamefont{G.~P.} \bibnamefont{{Holder}}},
  \bibinfo{author}{\bibfnamefont{L.}~\bibnamefont{{Knox}}}, \bibnamefont{and}
  \bibinfo{author}{\bibfnamefont{C.}~\bibnamefont{{Skordis}}},
  \bibinfo{journal}{\apj} \textbf{\bibinfo{volume}{583}}, \bibinfo{pages}{24}
  (\bibinfo{year}{2003}); 
\bibinfo{author}{\bibfnamefont{L.~P.~L.} \bibnamefont{{Colombo}}},
  \bibinfo{journal}{JCAP}
  \textbf{\bibinfo{volume}{3}}, \bibinfo{pages}{3} (\bibinfo{year}{2004}).

\bibitem{saul} 
\bibinfo{author}{\bibfnamefont{S.}~\bibnamefont{{Perlmutter}}}
 \bibnamefont{and}
 \bibinfo{author}{\bibfnamefont{B.~P.}~\bibnamefont{{Schmidt}}}, 
 \eprint{arXiv:astro-ph/0303428};
\bibinfo{author}{\bibfnamefont{S.}~\bibnamefont{{Nobili}}},
\bibinfo{author}{\bibfnamefont{R.}~\bibnamefont{{Amanullah}}},
\bibinfo{author}{\bibfnamefont{G.}~\bibnamefont{{Garavini}}},
\bibinfo{author}{\bibfnamefont{A.}~\bibnamefont{{Goobar}}},
\bibinfo{author}{\bibfnamefont{C.}~\bibnamefont{{Lidman}}},
\bibinfo{author}{\bibfnamefont{V.}~\bibnamefont{{Stanishev}}},
\bibinfo{author}{\bibfnamefont{G.}~\bibnamefont{{Aldering}}},
\bibinfo{author}{\bibfnamefont{P.}~\bibnamefont{{Antilogus}}},
\bibinfo{author}{\bibfnamefont{P.}~\bibnamefont{{Astier}}},
\bibinfo{author}{\bibfnamefont{M.~S.}~\bibnamefont{{Burns}}}
 \bibnamefont{et~al.}, \bibinfo{journal}{\aap} \textbf{\bibinfo{volume}{437}},
  \bibinfo{pages}{789} (\bibinfo{year}{2005}{\natexlab{a}}).

\end{thebibliography}
\end{document}